\documentclass[journal]{IEEEtran}
\usepackage{epsbox}
\usepackage[dvips]{graphicx,color}
\usepackage{latexsym}
\usepackage{array}
\begin{document}
\arraycolsep 0.5mm

\newcommand{\bfig}{\begin{figure}[t]}
\newcommand{\efig}{\end{figure}}
\setcounter{page}{1}
\newenvironment{indention}[1]{\par
\addtolength{\leftskip}{#1}\begingroup}{\endgroup\par}
%form: \begin{indention}{2.3cm}
%      \end{indention}
%
%namelist environment
%form: \begin{namelist}{width}
\newcommand{\namelistlabel}[1]{\mbox{#1}\hfill} 
\newenvironment{namelist}[1]{%
\begin{list}{}
{\let\makelabel\namelistlabel
\settowidth{\labelwidth}{#1}
\setlength{\leftmargin}{1.1\labelwidth}}
}{%
\end{list}}
%
%\input{ieeecom2e.tex}
%
%---------------------------- ieeecom2e.tex--------------------------
%
%
\newcommand{\bc}{\begin{center}}  %
\newcommand{\ec}{\end{center}}
\newcommand{\befi}{\begin{figure}[h]}  %
\newcommand{\enfi}{\end{figure}}
\newcommand{\bsb}{\begin{shadebox}\begin{center}}   %
\newcommand{\esb}{\end{center}\end{shadebox}}
\newcommand{\bs}{\begin{screen}}     %
\newcommand{\es}{\end{screen}}
\newcommand{\bib}{\begin{itembox}}   %
\newcommand{\eib}{\end{itembox}}
\newcommand{\bit}{\begin{itemize}}   %
\newcommand{\eit}{\end{itemize}}
\newcommand{\defeq}{\stackrel{\triangle}{=}}
\newcommand{\qed}{\hbox{\rule[-2pt]{3pt}{6pt}}}
\newcommand{\beq}{\begin{equation}}
\newcommand{\eeq}{\end{equation}}
\newcommand{\beqa}{\begin{eqnarray}}
\newcommand{\eeqa}{\end{eqnarray}}
\newcommand{\beqno}{\begin{eqnarray*}}
\newcommand{\eeqno}{\end{eqnarray*}}
\newcommand{\ba}{\begin{array}}
\newcommand{\ea}{\end{array}}
\newcommand{\vc}[1]{\mbox{\boldmath $#1$}}
\newcommand{\lvc}[1]{\mbox{\scriptsize \boldmath $#1$}}
\newcommand{\svc}[1]{\mbox{\scriptsize\boldmath $#1$}}

\newcommand{\wh}{\widehat}
\newcommand{\wt}{\widetilde}
\newcommand{\ts}{\textstyle}
\newcommand{\ds}{\displaystyle}
\newcommand{\scs}{\scriptstyle}
\newcommand{\vep}{\varepsilon}
\newcommand{\rhp}{\rightharpoonup}
\newcommand{\cl}{\circ\!\!\!\!\!-}
\newcommand{\bcs}{\dot{\,}.\dot{\,}}
\newcommand{\eqv}{\Leftrightarrow}
\newcommand{\leqv}{\Longleftrightarrow}
\newtheorem{co}{Corollary} 
\newtheorem{lm}{Lemma} 
\newtheorem{Ex}{Example} 
\newtheorem{Th}{Theorem}
\newtheorem{df}{Definition} 
\newtheorem{pr}{Property} 
\newtheorem{pro}{Proposition} 
\newtheorem{rem}{Remark} 

\newcommand{\lcv}{convex } 

\newcommand{\hugel}{{\arraycolsep 0mm
                    \left\{\ba{l}{\,}\\{\,}\ea\right.\!\!}}
\newcommand{\Hugel}{{\arraycolsep 0mm
                    \left\{\ba{l}{\,}\\{\,}\\{\,}\ea\right.\!\!}}
\newcommand{\HUgel}{{\arraycolsep 0mm
                    \left\{\ba{l}{\,}\\{\,}\\{\,}\vspace{-1mm}
                    \\{\,}\ea\right.\!\!}}
\newcommand{\huger}{{\arraycolsep 0mm
                    \left.\ba{l}{\,}\\{\,}\ea\!\!\right\}}}
\newcommand{\Huger}{{\arraycolsep 0mm
                    \left.\ba{l}{\,}\\{\,}\\{\,}\ea\!\!\right\}}}
\newcommand{\HUger}{{\arraycolsep 0mm
                    \left.\ba{l}{\,}\\{\,}\\{\,}\vspace{-1mm}
                    \\{\,}\ea\!\!\right\}}}

\newcommand{\hugebl}{{\arraycolsep 0mm
                    \left[\ba{l}{\,}\\{\,}\ea\right.\!\!}}
\newcommand{\Hugebl}{{\arraycolsep 0mm
                    \left[\ba{l}{\,}\\{\,}\\{\,}\ea\right.\!\!}}
\newcommand{\HUgebl}{{\arraycolsep 0mm
                    \left[\ba{l}{\,}\\{\,}\\{\,}\vspace{-1mm}
                    \\{\,}\ea\right.\!\!}}
\newcommand{\hugebr}{{\arraycolsep 0mm
                    \left.\ba{l}{\,}\\{\,}\ea\!\!\right]}}
\newcommand{\Hugebr}{{\arraycolsep 0mm
                    \left.\ba{l}{\,}\\{\,}\\{\,}\ea\!\!\right]}}
\newcommand{\HUgebr}{{\arraycolsep 0mm
                    \left.\ba{l}{\,}\\{\,}\\{\,}\vspace{-1mm}
                    \\{\,}\ea\!\!\right]}}

\newcommand{\hugecl}{{\arraycolsep 0mm
                    \left(\ba{l}{\,}\\{\,}\ea\right.\!\!}}
\newcommand{\Hugecl}{{\arraycolsep 0mm
                    \left(\ba{l}{\,}\\{\,}\\{\,}\ea\right.\!\!}}
\newcommand{\hugecr}{{\arraycolsep 0mm
                    \left.\ba{l}{\,}\\{\,}\ea\!\!\right)}}
\newcommand{\Hugecr}{{\arraycolsep 0mm
                    \left.\ba{l}{\,}\\{\,}\\{\,}\ea\!\!\right)}}

\newcommand{\hugepl}{{\arraycolsep 0mm
                    \left|\ba{l}{\,}\\{\,}\ea\right.\!\!}}
\newcommand{\Hugepl}{{\arraycolsep 0mm
                    \left|\ba{l}{\,}\\{\,}\\{\,}\ea\right.\!\!}}
\newcommand{\hugepr}{{\arraycolsep 0mm
                    \left.\ba{l}{\,}\\{\,}\ea\!\!\right|}}
\newcommand{\Hugepr}{{\arraycolsep 0mm
                    \left.\ba{l}{\,}\\{\,}\\{\,}\ea\!\!\right|}}

\newenvironment{jenumerate}
	{\begin{enumerate}\itemsep=-0.25em \parindent=1zw}{\end{enumerate}}
\newenvironment{jdescription}
	{\begin{description}\itemsep=-0.25em \parindent=1zw}{\end{description}}
\newenvironment{jitemize}
	{\begin{itemize}\itemsep=-0.25em \parindent=1zw}{\end{itemize}}
\renewcommand{\labelitemii}{$\cdot$}

\newcommand{\iro}[2]{{\color[named]{#1}#2\normalcolor}}
\newcommand{\irr}[1]{{\color[named]{Red}#1\normalcolor}}
\newcommand{\irg}[1]{{\color[named]{Green}#1\normalcolor}}
\newcommand{\irb}[1]{{\color[named]{Blue}#1\normalcolor}}
\newcommand{\irBl}[1]{{\color[named]{Black}#1\normalcolor}}
\newcommand{\irWh}[1]{{\color[named]{White}#1\normalcolor}}

\newcommand{\irY}[1]{{\color[named]{Yellow}#1\normalcolor}}
\newcommand{\irO}[1]{{\color[named]{Orange}#1\normalcolor}}
\newcommand{\irBr}[1]{{\color[named]{Purple}#1\normalcolor}}
\newcommand{\IrBr}[1]{{\color[named]{Purple}#1\normalcolor}}
\newcommand{\irBw}[1]{{\color[named]{Brown}#1\normalcolor}}
\newcommand{\irPk}[1]{{\color[named]{Magenta}#1\normalcolor}}
\newcommand{\irCb}[1]{{\color[named]{CadetBlue}#1\normalcolor}}
%\newcommand{\irDg}[1]{{\color[named]{DarkSlateGray}#1\normalcolor}}
%
%-----------------------notaion2.tex -----------------------------%
%
\newcommand{\lcov}{%lower
                     convex }

\newcommand{\cars}{s}
\newcommand{\sigNi}{\sigma_{N_i}^2}
\newcommand{\sigN}{\sigma_{N}^2}

\newcommand{\cef}{c_i}
\newcommand{\cefsq}{c_i^2}

\newcommand{\Iset}{L}
\newcommand{\Ntn}{\mbox{\boldmath $N$}}
\newcommand{\lNtn}{\mbox{\scriptsize\boldmath $N$}}
\newcommand{\Nitn}{\mbox{\boldmath $N$}_i}
\newcommand{\lNitn}{\mbox{\scriptsize\boldmath $N$}_i}
\newcommand{\Nost}{N_{0,t}}
\newcommand{\Nast}{N_{1,t}}
\newcommand{\Nlst}{N_{L,t}}

\newcommand{\tinbNs}{\tilde{N}%_S
                    }
\newcommand{\tiNs}{\tilde{\mbox{\boldmath $N$}}%_S
                  }
\newcommand{\tilNs}{\tilde{\mbox{\scriptsize\boldmath$N$}}%_S
                   }
\newcommand{\Xotn}{\mbox{\boldmath $X$}_0}
\newcommand{\lXotn}{\mbox{\scriptsize\boldmath $X$}_0}
\newcommand{\Xost}{X_{0,t}}

\newcommand{\Xatn}{\mbox{\boldmath $X$}_1}
\newcommand{\Xast}{X_{1,t}}

\newcommand{\Xbtn}{\mbox{\boldmath $X$}_2}
\newcommand{\Xbst}{X_{2,t}}

\newcommand{\Xltn}{\mbox{\boldmath $X$}_L}
\newcommand{\Xlst}{X_{L,t}}
\newcommand{\Xlsn}{X_{L,n}}

\newcommand{\Xitn}{\mbox{\boldmath $X$}_i}
\newcommand{\lXitn}{\mbox{\scriptsize\boldmath $X$}_i}
\newcommand{\Xisa}{X_{i,1}}
\newcommand{\Xisb}{X_{i,2}}
\newcommand{\Xist}{X_{i,t}}
\newcommand{\Xisn}{X_{i,n}}

\newcommand{\Xtn}{\mbox{\boldmath $X$}}

\newcommand{\hatXotn}{\hat{\mbox{\boldmath $X$}}_0}
\newcommand{\hatXost}{\hat{X}_{0,t}}

\newcommand{\tiXs}{%\tilde
                  {\mbox{\boldmath $Y$}}%_S
                  }
\newcommand{\tilXs}{%\tilde
                   {\mbox{\scriptsize\boldmath $Y$}}%_S
                   }

\newcommand{\Zatn}{\mbox{\boldmath $Z$}_1}
\newcommand{\lZatn}{\mbox{\scriptsize\boldmath $Z$}_1}
\newcommand{\Zbtn}{\mbox{\boldmath $Z$}_2}
\newcommand{\lZbtn}{\mbox{\scriptsize\boldmath $Z$}_2}

\newcommand{\rdf}{J}  

\newcommand{\Dff}{-}  
\newcommand{\coS}{S^{\rm c}} 

\newcommand{\piso}{\pi(k_1)} %{\pi_S(1)}
\newcommand{\pisi}{\pi(k_i)} %{\pi_S(i)}
\newcommand{\pisiadd}{\pi(k_{i+1})} %{\pi_S(i+1)}
\newcommand{\pisimo}{\pi(k_{i-1})} %{\pi_S(i-1)}
\newcommand{\pisj}{\pi(k_j)} %{\pi_S(j)}
\newcommand{\pisjmo}{\pi(k_{j-1})} %{\pi_S(j-1)}
\newcommand{\pisjadd}{\pi(k_{j+1})} %{\pi_S(j+1)}
\newcommand{\piss}{\pi(k_s)} %{\pi_S(s)}
\newcommand{\pissmo}{\pi(k_{s-1})} %{\pi_S(s-1)}
\newcommand{\pios}{\pi(S)} %{\pi_S}

\newcommand{\pibi}{\pi(B_i)} 
\newcommand{\picobi}{\pi(S\Dff B_{i})}
\newcommand{\pibimo}{\pi(B_{i-1})} 
\newcommand{\picobimo}{\pi(S\Dff B_{i-1})}
\newcommand{\pibj}{\pi(B_j)} 
\newcommand{\picobj}{\pi(S\Dff B_{j})}
\newcommand{\pibs}{\pi(B_s)} 
\newcommand{\pibsmo}{\pi(B_{s-1})} 

\newcommand{\maho}{many-help-one } %{Many-Help-One}
\newcommand{\Mho}{Many-help-one } %{Many-Help-One}
\newcommand{\oho}{one-helps-one } %{One-Helps-One}

\newcommand{\D}{\mbox{\rm d}} %
\newcommand{\E}{\mbox{\rm E}} %

\newcommand{\conv}{\mbox{\rm conv}} %
\newcommand{\rsub}{\empty}

\newcommand{\DisT}{\Sigma_D}
\newcommand{\Npre}{\preceq{\!\!\!\!\!|\:\:}}

\newcommand{\EP}[1]
{
\ts \frac{1}{2\pi{\rm e}}{\rm e}^{\scriptstyle \frac{2}{n}h(#1)}
}
\newcommand{\CdEP}[2]
{
\ts \frac{1}{2\pi{\rm e}}{\rm e}^{\frac{2}{n}
h(\scriptstyle #1|\scriptstyle #2)}
}
\newcommand{\MEq}[1]{\stackrel{%\mbox
{\rm (#1)}}{=}}

\newcommand{\MLeq}[1]{\stackrel{%\mbox
{\rm (#1)}}{\leq}}

\newcommand{\MGeq}[1]{\stackrel{%\mbox
{\rm (#1)}}{\geq}}

\newcommand{\MsL}[1]{\stackrel{%\mbox
{\rm (#1)}}{<}}

\newcommand{\MgL}[1]{\stackrel{%\mbox
{\rm (#1)}}{>}}

\newcommand{\MSub}[1]{\stackrel{%\mbox
{\rm (#1)}}{\subseteq}}

\newcommand{\MSup}[1]{\stackrel{%\mbox
{\rm (#1)}}{\supseteq}}

%                                                           %
%-----------------------------------------------------------%
%                                                           %
%\date{}
\title{
Distributed Source Coding 
of Correlated Gaussian Remote Sources %Observations
}
%Separate Coding 
%Rate Distortion Region for the Vector Gaussian CEO Problem
%
\author{Yasutada~Oohama%,~\IEEEmembership{Member,~IEEE,}
\thanks{Manuscript received xxx, 20XX; revised xxx, 20XX.}% 
\thanks{Y. Oohama is with the Department of Information Science 
        and Intelligent Systems, 
        University of Tokushima,
        2-1 Minami Josanjima-Cho, Tokushima 
        770-8506, Japan.}
}
\markboth{
%IEEE Transactions on Information Theory,~Vol.~XX,No.~Y, 
%~Month~20XX
}
{Oohama: Separate Source Coding of Correlated Gaussian Remote Sources
}
% *** Note that you probably will NOT want to include the author's name in ***
% *** the headers of peer review papers.                                   ***

% If you want to put a publisher's ID mark on the page
% (can leave text blank if you just want to see how the
% text height on the first page will be reduced by IEEE)
%\pubid{0000--0000/00\$00.00~\copyright~2002 IEEE}

%\specialpapernotice{(Invited Paper)}

\maketitle

\begin{abstract}
We consider the distributed source coding system for $L$ correlated
Gaussian remote sources $X_i, i=1,2,\cdots,L$, where
$X_i,i=1,2,\cdots, L$ are $L$ correlated Gaussian random variables. We
deal with the case where each of $L$ distributed encoders can not
directly observe $X_i$ but its noisy version $Y_i=X_i+N_i$.  Here
$N_i, i=1,2, \cdots, L$ are independent additive $L$ Gaussian noises
also independent of $X_i,i=1,2,\cdots, L$.  On this coding system the
determination problem of the rate distortion region remains open. In
this paper, we derive explicit outer and inner bounds of the rate
distortion region. We further find an explicit sufficient condition
for those two bounds to match. We also study the sum rate part of the rate
distortion region when the correlation has some symmetrical property
and derive a new lower bound of the sum rate part. We derive a
sufficient condition for this lower bound to be tight. The derived
sufficient condition depends only on the correlation property of the
sources and their observations.
%
%In the above setup $Y_i,
%i=1,2,\cdots, L$ can be regarded as correlated Gaussian observations
%of $X_i,i=1,2,\cdots,L$, respectively. This coding system can also be
%considered as a vector version of the Gaussian CEO problem investigated 
%by \cite{vb}, \cite{oh2}, and \cite{oh4}, where $X_i, i=1,2,\cdots, L$ 
%are identical.
%
%Let $X_i,i=1,2, \cdots, L$ be $L$ correlated Gaussian random variables
%and $N_i,$ $i=1,2,\cdots L$ be independent additive Gaussian random
%variables also independent of $X_i, i=1,2,\cdots, L$. We consider the
%case where for each $i=1,2,\cdots, L$, $Y_i=X_i+N_i$ is a noisy
%observation of $X_i$.  We consider the distributed source coding
%system for $L$ correlated Gaussian observations $Y_i, i=1,2,\cdots,
%L$.  
\end{abstract}


% Will be ignored for other modes.
\IEEEpeerreviewmaketitle

\section{Introduction}

In multi-user source networks distributed coding of correlated
information sources is a form of communication system which is
significant from both theoretical and practical point of view.  The
first fundamental theory in those coding systems was established by
Slepian and Wolf \cite{sw}. They considered a distributed source
coding system of two correlated information sources. Those two sources
are separately encoded and sent to a single destination, where the
decoder reconstruct the original sources. In this system, Slepian and
Wolf \cite{sw} determined the admissible rate region, the set that
consists of a pair of transmission rates for which two sources can be
decoded with an arbitrary small error probability.

In the above distributed source coding system we can consider the case
where the source outputs should be reconstructed with average
distortions smaller than prescribed levels. Such a situation suggests
the multiterminal rate-distortion theory.

%
%The rate-distortion theory for the separate coding system formulated by
%Slepian and Wolf has been studied by Wyner and Ziv \cite{wz}, Wyner
%\cite{w}, Berger \cite{bt}, Tung \cite{syt}, Berger {\it et al.} \cite{bh},
%Kaspi and Berger \cite{kb}, and Berger and Yeung \cite{by}. This problem,
%in general, remains an open problem and characterization of the rate
%distortion region has been unknown yet except for special cases. 
%

The rate distortion theory for the distributed source coding system
formulated by Slepian and Wolf has been studied by
\cite{wz}-\cite{oh1}. Recently, Wagner {\it et al.} \cite{wg3} have
given a complete solution in the case of Gaussian information sources
and mean squared distortion.

As a practical situation of the distributed source coding system, we
can consider a case where the separate encoders can not directly
observe the original source outputs but can observe their noisy
versions. This situation was first studied by Yamamoto and Ito
\cite{yam0}. Subsequently, a similar distributed source coding system
was studied by Flynn and R. M. Gray \cite{fg}.

In this paper we consider the distributed source coding system for $L$
correlated Gaussian remote sources $X_i, i=1,2,\cdots,L$, 
where $X_i,i=1,2,\cdots, L$ are $L$ correlated Gaussian 
random variables. We deal with
the case where each of $L$ distributed encoders can not directly 
observe $X_i$ but its noisy version $Y_i=X_i+N_i$. 
Here $N_i, i=1,2, \cdots, L$ are independent additive $L$ Gaussian noises
also independent of $X_i,i=1,2,\cdots, L$. In the above setup $Y_i,
i=1,2,\cdots, L$ can be regarded as correlated Gaussian observations
of $X_i,i=1,2,\cdots,L$, respectively. This coding system can also be
considered as a vector version of the Gaussian CEO problem investigated 
by \cite{vb}, \cite{oh2}, and \cite{oh4}, where $X_i, i=1,2,\cdots, L$ 
are identical.

The above distributed source coding system was first posed and
investigated by Pandya {\it et al. }\cite{pdya}. They derived upper
and lower bounds of the sum rate part of the rate distortion
region. Oohama \cite{oh5}, \cite{oh6} derived explicit outer and inner
bounds of the rate distortion region.  Wagner {\it et al.} \cite{wg3}
determined the rate distortion region in the case of $L=2$. 
%
%According
%to their result, the inner bound of Oohama \cite{oh5}, \cite{oh6} is
%tight in the case of $L=2$.  

In \cite{oh6}, Oohama also derived a
sufficient condition for his outer bound to coincide with the inner
bound.  Subsequently, Oohama \cite{oh7} derived a matching condition
which is simple and stronger than that of Oohama \cite{oh6}.

In this paper, we derive a new sufficient condition with respect to
the source correlation and the distortion 
under which the inner and outer bounds match. We show that
if the distortion is smaller than a threshold value which is a function 
of the source correlation, the inner and outer bounds
match and find an explicit form of this threshold value. 
This sufficient condition is a significant improvement of 
the condition derived by Oohama \cite{oh7}.
We also investigate the sum rate part of rate distortion region. The
optimal sum rate part of the outer bound derived by Oohama \cite{oh6}
serves as a lower bound of the sum rate part of the rate distortion
region. When the covariance matrix $\Sigma_{X^L}$ of 
the remote source $X^L=(X_1,X_2,\cdots,X_L)$ have a certain 
symmetrical property and the noise variances of 
$N_i,i=1,2,\cdots,L$ have an identical variance denoted by $\sigma^2$, 
we derive a new lower bound of the sum rate part. 
We further derive a sufficient
condition for this lower bound to be tight. The derived sufficient
condition depends only on $\Sigma_{X^L}$ and $\sigma^2$.
From this matching condition we can see 
that an explicit form of the sum rate part of the  rate distortion region can be
found when the noise variance $\sigma^2$ is relatively 
high compared with the eigen values of $\Sigma_{X^L}$.

In Oohama \cite{oh5}, \cite{oh6}, details of derivations of the inner
and outer bound were omitted.  In this paper we also present the
details of derivation of those two bounds.
%, \cite{oh6}, some parts of the proofs 
%of the results are omitted because of page constraint. 

The rest of this paper is organized as follows.  In Section II, we
present problem formulations and state the previous works on those
problems.  In Section III, we give our main result. We first derive
explicit inner and outer bounds of the rate distortion region. Next we
presented an explicit sufficient condition for the outer bound to
coincide with the inner bound. In Section IV, we explicitly compute
the matching condition for two examples of Gaussian sources.  In
Sections V and VI we give the proofs of the results. Finally, 
in Section VII, we conclude the paper.

\section{Problem Statement and Previous Results}

\subsection{
Formal Statement of Problem
}
\newcommand{\baseN}{\rm e}

In this subsection we present a formal statement of problem. Throughout 
this paper all logarithms are taken to the base natural. 
Let $\Lambda=\{1,2,\cdots,L\}$ and let $X_i, i\in \Lambda$ 
be correlated zero mean Gaussian 
random variables taking values in the real lines 
${\cal X}_i^n$. We write a $L$ dimensional 
random vector as $X^L=$ $(X_1,X_2,$ $\cdots, X_L)$ and use similar 
notation of other random variables. We denote the covariance matrix 
of $X^L$ by $\Sigma_{X^L}$. 
Let $\{(\Xast,$ $\Xbst, \cdots, \Xlst)\}_{t=1}^{\infty}$
be a stationary memoryless multiple Gaussian source. For each 
$t=1,$$2,\cdots, $ $(X_{1,t},X_{2,t},\cdots,$ $\!X_{L,t})\,$ obeys the
same distribution as $(X_1,$ $\!X_2,\cdots$, $\!X_L)\,$. 
Let a random vector consisting of $n$ independent copies of 
the random variable $X_i$ be denoted by   
$
{\vc X}_i=X_{i,1}$ $X_{i,2}$ $\cdots X_{i,n}.
$
Furthermore, let ${\vc X}^L$ denote the random 
vector $({\vc X}_1, {\vc X}_2,\cdots, {\vc X}_L)$. 

We consider the separate coding system for $L$ correlated sources, where 
$L$ encoders can only access noisy version $Y_i$ of $X_i$ for 
$i=1,2, \cdots, L$, that is,   
\beq 
Y_i=X_i+N_{i}, i\in\Lambda 
\eeq
where $N_{i},$ $i\in\Lambda$ are zero mean independent Gaussian 
random variables with variance $\sigma_{N_i}^2$. We assume that 
$X^L$ and $N^L$ are independent. The separate coding system 
for $L$ correlated Gaussian remote sources is shown in Fig. 1. 
For each  $i\in \Lambda$, the noisy 
version ${\vc Y}_{i}$ of ${\vc X}_{i}$ is separately encoded 
to $\varphi_i({\vc Y}_i)$. The $L$ encoded data 
$\varphi_i({\vc Y}_i)$, $i\in \Lambda$ are sent 
to the information processing center, 
where the decoder observes them and outputs the estimation 
$(\hat{\vc X}_1, \hat{\vc X}_2,\cdots,\hat{\vc X}_L)$ 
of $({\vc X }_1, {\vc X }_2,$ $\cdots,{\vc X }_L)$ 
by using the decoder function 
$\psi=(\psi_1,\psi_2,$ $\cdots,\psi_L)$. 

The encoder functions $\varphi_i\,,i\in\Lambda$ are defined by 
\beq
\varphi_i: {\cal X}_i^n \to {\cal M}_i=
\left\{1,2,\cdots, M_i\right\}  
\label{eqn:enc}
\eeq
and satisfy rate constraints 
\beq
\frac{1}{n}\log M_i \leq R_i+ \delta
\label{eqn:rate1}
\eeq
where $\delta$ is an arbitrary prescribed positive number. 
The decoder function $\psi=$ $(\psi_1,$ $\psi_2,$ $\cdots,\psi_L)$ 
is defined by  
\beq
\psi_i: {\cal M}_1 \times \cdots \times {\cal M}_L 
\to \hat{\cal X}_i^{n}\,,i=1,2,\cdots,L,
\label{eqn:decz}
\eeq
where $\hat{\cal X}_i$ is the real line in which 
a reconstructed random variable of $X_i$ takes values. 
Denote by ${\cal F}_{\delta}^{(n)}(R_1,$ $\!R_2,$ $\cdots,R_L)$ 
the set that consists of all the $(L+1)$ tuple of encoder 
and decoder functions 
$(\varphi_1,\varphi_2,$ $\cdots$, $\varphi_L,$ $\!\psi)$ 
satisfying (\ref{eqn:enc})-(\ref{eqn:decz}). 
For ${\vc X}^L$ $=({\vc X}_1,$ ${\vc X}_2,$ $\cdots,$ ${\vc X}_L)$ 
and its estimation 
\beqno
\hat{\vc X}^{L}&=&(\hat{\vc X}_1,\hat{\vc X}_2,\cdots,\hat{\vc X}_L)\\
           &\defeq&(\psi_1(\varphi_1({\vc Y}_1)),
               \psi_2(\varphi_2({\vc Y}_2)), 
               \cdots, 
               \psi_L(\varphi_L({\vc Y}_L)),
\eeqno
set 
\beqa 
d_{ii}%({\vc X}_i-\hat{\vc X}_i)
& \defeq & {\rm E}||{\vc X}_i-\hat{\vc X}_i||^2 
\,,
\nonumber\\
d_{ij}%({\vc X}_i-\hat{\vc X}_i,{\vc X}_j-\hat{\vc X}_j) 
& \defeq &
{\rm E} \langle {\vc X}_i-\hat{\vc X}_i,
        {\vc X}_j-\hat{\vc X}_j\rangle\,, 1 \leq i\ne j \leq L.
\nonumber
\eeqa
where $||{\vc a}||$ stands for the Euclid norm of $n$ 
dimensional vector ${\vc a}$ and $\langle {\vc a},{\vc b}\rangle$
stands for the inner product between 
%the $n$ dimensional vectors
${\vc a}$ and ${\vc b}$. Let $\Sigma_{{\lvc X}^L-\hat{\lvc X}^L}$ 
be a covariance matrix with $d_{ij}$ 
%({\vc X}_i,\hat{\vc X}_j)$  
in its $(i,j)$ element. 

%Let $\DisT$ be a given $L\times L$ 
%covariance matrix which serves 
%as a distortion criterion. We call this matrix 
%the distortion matrix. 

\bfig
\setlength{\unitlength}{1.00mm}
\begin{picture}(80,48)(2,0)
%(100,47)(-30,4)

%Information sources:
\put(3,35){\framebox(6,6){$X_1$}}
\put(3,20){\framebox(6,6){$X_2$}}
\put(5,11){$\vdots$}
\put(3,0){\framebox(6,6){$X_L$}}

%Random vectors X_1 and Y_1:
\put(10.5,40.6){${\vc X}_1$}
\put(20,40.6){${\vc Y}_1$}
\put(16,45){${\vc N}_1$}

%Random vectors X_2 and Y_2:
\put(10.5,25.6){${\vc X}_2$}
\put(20,25.6){${\vc Y}_2$}
\put(16,30){${\vc N}_2$}

%Random vectors X_L and Y_L:
\put(10.5,5.6){${\vc X}_L$}
\put(20,5.6){${\vc Y}_L$}
\put(16,10){${\vc N}_L$}

%Additive Noise 1:
\put(9,38){\vector(1,0){7.5}}
\put(19.5,38){\vector(1,0){7.5}}
\put(18,44){\vector(0,-1){4.5}}
\put(18,37){\line(0,1){2}}
\put(17,38){\line(1,0){2}}
\put(18,38){\circle{3.0}}

%Additive Noise 2:
\put(9,23){\vector(1,0){7.5}}
\put(19.5,23){\vector(1,0){7.5}}
\put(18,29){\vector(0,-1){4.5}}
\put(18,22){\line(0,1){2}}
\put(17,23){\line(1,0){2}}
\put(18,23){\circle{3.0}}

%Additive Noise L:
\put(9,3){\vector(1,0){7.5}}
\put(19.5,3){\vector(1,0){7.5}}
\put(18,9){\vector(0,-1){4.5}}
\put(18,2){\line(0,1){2}}
\put(17,3){\line(1,0){2}}
\put(18,3){\circle{3.0}}

%Encoder 1: 
\put(27,35){\framebox(6,6){$\varphi_1$}}
\put(35,40.6){$\varphi_1({\vc Y}_1)$}

%Encoder 2: 
\put(27,20){\framebox(6,6){$\varphi_2$}}
\put(35,25.6){$\varphi_2({\vc Y}_2)$}
\put(29,11){$\vdots$}
%Encoder L: 
\put(27,0){\framebox(6,6){$\varphi_L$}}
\put(35,5.6){$\varphi_L({\vc Y}_L)$}

%Transmission line 1:
\put(33,38){\line(1,0){15}}
\put(48,38){\vector(2,-3){10}}

%Transmission line 2:
\put(33,23){\line(1,0){15}}
\put(48,23){\vector(1,0){10}}

%Transmission line L:
\put(33,3){\line(1,0){15}}
\put(48,3){\vector(1,2){10}}

%Decoder: 
\put(58,20){\framebox(6,6){$\psi$}}
\put(64,23){\vector(1,0){4}}
\put(68,22){$ \ba[t]{l}(\hat{\vc X}_1,\hat{\vc X}_2,\\
              \qquad \cdots, \hat{\vc X}_L)
             \ea$ 
              }
\end{picture}
\begin{flushleft}
{\small Fig. 1. Separate coding system for $L$ correlated 
        Gaussian observations}
\end{flushleft}
\efig

In this communication system we can consider two distortion 
criterions. For each distortion criterion we define the 
determination problem of the rate distortion region. 
Those two problems are shown below.

{\it Problem 1. Vector Distortion Criterion: } Fix positive 
vector ${D}^L=(D_1,D_2,\cdots, D_L)$. 
For a given $D^L$, the rate vector $(R_1,R_2,\cdots, R_L)$ 
is {\it admissible} if for any positive $\delta>0$ and 
any $n$ with $n\geq n_0(\delta)$, 
there exists $(\varphi_1,\varphi_2, \cdots,$ $\varphi_L,$ 
$\psi)\in$ $\!{\cal F}_{\delta}^{(n)} (R_1,R_2$ $\cdots, R_L)$ 
such that
\beqno
\left[\ds \frac{1}{n}\Sigma_{{\lvc X}^L-\hat{\lvc X}^L}\right]_{ii} 
\leq D_i + \delta \,, 
\eeqno
where $[A]_{ii}$ stands for the $(i,j)$ entry of the matrix $A$. 
Let ${\cal R}_{\Iset}(D^L)$ denote the set of all the admissible 
rate vector. On a form of ${\cal R}_{\Iset}(D^L)$, we have 
a particular interest in its sum rate part. To examine 
this quantity, define 
$$ 
R_{{\rm sum}, L}(D^L)\defeq \min_{(R_1,R_2,\cdots,R_L)\in {\cal R}_{L}(D^L)}
\left\{\sum_{i=1}^{L}R_i \right\}\,.
$$
To determine $R_{{\rm sum}, L}(D^L)$ in an explicit form 
is also of our interest. 

{\it Problem 2. Sum Distortion Criterion: }
Fix positive $D$. For a given positive $D$, 
the rate vector $(R_1,R_2,\cdots, R_L)$ is {\it admissible} 
if for any positive $\delta>0$ and 
any $n$ with $n\geq n_0(\delta)$, 
there exists $(\varphi_1,\varphi_2, \cdots,$ $\varphi_L,$ $\psi)\in$ 
             $\!{\cal F}_{\delta}^{(n)} (R_1,R_2$ $\cdots, R_L)$ 
such that
\beqno
{\rm tr}\left[\ds \frac{1}{n}\Sigma_{{\lvc X}^L-\hat{\lvc X}^L}\right] 
\leq D +\delta \,, 
\eeqno
Let ${\cal R}_{\Iset}(D)$ denote the set of all the admissible 
rate vector. To examine 
the sum rate part of ${\cal R}_{\Iset}(D)$, define 
$$ 
R_{{\rm sum}, L}(D)\defeq \min_{(R_1,R_2,\cdots,R_L)\in {\cal R}_{L}(D)}
\left\{\sum_{i=1}^{L}R_i \right\}\,.
$$

We can easily show that we have the following relation
between 
${\cal R}_{L}(D)$
and 
${\cal R}_{L}^{({\rm in})}(D^L)$:
\beq
{\cal R}_{L}(D)
=\bigcup_{\sum_{i=1}^LD_i\leq D}
{\cal R}_{L}(D^L)\,. 
\label{eqn:Eq1}
\eeq
In this paper our argument is concentrated on the study 
of Problem 2. It is well known that when 
$D\geq $ ${\rm tr}[\Sigma_{X^L}],$
$R_1=R_2=$$\cdots=R_L=0$ is admissible. 
In this case, we have  
$$
{\cal R}_L(D)=
\left\{(R_1,\cdots,R_L):R_i\geq 0, i\in \Lambda \right\}.
$$
In the subsequent arguments we focus on our arguments 
in the case of $D<{\rm tr}[\Sigma_{X^L}]$.

\subsection{
Previous Results
%Inner Bound of the Rate Distortion Region
}
In this subsection we state previous results on the determination
problem of ${\cal R}_L(D)$. We first state a previous result on an
inner bound of ${\cal R}_L(D)$ and ${\cal R}_L(D^L)$. Let ${U}_i,
i\in\Lambda$ be random variables taking values in real lines ${\cal U}_i$. 
For any subset $S\subseteq \Lambda$, we introduce the notation
$U_S$$\defeq$$(U_i)_{i\in S}$. In particular,
$U_{\Lambda}$$=U^L=$$(U_1,U_2,$ $\cdots,U_L)$. Similar notations are
used for other random variables. Define
\beqno
{\cal G}(D^L)
&\defeq& \ba[t]{l}
 \left\{U^L \right.:
  \ba[t]{l} 
  U^L\mbox{ is a Gaussian }
  \vspace{1mm}\\
  \mbox{random vector that satisfies}
  \vspace{1mm}\\
  U_S\to Y_S \to X^L \to Y_{S^{\rm c}} \to U_{S^{\rm c}}\,, 
  \vspace{1mm}\\
  U^L \to Y^L \to X^L
  \vspace{1mm}\\
  \mbox{for any $S\subseteq \Lambda$ and }\\ 
  \ds {\rm E}\left[X_i-{\tilde{\psi}}_i(U^L)\right]^2 \leq D_i
  \vspace{1mm}\\
  \mbox{for some linear mapping }
  \vspace{1mm}\\
  {\tilde{\psi}}_i: {\cal U}^L 
  \to \hat{\cal X}_i, i\in\Lambda\,. 
  \left. \right\}
  \ea
\ea
\eeqno
and set 
\beqno
\hat{\cal R}_{L}^{({\rm in})}(D^L)
&\defeq&{\rm conv}\ba[t]{l}
\left\{R^L \right. : 
 %\vspace{1mm}\\
  \ba[t]{l}
  \mbox{There exists }U^L\in {\cal G}(D^L) 
  \vspace{1mm}\\
  \mbox{ such that }
  \vspace{1mm}\\
  \ds \sum_{i \in S} R_i \geq I(U_S;Y_S|U_{S^{\rm c}})
  \vspace{1mm}\\
  \mbox{ for any } S\subseteq \Lambda\,.
  \left. \right\}\,,
  \ea
\ea
\\
%& &
\hat{\cal R}_{L}^{({\rm in})}(D)
%\nonumber\\
&\defeq&{\rm conv}\ba[t]{l}
\left\{R^L \right. : 
 %\vspace{1mm}\\
  \ba[t]{l}
  \mbox{There exist }D^L\mbox{ and }
  \vspace{1mm}\\ 
  U^L\in {\cal G}(D^L)\mbox{ such that} 
  \vspace{1mm}\\
  \ds \sum_{i \in S} R_i \geq I(U_S;Y_S|U_{S^{\rm c}})
  \vspace{1mm}\\
  \mbox{ for any } S\subseteq \Lambda \mbox{ and }
  \vspace{1mm}\\
  \ds \sum_{i=1}^LD_i\leq D\,.
  \left. \right\}\,,
  \ea
\ea
\eeqno 
where $\conv\{A\}$ denotes a convex hull of the set $A$.
We can easily show that we have the following relation
between 
$\hat{\cal R}_{L}^{({\rm in})}(D)$
and 
$\hat{\cal R}_{L}^{({\rm in})}(D^L)$:
\beq
\hat{\cal R}_{L}^{({\rm in})}(D)
=\bigcup_{\sum_{i=1}^LD_i\leq D}
\hat{\cal R}_{L}^{({\rm in})}(D^L)\,. 
\label{eqn:Eq2}
\eeq
Then, we have the following result.
%Oohama \cite{oh5} obtained the following result. 
\begin{Th}[Berger \cite{bt} and Tung \cite{syt}]\label{th:direct}
$$
\hat{\cal R}_{L}^{({\rm in})}(D)\subseteq {\cal R}_{L}(D)\,,
\hat{\cal R}_{L}^{({\rm in})}(D^L)\subseteq {\cal R}_{L}(D^L)\,.
$$
\end{Th}

The inner bound $\hat{\cal R}_{L}^{({\rm in})}(D^L)$ 
is well known as the inner bound of Berger \cite{bt} 
and Tung \cite{syt}. The inner bound 
$\hat{\cal R}_{L}^{({\rm in})}(D)$ 
can be regarded as a variant of their inner bound. 
%
%Oohama \cite{oh6} obtained the following result. 
%\begin{Th}[Oohama \cite{oh6}] \label{th:conv2}
%$$
%{\cal R}_{L}^{({\rm in})}(D) \subseteq 
%       {\cal R}_{L}(D) \subseteq 
%{\cal R}_{L}^{({\rm out})}(D) \,.
%$$
%\end{Th}
%------------------ This part needs consideration  ----------------%
%As Zhang and Wicker \cite{zh} pointed out in their paper, since $A$ is
%invertible we can assume without loss of generality that $A$ is an
%identity matrix. 
%------------------------------------------------------------------%
%This implies that the coding system Pandya {\it et al.}
%Here, we state the previous results on the above problem. 
%Distributed source coding for Gaussian observation  
%------------------------------------------------------------------%
%

The source coding problem considered in this paper 
was first posed and investigated by Pandya {\it et al.}\cite{pdya}. 
They dealt with the case that 
$
Y^L=X^LA+N^L\,,
$
where $A$ is $L \times L$ a positive definite attenuation matrix.
When $A$ is an identity matrix, the problem studied by Pandya {\it et
al.} is the same as the problem considered here. They derived upper
and lower bounds of $R_{\rm sum,L}(D)$. 

%
%Subsequently, Zhang and Wicker \cite{zh}
%studied the same coding problem derived an explicit inner bound of
%${\cal R}_{L}(D)$.

Recently, Wagner {\it et al.} \cite{wg3} 
have determined ${\cal R}_{2}($$D_1,D_2)$. 
Their result is as follows. 
\begin{Th}[Wagner {\it et al.} \cite{wg3}]\label{th:conv3}
For any positive $D_1$ and $D_2$, we have  
$$
{\cal R}_{2}(D_1,D_2)=\hat{\cal R}_{2}^{({\rm in})}(D_1,D_2)\,.
$$
\end{Th}

From the above theorem, (\ref{eqn:Eq1}) and (\ref{eqn:Eq2}), 
we immediatly obtain the following corollary.
\begin{co}
[Wagner {\it et al.} \cite{wg3}]\label{co:conv3z}
For any positive $D$, we have 
$$
{\cal R}_{2}(D)=\hat{\cal R}_{2}^{({\rm in})}(D)\,.
$$
\end{co}

According to Wagner {\it et al.} \cite{wg3}, the results of Oohama 
\cite{oh1}, \cite{oh2}, and \cite{oh4} play an essential role 
in deriving the above result. The determination problems 
of ${\cal R}_L(D^L)$ 
and ${\cal R}_L(D)$ for $L\geq 3$ still remains to be solved. 
Their method for the proof depends heavily on the specific 
property of $L=2$. It is hard to generalize it to 
the case of $L\geq 3$. 

%Oohama \cite{oh5}, \cite{oh6} derived inner and outer 
%bounds of ${\cal R}_{L}(D)$. The outer bounds of Oohama \cite{oh5}, 
%\cite{oh6} have forms very close to the inner bound. 
%Form their form the 
%outer bounds of Oohama \cite{oh5}, \cite{oh6} are tighter than that of 
%Pandya {\it et al.} \cite{pdya}. Oohama \cite{oh6} also derived sufficient 
%conditions for inner and outer bounds to match and found examples of 
%information source for which rate distortion region are 
%explicitly determined. 
%Wagner {\it et al.}\cite{wg}, \cite{wg2} 
%determined the rate distortion region 
%in the case of $L$. 
%\end{document}
%s of Oohama \cite{oh5} 
%\subsection{Separate Coding of Two Gaussian Remote Sources} 
%Throughout this paper we use the notations which do not distinguish
%differential (conditional) entropies from ordinary (conditional)
%entropies. 
%as Oohama \cite{oh5} and 
%In this paper we derive explicit inner and outer 
%bounds of the rate distortion region ${\cal R}_{L}(D)$. 

\section{Main Results}

%\subsection{Inner and Outer Bounds}

In this section we state our results on ${\cal R}_L(D)$ 
and ${R}_{{\rm sum}, L}($ $D)$. 

\subsection{Definition of Functions and their Properties}

%To describe the results we define several 
%functions and sets. 

In this subsection we define several functions which 
are necessary to describe our results and present their 
properties. For $r_i\geq 0, i\in \Lambda$, let $N_{i}(r_i),$ 
$i\in \Lambda$ be $L$ independent Gaussian random 
variables with mean 0 and variance 
$\sigma_{N_i}^2/(1-{\baseN}^{-2r_i})$. 
Let $\Sigma_{N^L(r^L)}$ be a covariance matrix for the random 
vector $N^L(r^L)$. For any subset $S\subseteq \Lambda$, 
we set $r_S\defeq (r_i)_{i\in S}$. 
In particular, $r_\Lambda=$ $r^L=$ 
$(r_1,$ $r_2,$ $\cdots, r_L)$. 
Fix nonnegative vector $r^L$. 
Let $\alpha_i = \alpha_i(r^L), i\in\Lambda$ 
be $L$ eigen values of the matrix $\Sigma_{X^L}^{-1}$ 
$+\Sigma_{N^L(r^L)}^{-1} \,.$
For $S \subseteq  \Lambda $, and $\theta>0$, define 
\beqno
%\hspace*{-2mm}
\Sigma_{N^L(r_{S^{\rm c}})}^{-1} 
&\defeq& 
\left. \Sigma_{N^L(r^L)}^{-1} \right|_{r_{S}={\lvc 0}} \,,
\\
{\underline{J}}_{S}(\theta, r_S|r_{\coS})
&\defeq &\frac{1}{2}\log^{+}
   \left[\ts 
   \frac{\ds \prod_{i\in S} {\baseN}^{2r_i} }
        {\ds  \theta\left|\Sigma_{X^L}^{-1}
                 +\Sigma_{N^L(r_{S^{\rm c}})}^{-1}\right|
        }
  \right],
\\
J_{S}\left(r_S|r_{\coS}\right)
&\defeq &\frac{1}{2}\log
   \left[\ts 
   \frac{\ds \left|\Sigma_{X^L}^{-1}+\Sigma_{N^L(r^L)}^{-1}\right|
             \left\{\prod_{i\in S} {\baseN}^{2r_i}\right\}
        }
        {\ds  \left|\Sigma_{X^L}^{-1}
                 +\Sigma_{N^L(r_{S^{\rm c}})}^{-1}\right|
        }
  \right],
\eeqno
where $S^{\rm c}=\Lambda-S$ and $\log^{+}x\defeq\max\{ \log x, 0 \}\,.$
Let ${\cal B}_L(D)$  be the set of all nonnegative 
vectors $r^L$ that satisfy
\beq
{\rm tr}\left[\left(\Sigma_{X^L}^{-1}+
\Sigma_{N^L(r^L)}^{-1}\right)^{-1}
             \right]\leq D\,.
\label{eqn:trace0}
\eeq
Let $\partial{\cal B}_L(D)$ be the boundary of ${\cal B}_L(D)$, 
that is, the set of all nonnegative vectors $r^L$ that satisfy
$$
{\rm tr}\left[\left(
        \Sigma_{X^L}^{-1}+\Sigma_{N(r^L)}^{-1}
        \right)^{-1}\right]=D\,. 
$$
Let $\xi$ be nonnegative number that satisfy 
$$
\sum_{i=1}^L\left\{[\xi- \alpha_i^{-1}]^{+}+\alpha_i^{-1}\right\}=D.
$$   
Define  
$$
\theta(D,r^L)\defeq
\prod_{i=1}^L\left\{[\xi-\alpha_i^{-1} ]^{+}+\alpha_i^{-1} \right\}.
$$
We can show that for $S\subseteq \Lambda$, 
$\underline{J}_S(\theta(D,r^L),r_S|r_{\coS})$ 
and $J_S(r_S|r_{\coS})$ 
satisfy the following two properties.
\begin{pr}{
\label{pr:prz01z}
$\quad$
\begin{itemize}
\item[{\rm a)}] If $r^L\in {\cal B}_L(D)$, then, for any 
$S\subseteq \Lambda$, 
$$
\underline{J}_S(\theta(D,r^L),r_S|r_{\coS})\leq J_S(r_S|r_{\coS})\,.
$$
The equality holds when $r^L\in \partial{\cal B}_L(D)$.
\item[{\rm b)}] Suppose that $r^L\in {\cal B}_L(D)$. 
If $\left. r^L\right|_{r_S={\lvc 0}}$ still belongs to 
${\cal B}_L(D)$, then, 
\beqno
& &\left. \underline{J}_S(\theta(D,r^L), r_S|r_{\coS})
   \right|_{r_S={\lvc 0}}
  =\left. J_S(r_S|r_{\coS})\right|_{r_S={\lvc 0}}
\\
& &=0\,.
\eeqno
\end{itemize}
}\end{pr}

\begin{pr}\label{pr:matroid}{\rm 
Fix $r^L\in {\cal B}_L(D)$. 
For $S \subseteq \Lambda$, set 
\beqno
{f}_S&=&{f}_S(r_S|r_{\coS})
\defeq 
\underline{J}_S(\theta(D,r^L),r_S|r_{\coS})\,.
%\\
%{\rho}_S&=&{\rho}_S(r_S)\defeq J_S(r_S|r_{\coS})\,.
\eeqno
By definition it is obvious that 
${f}_S,S \subseteq \Lambda$ 
are nonnegative. We can show that
$f\defeq \{{f}_S\}_{S \subseteq \Lambda}$ 
satisfies the followings:
\begin{itemize}
\item[{\rm a)}]${f}_{\emptyset}=0$. 
\item[{\rm b)}] 
${f}_A\leq {f}_B$ 
for $A\subseteq B\subseteq \Lambda$.  
\item[{\rm c)}] ${f}_A+{f}_B \leq {f}_{A \cap B}+{f}_{A\cup B}\,.$
\end{itemize}
In general $(\Lambda,f)$ is called a {\it co-polymatroid} 
if the nonnegative function $f$ on $2^{\Lambda}$ satisfies 
the above three properties. Similarly, we set
\beqno
\tilde{f}_S&=&\tilde{f}_S(r_S|r_{\coS})\defeq J_S(r_S|r_{\coS})\,,
\quad \tilde{f}=\left\{\tilde{f}_S\right\}_{S \subseteq \Lambda}\,.
\eeqno
Then, $(\Lambda,\tilde{f})$ also has the same three properties 
as those of $(\Lambda,f)$ and becomes a co-polymatroid. 
}\end{pr}
 
\subsection{Results}

In this subsection we present our results on ${\cal R}_L(D)$. 
To describe our result on inner and outer bounds of 
${\cal R}_L(D)$, set 
\beqno
%& &
{\cal R}_L^{({\rm out})}(D,r^L)
%\\
&\defeq&
\ba[t]{l}
  \left\{R^L \right. : 
  \ba[t]{l}
  \ds \sum_{i \in S} R_i 
  \geq {\underline{J}}_{S}\left(\theta(D,r^L),r_S|r_{\coS}\right)
  \vspace{1mm}\\
  \mbox{ for any }S \subseteq \Lambda\,. 
  \left. \right\}\,,
  \ea
\ea
\nonumber\\
{\cal R}_{L}^{({\rm out})}(D)
&\defeq& 
\bigcup_{r^L \in {\cal B}_L(D)} {\cal R}_L^{({\rm out})}(D,r^L)\,, 
\nonumber\\
%& &
{\cal R}_L^{({\rm in})}(r^L)
%\\
&\defeq&
\ba[t]{l}
  \left\{R^L \right.: 
  \ba[t]{l}
  \ds \sum_{i \in S} R_i 
  \geq J_{S}\left(r_S|r_{\coS}\right)
  \vspace{1mm}\\
  \mbox{ for any }S \subseteq \Lambda\,. 
  \left. \right\}\,,
  \ea
\ea
\nonumber\\
{\cal R}_L^{({\rm in})}(D)
&\defeq&{\rm conv}
        \left\{
        \bigcup_{r^L \in %\partial
                 {\cal B}_L(D)}
        {\cal R}_L^{({\rm in})}(r^L)
        \right\}
\,. 
\eeqno
Our main result is as follows.
\begin{Th}\label{th:conv2}%(Oohama \cite{oh5}) 
$$
{\cal R}_{L}^{({\rm in})}(D) 
\subseteq 
\hat{\cal R}_L^{({\rm in})}(D)
\subseteq 
{\cal R}_{L}(D) 
\subseteq 
{\cal R}_{L}^{({\rm out})}(D) \,.
$$
\end{Th}

%Oohama \cite{oh6} obtained the following result. 
%\begin{Th}[Oohama \cite{oh6}] \label{th:conv2}
%$$
%{\cal R}_{L}^{({\rm in})}(D) \subseteq 
%       {\cal R}_{L}(D) \subseteq 
%{\cal R}_{L}^{({\rm out})}(D) \,.
%$$
%\end{Th}
Proof of this theorem will be given in Section V. 

An essential gap between 
${\cal R}_L^{({\rm out})}(D)$ and ${\cal R}_L^{({\rm in})}(D)$  
is the difference between 
$\underline{J}_S(\theta(D,$ $r^L),r_S|r_{\coS})$
in the definition of ${\cal R}_L^{({\rm out})}(D)$
and $J_{S}\left(r_S|\right.$ $\left. r_{\coS}\right)$ 
in the definition 
of ${\cal R}_L^{({\rm in})}(D)$. 
By Property \ref{pr:prz01z} part a) 
and the definitions 
of ${\cal R}_L^{({\rm out})}(D, r^L)$ and 
${\cal R}_L^{({\rm in})}($ $r^L)$, if 
$r^L \in\partial{\cal B}_L(D)$, then, 
$$
{\cal R}_L^{({\rm out})}(D, r^L)={\cal R}_L^{({\rm in})}(r^L)\,,
$$
which suggests a possibility that in some nontrivial 
cases ${\cal R}_L^{({\rm out})}(D)$ and ${\cal R}_L^{({\rm in})}(D)$  
match. For $L\geq 3$, we present a sufficient 
condition for ${\cal R}^{{(\rm out)}}_L(D)$ $\subseteq$ 
${\cal R}_L^{({\rm in})}($$D)\,.$ We consider 
the following condition on 
$\theta(D,r^L)$.

{\it Condition: } For any $i\in \Lambda$, 
${\baseN}^{-2r_i}\theta(D,r^L)$ is a monotone 
decreasing function of $r_i\geq 0$.

We call this condition the MD condition. The following 
is a key lemma to derive the matching condition. 
\begin{lm}\label{lm:lem1}
If $\theta(D,r^L)$ satisfies 
the MD condition on ${\cal B}_L($ $D)$, then, 
$$
 {\cal R}_L^{({\rm in})}(D)=\hat{\cal R}_L^{({\rm in})}(D)
={\cal R}_L(D)={\cal R}_L^{({\rm out})}(D).
$$ 
\end{lm}

Proof of this lemma will be given in Section VI. 
Based on Lemma \ref{lm:lem1}, we derive a sufficient 
condition for $\theta(D,r^L)$ to satisfy the MD condition. 

Let $a_{ii},i=1,2, \cdots, L$ be $(i,i)$-element 
of $\Sigma_{X^L}^{-1}$ and set 
$c_i \defeq \frac{1}{\sigma_{N_i}^2}\,.$  
Let $\alpha_{\min}=\alpha_{\min}(r^L)$ and
$\alpha_{\max}=\alpha_{\max}(r^L)$ be the minimum 
and maximum eigen values of 
$\Sigma_{X^L}^{-1}+$ $\Sigma_{N^L(r^L)}^{-1}$, respectively.
The following is a key lemma to derive a sufficient 
condition for the MD condition to hold. 
\begin{lm}\label{lm:pro3}
If 
$\alpha_{\min}(r^L)$ and $\alpha_{\max}(r^L)$ satisfy 
$$
\frac{1}{\alpha_{\min}(r^L)}-\frac{1}{\alpha_{\max}(r^L) }
\leq \frac{1}{a_{ii}+c_i}\,, 
\quad \mbox{ for } i\in \Lambda 
$$
on ${\cal B}_L(D)$, then, $\theta(D,r^L)$ satisfies 
the MD condition on ${\cal B}_L(D)$.
\end{lm}

Set
\beqno 
{\cal C}
&\defeq & \{(D,\Sigma_{X^L},\Sigma_{N^L}): 
\ba[t]{l} r^L \in {\cal B}_L(D)\\ 
\mbox{ for some nonnegative }r^L. \}.   
\ea
\eeqno
When $r^L\geq s^L$, we have 
\beqa
& & 
\Sigma_{X^L}^{-1}+\Sigma_{N^L(r^L)}^{-1}
\succeq \Sigma_{X^L}^{-1}+\Sigma_{N^L(s^L)}^{-1}\,,
\nonumber\\
&\Rightarrow &
\left(\Sigma_{X^L}^{-1}+\Sigma_{N^L(r^L)}^{-1}\right)^{-1}
\preceq 
\left(\Sigma_{X^L}^{-1}+\Sigma_{N^L(s^L)}^{-1}\right)^{-1},
\label{eqn:za000}
\eeqa
where $B \succeq A$ stands for that $B-A$ 
is positive semi-definite. 
The equation (\ref{eqn:za000}) implies that 
$
{\rm tr}\left[\left(\Sigma_{X^L}^{-1}
+\Sigma_{N^L(r^L)}^{-1}\right)^{-1}\right]
$
is a monotone decreasing function of $r^L$. 
Hence, we have
$$
{\cal C}
=
\ba[t]{l}
\left\{(D,\Sigma_{X^L},\Sigma_{N^L}):
D>{\rm tr}\left[\left(\Sigma_{X^L}^{-1}
+\Sigma_{N^L}^{-1}\right)^{-1}\right]\right\}\,.  
\ea
$$
From Lemmas \ref{lm:lem1}, \ref{lm:pro3} and an elementary 
computation we obtain the following. 

\begin{Th}\label{th:matchTh}
Let $\alpha_{\max}^{\ast}$ be the maximum eigen value of
$\Sigma_{X^L}^{-1}+$ $\Sigma_{N^L}^{-1}$. 
If 
$$
{\rm tr}\left[\left(\Sigma_{X^L}^{-1}+\Sigma_{N^L}^{-1}\right)^{-1}\right]  
< D \leq \ts \frac {L+1}{\alpha_{\max}^{\ast}}\,,
$$
then,  
$$
 {\cal R}_L^{({\rm in})}(D)
=\hat{\cal R}_L^{({\rm in})}(D)
={\cal R}_L(D)={\cal R}_L^{({\rm out})}(D).
$$
In particular,
\beqa
& &R_{{\rm sum},L}(D)
\nonumber\\
&=& \min_{r^L\in{\cal B}_L(D)}
\left\{
\sum_{i=1}^Lr_i
+\frac{1}{2}\log
           %\left[ 
      \frac{\left|\Sigma_{X^L}^{-1}+\Sigma_{N^L(r^L)}^{-1}\right|}            
           {\left|\Sigma_{X^L}^{-1}\right|}
           %\right] 
\right\}\,.
\label{eqn:miniz}
\eeqa
\end{Th}

Proofs of Lemma \ref{lm:pro3} and Theorem \ref{th:matchTh} will be
stated in Section VI. From Theorem \ref{th:matchTh}, we can see that
we have several nontrivial cases where ${\cal R}^{(\rm in)}_L(D)$ and
${\cal R}^{({\rm out})}_L(D)$ match.
In Oohama \cite{oh7}, the author derived the 
sufficient matching condition 
$ 
D\leq \ts \frac{L+\frac{1}{L-1}}{\alpha_{\max}^{*}}
$  
on upper bound of $D$. Thus the matching condition 
presented here provides a significant improvement 
of that of Oohama \cite{oh7} for large $L$.

%\subsection{Results on the Sum Rate Characterization}
%
%In this subsection, we examine sum rate part of the rate distortion 
%region ${\cal R}_L(D)$. We have the following 
%as a corollary of Theorem \ref{th:matchTh}\,.
%\begin{co}
%If $(D,\Sigma_{X^L}, \Sigma_{N^L})\in {\cal C}$ 
%and $\eta_{\max}^{\ast}\leq L+1\,$, then, 
%\beqa
%& & 
%R_{{\rm sum},L}(D)
%\nonumber\\
%&=& \min_{r^L\in{\cal B}_L(D)}
%\left\{
%\sum_{i=1}^Lr_i
%+\frac{1}{2}\log
%           %\left[ 
%      \frac{\left|\Sigma_{X^L}^{-1}+\Sigma_{N^L(r^L)}^{-1}\right|}            
%           {\left|\Sigma_{X^L}^{-1}\right|}
%           %\right] 
%\right\}\,.
%\label{eqn:miniz}
%\eeqa
%\end{co}

We further examine an explicit characterization 
of $R_{{\rm sum},L}($ $D)$ when the source has 
a certain symmetrical property. Let %$\tau$ be 
\beqno
\tau&=&
\left(
\ba{cccccc}
    1 &2&\cdots&      i &\cdots&    L\\ 
\tau(1)&\tau(2)&\cdots&\tau(i)&\cdots&\tau(L)
\ea
\right)
\eeqno
be a cyclic shift on 
$\Lambda$, that is,
$$
\tau(1)=2,\tau(2)=3,\cdots,\tau(L-1)=L,\tau(L)=1\,.
$$
Let
$
p_{X_{\Lambda}}(x_{\Lambda})
=p_{X_1X_2\cdots X_L}(x_1,x_2,\cdots,x_L)
$     
be a probability density function of $X^L$. 
The source ${X^L}$ is said to be 
cyclic shift invariant if we have
\beqno
p_{X_{\Lambda}}(x_{\tau(\Lambda)})
&=&p_{X_1X_2\cdots X_L}(x_2,x_3,\cdots,x_L,x_1)
\\
&=&p_{X_1X_2\cdots X_L}(x_1,x_2,\cdots,x_{L-1},x_{L})
\eeqno 
for any $(x_1,x_2,$ $\cdots,x_L)$$\in {\cal X}^L$. 
In the following argument we assume that $X^L$ satisfies 
the cyclic shift invariant property. 
We further assume that $N_i, i\in \Lambda$ are 
independent identically distributed (i.i.d.) Gaussian 
random variables with mean 0 and variance $\sigma^2$.
Then, the observation $Y^L=X^L+N^L$ also satisfies 
the cyclic shift invariant property. 

Fix $r>0$, let $N_{i}(r),$ $i\in \Lambda$ be $L$ 
i.i.d. Gaussian random variables with mean 0 and 
variance $\sigma^2/(1-{\baseN}^{-2r})$. 
Let $\Sigma_{N^L(r)}$ be a covariance matrix for the random 
vector $N^L(r)$. Let $\lambda_i, i\in\Lambda$ be 
$L$ eigen values of the matrix $\Sigma_{X^L}$ and let 
$\beta_i=\beta_i(r), i\in\Lambda$ 
be $L$ eigen values of the matrix 
$\Sigma_{X^L}^{-1}$ $+\Sigma_{N^L(r)}^{-1}\,.$ 
Using the eigen values of $\Sigma_{X^L}$, 
$\beta_i(r), i\in\Lambda$ can be written as
$$
\beta_i(r)=\frac{1}{\lambda_i}+\frac{1}{\sigma^2}(1-{\rm e}^{-2r})\,.
$$
Let $\xi$ be a nonnegative number that satisfies 
$
\sum_{i=1}^L\{[\xi-\beta_i^{-1}]^{+}$ $+\beta_i^{-1}\}=D.
$ 
Define  
\beqno
\theta(D,r)&\defeq&
\prod_{i=1}^L\left\{[\xi-\beta_i^{-1} ]^{+}
+\beta_i^{-1}\right\},
\\
\underline{J}(\theta(D,r),r)
&\defeq &\frac{1}{2}\log
   \left[\ts 
   \frac{\ds {\baseN}^{2Lr}\left|\Sigma_{X^L}\right|}
        {\ds \theta(D,r)}
  \right],
\eeqno
and set
$$
\phi(r)\defeq {\rm tr}
\left[\left(
\Sigma_{X^L}^{-1}+\Sigma_{N^L(r)}^{-1}\right)^{-1}\right]  
=\sum_{i=1}^{L}\frac{1}{\beta_i(r)}\,.
$$
Since $\phi(r)$ is a monotone decreasing function 
of $r$, there exists a unique $r$ such that 
$\phi(r)=D$, we denote it by $r^{\ast}(D)$.  
Note that
\beqno
& &(\underbrace{r,r,\cdots,r}_L) \in {\cal B}_L(D)
\Leftrightarrow \phi(r)\leq D
\Leftrightarrow r\geq r^{\ast}(D)\,,
\\
& &
{\theta}(D,r^\ast)
=\left|\Sigma_{X^L}^{-1}+\Sigma_{N^L(r^\ast)}^{-1}\right|^{-1}\,.
\eeqno
Set 
$$
R_{{\rm sum},L}^{(\rm l)}(D)\defeq 
\min_{r\geq r^*(D)}\underline{J}(\theta(D,r),r)\,.
$$
Then, we have the following. 
\begin{Th} \label{th:sr0} Assume that the source 
$X^L$ and its noisy version $Y^L=X^L+N^L$ are 
cyclic shift invariant. Then, we have    
$$
R_{{\rm sum},L}(D)\geq R_{{\rm sum},L}^{(\rm l)}(D)\,.
$$
\end{Th}
%Proof of this theorem will be stated in Section V.
%

Proof of this theorem will be stated in Section V.

Next, we examine a sufficient condition for  
$R_{{\rm sum},L}^{(\rm l)}(D)$ to coincide with 
$R_{{\rm sum},L}(D)$. It is obvious from the definition 
of $\underline{J}(\theta(D,r),r)$ 
that when ${\rm e}^{-2Lr}\theta(D,r)$ is a monotone decreasing 
function of $r\in [r^*(D),+\infty)$, 
we have $R_{{\rm sum},L}^{(\rm l)}(D)$ 
$=R_{{\rm sum},L}(D)$. 

\begin{lm}\label{lm:pro3b}
Let $a$ be an identical diagonal element 
of $\Sigma_{X^L}^{-1}$. Set 
$c \defeq \frac{1}{\sigma^2}\,.$  
Let $\lambda_{\min}$ and $\lambda_{\max}$ 
be the minimum and maximum eigen values 
of $\Sigma_{X^L}$, respectively.
Let the minimum and maximum eigen values 
of $\Sigma_{X^L}^{-1}+$ $\Sigma_{N^L(r)}^{-1}$
be denoted by $\beta_{\min}=\beta_{\min}(r)$
and $\beta_{\max}=\beta_{\max}(r)$, respectively. 
Those are given by 
\beqno
\beta_{\min}(r)=
\frac{1}{\lambda_{\max}}+\frac{1}{\sigma^2}
(1-{\rm e}^{-2r})\,,
\\
\beta_{\max}(r)=
\frac{1}{\lambda_{\min}}+\frac{1}{\sigma^2}
(1-{\rm e}^{-2r})\,.
\eeqno
If $\beta_{\min}(r)$ and $\beta_{\max}(r)$ satisfy 
$$
\frac{1}{\beta_{\min}(r)}-\frac{1}{\beta_{\max}(r)}
\leq \frac{L\sigma^2{\rm e}^{2r}}{L-1}
\cdot \frac{\beta_{\min}(r)}{\beta_{\max}(r)}
$$
for $r\geq r^*(D)$, then, ${\rm e}^{-2Lr}\theta(D,r)$ 
is a monotone decreasing function of 
$r\in [r^*(D),\infty)$.
\end{lm}

From Lemma \ref{lm:pro3b} and an elementary 
computation we obtain the following. 
\begin{Th}\label{th:matchTh2}
Assume that $X^L$ and $Y^L=X^L+N^L$ are cyclic shift invariant. 
%The quantities $a$,$c$,$\lambda_{\min}$, and $\lambda_{\max}$    
%are the same as those of previous definitions.
If 
\beq
\sigma^2\geq \frac{L-1}{L}\cdot 
\frac{\lambda_{\max}}{\lambda_{\min}} 
(\lambda_{\max}-\lambda_{\min})\,,
\label{eqn:zaa0}
\eeq
then, 
$
{R}_{{\rm sum},L}^{\rm (l)}(D)={R}_{{\rm sum},L}(D).
$ 
Furthermore, the curve $R=R_{{\rm sum},L}(D)$ has 
the following parametric form:
$$
\left.
\ba{rcl}
R&=&\ds\frac{1}{2}
\log\left[{|\Sigma_{X^L}|}{\rm e}^{2Lr}
\prod_{i=1}^L\beta_{i}(r)\right]\,,
\vspace*{2mm}\\
D&=&\ds\sum_{i=1}^L\frac{1}{\beta_i(r)}\,.
\ea
\right\}
$$
\end{Th}

Proofs of Lemma \ref{lm:pro3b} and Theorem \ref{th:matchTh2} 
will be stated in Section VI. Note that the condition 
(\ref{eqn:zaa0}) depends only on the correlation 
property of $X^L$ and $N^L$. From Theorem \ref{th:matchTh2} 
we can see that for $(X^N,N^N)$ satisfying the cyclic 
shift invariant property the determination problem of 
$R_{\rm sum,L}(D)$ is solved if the identical varaince $\sigma^2$ 
of $N_i,i\in {\Lambda}$ is relatively high compared 
with the eigen values of $\Sigma_{X^L}$.  

\section{Computation of Matching Conditions} 

In this section we explicitly compute the matching 
condition for some class of Gaussian information sources. 
Define
\beq
u_i \defeq 
a_{ii}+\ts c_i (1-{\baseN}^{-2r_i})
\,, i\in\Lambda\,. 
\label{eqn:tr}
\eeq
From (\ref{eqn:tr}), we have 
$$
2r_i=\log \frac{c_i}{a_{ii}+c_i-u_i}\,. 
$$
By the above transformation we regard 
$\theta(D,r^L)$ and $\Sigma_{X^L}^{-1}+\Sigma_{N^L(r^L)}^{-1}$
as functions of $u^L$, that is, 
$\theta(D,r^L)=\theta(D,u^L)$ and 
$$
\Sigma_{X^L}^{-1}+\Sigma_{N^L(r^L)}^{-1}
=\Sigma_{X^L}^{-1}+\Sigma_{N^L(u^L)}^{-1}\,.
$$
We consider the case where $\Sigma_{X^L}$ have identical 
diagonal and nondiagonal elements, that is, 
\beqno
{\rm Var}[X_i] &=&\sigma_{X_i}^2
=1\,,\mbox{ for }i\in \Lambda,\\ 
{\rm Cov}[X_i,X_j]&=&\rho \sigma_{X_i}\sigma_{X_j}
=\rho 
\mbox{ for }i,j\in \Lambda,i \ne j. 
\eeqno
In this identical variance case, $(i,j)$ elements 
$a_{ij}$ of $\Sigma_{X^L}^{-1}$ is given by 
\beqno
a_{ij}
&=&
\left\{ 
\ba{c}
\ts\frac{1+(L-2)\rho}{(1-\rho)(1+(L-1)\rho)}%\cdot\frac{1}{\sigma^2}
\mbox{ if } i=j\,,
\\
\ts\frac{-\rho}{(1-\rho)(1+(L-1)\rho)}%\cdot\frac{1}{\sigma^2}
\mbox{ if } i \ne j\,. 
\ea
\right. 
\eeqno
For simplicity of notations we set 
$a\defeq a_{ii}, b\defeq -a_{ij}.$ 
We first derive an explicit form of the set ${\cal B}_L(D)$. 
To this end we use the following formula  
\beqa
%& &
\hspace*{-3mm}
\left|
\begin{array}{cccc}                   
  z_{1}  & \delta & \ldots & \delta \\
  \delta & z_{2}  & \ldots & \delta \\ 
  \vdots & \vdots & \ddots & \vdots \\
  \delta & \delta & \ldots & z_{L}  
  \end{array}
  \right|
%\nonumber\\
&=&\left\{\prod_{i=1}^L(z_i-\delta)\right\}
   \left\{1+\delta\sum_{i=1}^L
     \frac{1}{z_{i}-\delta}  
   \right\}\,. 
\label{eqn:formula}
\eeqa
Using (\ref{eqn:formula}), the condition 
%(\ref{eqn:aaz0z})
\beq
{\rm tr}\left[\left(\Sigma_{X^L}^{-1}+\Sigma_{N^L(u^L)}^{-1}
\right)^{-1}\right] \leq D
%\label{eqn:aaz0z}
\eeq
is explicitly given by the following:
\beqa
&&\sum_{i\ne j}
  \frac{b^2}{(u_i+b)(u_j+b)}
  \nonumber\\
&&-(1+Db)\sum_{i=1}^L \frac{b}{u_i+b}+Db
\geq 0\,.
\label{eqn:det00} 
\eeqa
Set 
\beqno
\kappa_1&\defeq&\frac{1}{2}\cdot\frac{1+Db}{L-1}\,,
%\nonumber\\
\kappa_2\defeq \frac{L}{4(L-1)}(1+Db)^2-Db\,.
\eeqno
Then, the above condition is rewritten as 
\beq
\sum_{i\ne j}
\left(\kappa_1-\ts\frac{b}{u_i+b}\right)
\left(\kappa_1-\ts\frac{b}{u_j+b}\right)
\geq \kappa_2\,.
\label{eqn:eigeq110}
\eeq
From (\ref{eqn:eigeq110}), we can see that 
the region ${\cal C}$ is given by the set of 
all $(a,b,c^L,D)$ satisfying 
\beq
\sum_{i\ne j}
\left(\kappa_1-\ts\frac{b}{a+b+c_i}\right)
\left(\kappa_1-\ts\frac{b}{a+b+c_j}\right)
\geq \kappa_2\,.
\label{eqn:ast0}
\eeq
The above condition is equivalent to
\beqa
&&\sum_{i\ne j}
  \frac{b^2}{(a+b+c_i)(a+b+c_j)}
  \nonumber\\
&&-(1+Db)\sum_{i=1}^L \frac{b}{a+b+c_i}+Db
\geq 0\,.
\label{eqn:det000} 
\eeqa
Solving (\ref{eqn:det000}) with respect to $D$, we obtain
\beq
D\geq
\frac{\ds \sum_{i=1}^L{\ts\frac{1}{a+b+c_i}}-\sum_{i\ne j}
{\ts \frac{b}{(a+b+c_i)(a+b+c_j)}}}
{\ds 1-\sum_{i=1}^L {\ts \frac{b}{a+b+c_i}}}\,. 
\eeq
From Theorem \ref{th:matchTh}, we obtain 
the following corollary. 
\begin{co} \label{co:za}
If $D$ satisfy 
\beqno
\frac{\ds \sum_{i=1}^L{\ts\frac{1}{a+b+c_i}}-\sum_{i\ne j}
{\ts \frac{b}{(a+b+c_i)(a+b+c_j)}}}
{\ds 1-\sum_{i=1}^L {\ts \frac{b}{a+b+c_i}}}
&\leq& D\leq \ts \frac{L+1}{\alpha_{\max}^*},  
\eeqno
then 
$$
{\cal R}_L^{({\rm in})}(D)
={\cal R}_L(D)={\cal R}_L^{({\rm out})}(D).
$$ 
\end{co}

Next we derive a more explicit sufficient condition. 
%Let $L_0$ be the solution of the following 
%equation 
%$$
%L(L-1)\left(\kappa_1-\ts\frac{1}{L_0+Db}\right)^2
%=\kappa_2\,.
%$$
%$L_0$ is given by 
%\beqno
%L_0
%&=&\frac{L}{2}
%\left[
%1+ Db +\sqrt{(1-Db)^2+{\ts\frac{4Db}{L}}}
%\right]-Db\,.
%\eeqno
Set 
$$
c_{\rm min}\defeq\min_{1\leq i\leq L}c_i,\quad
c_{\rm max}\defeq\max_{1\leq i\leq L}c_i\,.
$$
Then, the condition 
\beq
L(L-1)\left(\kappa_1-\ts\frac{b}{a+b+c_{\rm min}}\right)^2 
> \kappa_2
\label{eqn:cond10}
\eeq
is a sufficient condition for $(a,b,c^L,D)$ $\in{\cal C}$.
The above condition is equivalent to
\beq
D\geq \ts \frac{L}{a+b+c_{\rm min}}\cdot
\left(1+\frac{b}{a+b+c_{\rm min}-Lb}\right)\,.
\label{eqn:condxx}
\eeq
On the other hand, the maximum eigen value 
of $\Sigma_{X^L}^{-1}$
$+\Sigma_{N^L(u^L)}^{-1}$   
satisfies  
\beq
\alpha^{\ast}_{\max} 
\leq \max_{1\leq j\leq L}\{u_j+b\}
\leq a+b+c_{\rm max}\,.
\label{eqn:azz0}
\eeq
Properties on bounds of the eigen values 
of $\Sigma_{X^L}^{-1}+$ $\Sigma_{N^L(u^L)}^{-1}$
including the property stated in (\ref{eqn:azz0}) and their proofs 
are given in Appendix C. From (\ref{eqn:condxx}), (\ref{eqn:azz0}), 
and Corollary \ref{co:za}, we obtain the 
following theorem.

\begin{Th}\label{th:exth3z} 
If $(a,b,c_{\rm min},c_{\rm max},D)$ satisfies   
\beqa
\ts\frac{L}{a+b+c_{\rm min}}\cdot
   \left(1+\frac{b}{a+b+c_{\rm min}-Lb}\right)
\leq D\leq \ts \frac{L+1}{a+b+c_{\rm max}}
\label{eqn:thcondd}
\eeqa
then, 
$$
{\cal R}_L^{({\rm in})}(D)={\cal R}_L(D)
={\cal R}_L^{({\rm out})}(D).
$$ 
In particular,
\beqa
& & 
R_{{\rm sum},L}(D)
\nonumber\\
&=& \min_{r^L\in{\cal B}_L(D)}
\left\{
\sum_{i=1}^Lr_i
+\frac{1}{2}\log
           %\left[ 
      \frac{\left|\Sigma_{X^L}^{-1}+\Sigma_{N^L(r^L)}^{-1}\right|}            
           {\left|\Sigma_{X^L}^{-1}\right|}
           %\right] 
\right\}\,.
\label{eqn:minid}
\eeqa
\end{Th}

It can be seen from (\ref{eqn:thcondd}) 
that the matching condition holds for sufficiently small $b$ 
and $c_{\max}$. This implies that the determination problem 
of ${\cal R}_L(D)$ is solved if the correlation of $X^L$      
is relatively small and the noise variance of $N^L$ is relatively 
large.

Now we derive an explicit form of $R_{{\rm sum},L}(D)$ 
in the case where $c=c_{\rm min}=c_{\rm max}$. 
In this case, we have 
\beqno
{\cal C}&=&\{(a,b,c,D): 
\\
&&\qquad D\geq \ts \frac{L}{a+b+c}\cdot
   \left(1+\frac{b}{a+b+c-Lb}\right)\}\,.
\eeqno
Set
\beqa
L_1&\defeq&
\frac{L}{2}
\left[
1+ Db +\sqrt{(1-Db)^2+\ts\frac{4Db}{L}}
\right]\,.
\label{eqn:defz}
\eeqa
Solving the minimization problem in the right member
of (\ref{eqn:minid}), we have the following. 
\begin{Th}\label{th:exth4zz}
If $(a,b,c,D)$ satisfies 
\beqa
\ts \frac{L}{a+b+c}\cdot
   \left(1+\frac{b}{a+b+c-Lb}\right)
&\leq& \ts 
D\leq \frac{L}{a+b+c}\left(1+\frac{1}{L}\right)
\label{eqn:thcondz}
\eeqa
then, 
\beqno
R_{{\rm sum},L}(D)
&=& 
\frac{L}{2}\log
\left(\frac{(1-\rho)L_1 c}{D(a+b+c)-L_1}\right)
\nonumber\\
& &+\frac{1}{2}\log
\left\{\frac{1+(L-1)\rho}{1-\rho}
\left(1-\frac{LDb}{L_1}\right)
\right\}\,.
\eeqno
\end{Th}

Proof of this theorem is given in Appendix A. 

Next, we consider another example where the source and its 
noisy observation are cyclic shift invariant. 
Let $L=4$ and  

\beq
\Sigma_{X^4}=
\left[
\ba{cccc}
    1&\rho&   0&\rho\\     
 \rho&   1&\rho&0\\
    0&\rho&   1&\rho\\
 \rho&   0&\rho&   1\\
\ea 
\right],\:|\rho|<\frac{1}{2},\:
\Sigma_{N^4}=
\left[
\ba{cccc}
    1&   0&   0&0\\     
    0&   1&   0&0\\
    0&   0&   1&0\\
    0&   0&   0&1\\
\ea 
\right]\,.
\label{eqn:covMatrix1}
\eeq
In this case, we have   
\beqno
& &|\Sigma_{X^4}|=1-4\rho^2, a=\frac{1-2\rho^2}{1-4\rho^2},
\\
& &
\lambda_{1}=1-2\rho,
\lambda_{2}=\lambda_{3}=1,
\lambda_{4}=1+2\rho\,. 
\eeqno
Four eigen valules $\beta_i(r),i=1,2,3,4$ are given by
\beq
\left.
\ba{rcl}
& &\beta_1(r)=1-2\rho+\frac{1}{\sigma^2}(1-{\rm e}^{-2r})\,,
\vspace*{1mm}\\
& &\beta_2(r)=\beta_3(r)=1+\frac{1}{\sigma^2}(1-{\rm e}^{-2r})\,,
\vspace*{1mm}\\
& &\beta_4(r)=1+2\rho+\frac{1}{\sigma^2}(1-{\rm e}^{-2r})\,.
\ea
\right\}
\label{eqn:dod}
\eeq
The matching condition is 
\beqno
\sigma^2\geq 3|\rho|\frac{1+2|\rho|}{1-2|\rho|}\,.
%\label{eqn:cond10z}
\eeqno
%Solving (\ref{eqn:cond10z}) with respect to $\sigma^2$, 
%we obtain
%$$
%
%$$
Summerizing the above argument, we obtain the following.
\begin{Th} We consider the case where $L=4$,
$\Sigma_{X^4}$ and $\Sigma_{N^4}$ are given by (\ref{eqn:covMatrix1}).
If
\beqno
\sigma^2\geq 3|\rho|\frac{1+2|\rho|}{1-2|\rho|},
\eeqno
%$$
%\sigma^2\geq \frac{4|\rho|(1-4\rho^2)}{1-4(1-2|\rho|)\rho^2}\,,
%$$
then, the rate distortion curve $R=R_{{\rm sum},4}(D)$ has 
the following parametric form:
$$
\left.
\ba{rcl}
R&=&\ds\frac{1}{2}
\log\left[(1-4\rho^2){\rm e}^{8r}
\prod_{i=1}^4\beta_{i}(r)\right]\,,
\vspace*{2mm}\\
D&=&\ds\sum_{i=1}^4\frac{1}{\beta_i(r)}\,,
\ea
\right\}
$$
where $\beta_{i}(r),i=1,2,3,4$ are definded by (\ref{eqn:dod}).
\end{Th}

From this theorem we can see that for the above example 
of $(X^4,N^4)$ satisfying the cyclic 
shift invariant property the determination problem of 
$R_{\rm sum,4}(D)$ is solved if the identical varaince $\sigma^2$ 
is relatively high or correlation coefficient $\rho$ 
is relatively small. 

The determination problem 
of $R_{{\rm sum},L}(D)$ was first investigated 
by Pandya {\it et al.} \cite{pdya}. They derived upper and lower bound 
of $R_{{\rm sum},L}(D)$. Pandya {\it et al.} \cite{pdya} also numerically 
compared those two bounds to show that the gap between them 
is relatively small for some examples. In this paper we have determined 
$R_{{\rm sum},L}(D)$ for some nontrivial case of Gaussian sources.

\section{Derivation of Outer and Inner Bounds}

In this section we state the proofs of Theorems \ref{th:conv2} 
and \ref{th:sr0} stated in Section III.

\renewcommand{\irb}[1]{{\color[named]{Black}#1\normalcolor}}
\renewcommand{\irg}[1]{{\color[named]{Black}#1\normalcolor}}
\renewcommand{\irBr}[1]{{\color[named]{Black}#1\normalcolor}}
\renewcommand{\irBw}[1]{{\color[named]{Black}#1\normalcolor}}

\subsection{Derivation of the Outer Bound}

In this subsection we prove the inclusion
${\cal R}_L(D)$ $\subseteq $ ${\cal R}_L^{ ({\rm out})}(D)$
stated in Theorem \ref{th:conv2}. We use the following 
two well known lemmas to prove this inclusion. 

\begin{lm}[{Water Filling Lemma}] \label{lm:lm0z} Let 
$a_i, i=1,2,\cdots, L$ be $L$ positive numbers. The maximum of 
$\prod_{i=1}^L \xi_i$ subject to $\sum_{i=1}^{L} {\xi_i}\leq D $ 
and $\xi_i \geq a_i, i=1,2,\cdots, L $ is given by 
$$
\prod_{i=1}^L\left\{ [\xi- a_i]^{+} + a_i\right\}\,,
$$ 
where $\xi$ is determined by 
$\sum_{i=1}^L\left\{ [\xi - a_i]^{+} + a_i\right\}=D$.
\end{lm}
\begin{lm}\label{lm:lm1}
For any $n$ dimensional random vector ${\vc U}_i, i=1,2$, we have 
\beq
\frac{1}{n}h({\vc U}_1| {\vc U}_2 ) 
\leq \frac{1}{2} 
\log 
\left[
(2\pi{\rm e}) \cdot \frac{1}{n}{\rm E}||{\vc U}_1 - {\vc U}_2||^2 
\right]\,,
\eeq
where $h(\cdot)$ stands for the differential entropy. 
\end{lm}
%{\it Proof:} The proof of this lemma is standard argument on the 
%rate distortion theory for single source. We omit the detail. 
%\hfill \IEEEQED

Next, we state an important lemma which is a mathematical core of 
the converse coding theorem. For $i=1,2,\cdots, L$, set 
\beq
\irb{W_i}\defeq\varphi_i(\irBw{\vc Y}_i), 
\irb{r_i^{(n)}}\defeq\frac{1}{n}I(\irBw{\vc Y}_i;\irb{W_i}|{\vc X}_i)\,. 
\eeq
For $\irBr{S}\subseteq \Lambda$, let ${Q}_\irBr{S}$ 
be a unitary matrix which transforms $X_{\irBr{S}}$ 
into $\irg{Z}_{\irBr{S}} = X_{\irBr{S}}{Q}_{\irBr{S}}$. 
For ${\vc X}_S=(X_{S,1},$ $X_{S,2},$ $\cdots, X_{S,n})$, we set 
$$
{\vc Z}_S={\vc X}_S Q_S=(X_{S,1}Q_S, X_{S,2}Q_S, \cdots,X_{S,n}Q_S)\,.
$$
Then, we have the following lemma. 
\begin{lm}\label{lm:lm2} 
For any $S\subseteq \Lambda$, we have
\beqno
%\frac{1}{n}
& &h
\left.\left(
\irg{\vc Z}_i\right|\irg{\vc Z}_{\irBr{S}-\{i\}}\irb{W}_{\irBr{S}}
\right)
\\
&\geq& 
\frac{n}{2}\log 
\left\{
(2\pi{\rm e})\left[
Q^{-1}_{\irBr{S}}\left(\Sigma_{X_\irBr{S}}^{-1}
+\irBr{\Sigma_{N_S( \irb{r_{\irBr{S}}^{(n)}})}^{-1}}\right)Q_{\irBr{S}}
\right]_{ii}^{-1}
\right\}\,,
\eeqno
where $[C]_{ij}$ stands for the $(i,j)$ element of the matrix $C$.
\end{lm}

%The above lemma is mathematical core of the proof of the converse 
%coding theorem. 
Proof of this lemma will be stated in Appendix B. 
This lemma provides a strong result on outer bound 
of the rate distortion region. From Lemma \ref{lm:lm2}, we 
obtain the following corollary. 
\begin{co}\label{co:co1} For any $S\subseteq \Lambda $, we have  
\beq
%\frac{1}{n}
I({\vc X}_S;W_S) \leq 
\frac{n}{2}
\log  \left|I+\Sigma_{X_{S}}\Sigma_{N_{S}(r_S^{(n)})}^{-1}\right|\,.
\eeq
\end{co}

{\it Proof:} We choose unitary matrix $Q_S$ so that 
$$
Q_S^{-1}\left(\Sigma_{X_S}^{-1}+\Sigma_{N_S(r_S^{(n)})}^{-1}\right)Q_S
$$ 
becomes the following diagonal matrix: 
\beq
Q_S^{-1}\left(\Sigma_{X_S}^{-1}+\Sigma_{N_S(r_S^{(n)})}^{-1}\right)Q_S
=\left[
\begin{array}{cccc}
    \nu_1 &           &        & \mbox{\huge 0}\\
          & \nu_2     &        &          \\
          &           & \ddots &          \\
\mbox{\huge 0} &      &        & \nu_{|S|}\\
\end{array}
\right]\,.
\label{eqn:diag}
\eeq
Then, we have the following chain of inequalities:  
\beqa
 &    &   
I({\vc X}_S;W_S)
\nonumber\\ 
&\MEq{a} & h\left({\vc X}_S \right) 
          -h\left({\vc Z}_S|W_S \right)
\nonumber\\
&\leq & h\left({\vc X}_S \right) 
        -\sum_{i=1}^{|S|} 
        h\left({\vc Z}_i|{\vc Z}_{S-\{i\} }W_S \right)
\nonumber\\
&\MLeq{b}& 
\frac{n}{2} 
\log \left[(2\pi{\rm e})^{|S|}\left|\Sigma_{X_S}\right|\right]
\nonumber\\
& &\hspace*{-2mm} + \sum_{i=1}^{|S|} \frac{n}{2} \log 
              \left\{
              \frac{1}{2\pi{\rm e}}%\cdot 
              \left[
              Q_S^{-1}\left(\Sigma_{X_S}^{-1}
              +\Sigma_{N_S(r_S^{(n)})}^{-1}\right)Q_S
              \right]_{ii}
              \right\}
\nonumber\\
&\MEq{c} &  \frac{n}{2} \log \left| \Sigma_{X_S} \right| 
        + \sum_{i=1}^{|S|} \frac{n}{2}\log \left[ \nu_i \right]
\nonumber\\
&=&\frac{n}{2}\log \left|\Sigma_{X_S} \right| 
  +\frac{n}{2}\log \left|\Sigma_{X_S}^{-1}
  +\Sigma_{N_S(r_S^{(n)})}^{-1}\right| 
\nonumber\\
&=& \frac{n}{2} \log \left|I
    +\Sigma_{X_{S}}\Sigma_{N_{S}(r_S^{(n)})}^{-1}\right|\,. 
\eeqa
Step (a) follows from the rotation invariance of 
the (conditional) differential entropy. 
Step (b) follows from Lemma \ref{lm:lm2}. 
Step (c) follows from (\ref{eqn:diag}).
\hfill \IEEEQED 

Using Lemmas \ref{lm:lm0z}-\ref{lm:lm2}, 
Corollary \ref{co:co1} and a standard 
argument on the proof of converse coding theorems, 
we can prove 
$
{\cal R}_L(D)\subseteq {\cal R}_L^{ ({\rm out})}(D)\,.
$

{\it Proof of ${\cal R}_L(D)\subseteq {\cal R}_L^{ ({\rm out})}(D)$:}
Assume that
$(R_1, R_2,$ $\!\cdots, R_L) \in {\cal R}_{L}(D)$. 
Then, for any $\delta>0$ and any $n$ with $n\geq n_0(\delta)$, 
there exists $(\varphi_1,\varphi_2, \cdots,$ $\varphi_L,$ $\psi)\in$ 
$\!{\cal F}_{\delta}^{(n)} (R_1,R_2$ $\cdots, R_L)$
such that
$$
\sum_{i=1}^L{\rm E}||{\vc X}_i-\hat{\vc X}_i||^2\leq D+\delta\,. 
$$
We set ${\vc Z}_{\Lambda}\defeq{\vc X}_{\Lambda}Q$, $\hat{\vc Z}_{\Lambda}
\defeq\hat{\vc X}_{\Lambda}Q$. Furthermore, for $i\in\Lambda$, set 
\beqno
& &\xi_i^{(n)}\defeq\frac{1}{n}{\rm E}||{\vc Z}_i-\hat{\vc Z}_i||^2\,.
\eeqno
By rotation invariance of the squared norm, we have
\beqa
\sum_{i=1}^L\xi_i^{(n)}
&=&
\sum_{i=1}^L\frac{1}{n}{\rm E}||{\vc Z}_i-\hat{\vc Z}_i||^2
\nonumber\\
&=&
\sum_{i=1}^L
\frac{1}{n}{\rm E}||{\vc X}_{i}- \hat{\vc X}_{i}||^2
\leq D+\delta\,.
\label{eqn:prthy}
\eeqa
By Lemmas \ref{lm:lm1} and \ref{lm:lm2}, for $i=1,2,\cdots, L$, 
we have 
\beqa
&  & \frac{n}{2}\log \left[(2\pi{\rm e}) \xi_i^{(n)}\right]
\geq h({\vc Z}_i-\hat{\vc Z}_{i})  
%\nonumber\\
\geq h({\vc Z}_i|\hat{\vc Z}_{i}) 
\nonumber\\
&\geq & h({\vc Z}_i|W_{\Lambda}) 
\geq h({\vc Z}_i|{\vc Z}_{\Lambda -\{i\}}W_{\Lambda}) 
\nonumber\\
&\geq& 
\frac{n}{2}\log 
\left\{
(2\pi{\rm e})\left[
Q^{-1}\left(\Sigma_{X_\irBr{\Lambda}}^{-1}
+\irBr{\Sigma_{N_\Lambda(\irb{r_{\irBr{\Lambda}}^{(n)}})}^{-1}}\right)Q
\right]_{ii}^{-1}
\right\}\,,
\nonumber
%\frac{n}{2}\log \left[(2\pi{\rm e})\alpha_i^{-1}\right] \,. 
%\label{eqn:prtha}
\eeqa
from which we have  
\beqa
\xi_i^{(n)}
&\geq& \left[Q^{-1}\left(\Sigma_{X_\irBr{\Lambda}}^{-1}
+\irBr{\Sigma_{N_\Lambda( \irb{r_{\irBr{\Lambda}}^{(n)}})}^{-1}}\right)Q
\right]_{ii}^{-1}
\nonumber\\
& &\mbox{ for }i\in\Lambda\,.
\label{eqn:prth96z} 
\eeqa
Now we proceed to the derivation of the outer bound. 
We first observe that 
\beq
 W_S\to {\vc X}_S \to {\vc X}_{\coS} \to W_{\coS}
\label{eqn:MkPr1q}
\eeq
hold for any subset $S$ of $\Lambda$. 
For any subset $S \subseteq \Lambda$, 
we obtain the following chain of inequalities:
\beqa
& &
\sum_{i\in S}n(R_i+\delta)
%\nonumber\\
\geq  
\sum_{i\in S}\log M_i
\nonumber\\
&\geq & \sum_{i\in S}H(W_i)
%\nonumber\\
\geq H(W_S|W_{\coS})
\nonumber\\
&=&I({\vc X}_{\Lambda};W_S|W_{\coS})+ H(W_S|W_{\coS}{\vc X}_{\Lambda})
\nonumber\\
&=&I({\vc X}_{\Lambda};W_S|W_{\coS})+\sum_{i\in S}H(W_i|{\vc X}_{\Lambda}) 
\nonumber\\
&=&I({\vc X}_{\Lambda};W_S|W_{\coS})+\sum_{i\in S}H(W_i|{\vc X}_i) 
\nonumber\\
&\MEq{a}&I({\vc X}_{\Lambda};W_S|W_{\coS})+n\sum_{i\in S}r_i^{(n)}. 
\label{eqn:conv1}
\eeqa
Step (a) follows from (\ref{eqn:MkPr1q}). We estimate 
a lower bound of $I({\vc X}_{\Lambda};W_S|W_{\coS})$. 
Observe that
\beqa
%& & 
I({\vc X}_{\Lambda};W_S|W_{S^{\rm c}})
%\nonumber\\
& =& I({\vc X}_{\Lambda};W_{\Lambda})-I({\vc X}_{\Lambda};W_{\coS}) 
\nonumber\\
&= &  I({\vc X}_{\Lambda};W_{\Lambda})-I({\vc X}_{\coS};W_{\coS})\,.
\label{eqn:prthc}
\eeqa
Since an upper bound of $I({\vc X}_{\coS};W_{\coS})$ 
is derived by Corollary \ref{co:co1}, it suffices 
to estimate a lower bound of $I({\vc X}_{\Lambda};$ $W_{\Lambda})$.
On a lower bound of this quantity we have the following 
chain of inequalities:
\beqa
& &
I({\vc X}_{\Lambda};W_{\Lambda})
\nonumber\\
&=&h({\vc X}_{\Lambda})-h({\vc X}_{\Lambda}|W_{\Lambda})
\MEq{a} h({\vc X}_{\Lambda})-h({\vc Z}_{\Lambda}|W_{\Lambda})
\nonumber\\
&=&h({\vc X}_{\Lambda})-\sum_{i=1}^Lh({\vc Z}_i|{\vc Z}^{i-1}W_{\Lambda})
\nonumber\\
&\geq &h({\vc X}_{\Lambda})-\sum_{i=1}^Lh({\vc Z}_i|\hat{\vc Z}_i)
\nonumber\\
&\MGeq{b} &\frac{n}{2}\log \left[(2\pi{\rm e})^{L}|\Sigma_{X_{\Lambda}}|\right]
       -\sum_{i=1}^L \frac{n}{2}\log\left[(2\pi{\rm e})\xi_i^{(n)}\right]
\nonumber\\
&=&  \frac{n}{2}\log |\Sigma_{X_{\Lambda}}|
    -\frac{n}{2}\log\left[\prod_{i=1}^L\xi_i^{(n)}\right]\,.
\label{eqn:prthz0}
\eeqa
Step (a) follows from the rotation invariance 
of the differential entropy.
Step (b) follows from Lemma \ref{lm:lm1}. 
Combining (\ref{eqn:prthc}), (\ref{eqn:prthz0}) and 
Corollary \ref{co:co1}, we have 
\beqno
& & 
I({\vc X}_{\Lambda};W_S|W_{S^{\rm c}}) +n\sum_{i\in S}r_i^{(n)}
\nonumber\\
&\geq & \frac{n}{2}\log 
\left[
\frac{\prod_{i\in S}{\baseN}^{2r_i^{(n)}}
\left|\Sigma_{X_{\Lambda}}\right|}
{\left|I+ \Sigma_{X_{S^{\rm c}}}\Sigma_{N_{S^{\rm c}}(r_{S^{\rm c}}^{(n)})}^{-1}\right|
\prod_{i=1}^L\xi_i^{(n)}}
\right]
\\
&=& \frac{n}{2}\log 
\left[
\frac{\prod_{i\in S}{\baseN}^{2r_i^{(n)}} \left|\Sigma_{X_{\Lambda}}\right|}
{
\left|I+ \Sigma_{X_{\Lambda}}\Sigma_{N_{\Lambda}(r_{S^{\rm c}}^{(n)})}^{-1}\right|
\prod_{i=1}^L\xi_i^{(n)}
}
\right]
\\
&=& \frac{n}{2}\log 
\left[
\frac{\prod_{i\in S}{\baseN}^{2r_i^{(n)}} }
{
\left|\Sigma_{X_{\Lambda}}^{-1}+\Sigma_{N_{\Lambda}(r_{S^{\rm c}}^{(n)}) }^{-1}\right|
\prod_{i=1}^L\xi_i^{(n)}
}
\right]\,.
\eeqno
Note here that 
$$
I({\vc X}_{\Lambda};W_S|W_{\coS})+n\sum_{i\in S}r_i^{(n)}
$$
is nonnegative. Hence, we have 
\beqa
& &I({\vc X}_{\Lambda};W_S|W_{\coS})+n\sum_{i\in S}r_i^{(n)}
\nonumber\\
&\geq& n\underline{J}_{S}
\left(\left.\prod_{i=1}^L\xi_i^{(n)},r_S^{(n)}\right|r_{\coS}^{(n)}\right)\,.
\label{eqn:prthd}
\eeqa
Combining (\ref{eqn:conv1}) and (\ref{eqn:prthd}), we obtain
\beqa
\sum_{i\in S}(R_i+\delta) 
&\geq & 
\underline{J}_{S}
\left(\left.\prod_{i=1}^L\xi_i^{(n)},
r_S^{(n)}\right|r_{\coS}^{(n)}\right)\,.
\label{eqn:prth100z}
\eeqa
for $S\subseteq \Lambda$. For $i\in \Lambda$, set
\beqno
r_i&\defeq&\limsup_{n\to\infty}r_i^{(n)}
   =\limsup_{n\to\infty}\frac{1}{n}I({\vc Y}_i;W_i|{\vc X}_i)\,,
\\
\xi_i&\defeq&\limsup_{n\to\infty}\xi_i^{(n)}
=\limsup_{n\to\infty}
 \frac{1}{n}{\rm E}||{\vc Z}_{i}- \hat{\vc Z}_{i}||^2\,.
\eeqno
Then, by letting $n\to\infty $ in 
    (\ref{eqn:prthy}), 
    (\ref{eqn:prth96z}), 
and (\ref{eqn:prth100z}), we obtain 
\beq
\left.
\ba{rcl}
\ds\sum_{i=1}^L\xi_i
&\leq& D+\delta\,,
\vspace{1mm}\\
\xi_i&\geq &
\left[Q^{-1}\left(\Sigma_{X_\irBr{\Lambda}}^{-1}
+\irBr{\Sigma_{N_\Lambda( \irb{r_{\irBr{\Lambda}}})}^{-1}}\right)Q
\right]_{ii}^{-1}\,, 
\vspace{1mm}\\
& &\mbox{ for }i\in\Lambda\,,  
\vspace{1mm}\\
\ds\sum_{i\in S}(R_i+\delta) 
&\geq &\underline{J}_{S}
\left(\left.\prod_{i=1}^L\xi_i,r_S\right|r_{\coS}\right)
\vspace{1mm}\\
& &\mbox{ for }S\subseteq \Lambda.
\ea
\right\}
\eeq
Since $\delta$ can be made arbitrary small, we obtain  
\beq
\left.
\ba{rcl}
\ds\sum_{i=1}^L\xi_i
&\leq& D\,,
\vspace{1mm}\\
\xi_i&\geq &
\left[Q^{-1}\left(\Sigma_{X_\irBr{\Lambda}}^{-1}
+\irBr{\Sigma_{N_\Lambda( \irb{r_{\irBr{\Lambda}}})}^{-1}}\right)Q
\right]_{ii}^{-1}\,, 
\vspace{1mm}\\
& &\mbox{ for }i\in\Lambda\,,  
\vspace{1mm}\\
\ds\sum_{i\in S}R_i 
&\geq &\underline{J}_{S}
\left(\left.\prod_{i=1}^L\xi_i,r_S\right|r_{\coS}\right)
\vspace{1mm}\\
& &\mbox{ for }S\subseteq \Lambda.
\ea
\right\}
\label{eqn:prth118z}
\eeq
Here we choose unitary matrix $Q$ so that 
$Q^{-1}(\Sigma_{X_{\Lambda}}^{-1}$ $+$ 
$\Sigma_{N_{\Lambda}(r_{\Lambda})}^{-1})Q$ 
becomes the following diagonal matrix: 
\beq
Q^{-1}\left(\Sigma_{X_{\Lambda}}^{-1}
+\Sigma_{N_{\Lambda}(r_{\Lambda})}^{-1}\right)Q
=\left[
\begin{array}{cccc}
\alpha_1  &          &        & \mbox{\huge 0} \\
          & \alpha_2 &        &                \\
          &          & \ddots &                \\
\mbox{\huge 0} &     &        & \alpha_{L}     \\
\end{array}
\right]\,.
\label{eqn:diagg}
\eeq
From the second inequality of (\ref{eqn:prth118z}), we have 
\beq
 \xi_i \geq \alpha_i^{-1}=\alpha_i^{-1}(r_{\Lambda}), 
\quad i=1,2,\cdots, L\,, 
\label{eqn:prthw}
\eeq  
which together with the first inequality of (\ref{eqn:prth118z})
yields that
\beqa 
& &\sum_{i=1}^L \alpha_i^{-1}(r_{\Lambda})
\nonumber\\
&=&{\rm tr}\left[\left(\Sigma_{X_{\Lambda}}^{-1}
+\Sigma_{N_{\Lambda}(r_{\Lambda})}^{-1}\right)^{-1}\right]
\leq \sum_{i=1}^L\xi_i \leq D\,.
\label{eqn:prth101z}
\eeqa
On the other hand, 
by the first inequality of (\ref{eqn:prth118z}), 
(\ref{eqn:prthw}), and Lemma \ref{lm:lm0z}, 
we have 
\beq
\prod_{i=1}^{L}\xi_i\leq \theta(D,r_{\Lambda})\,, 
\label{eqn:prthzb}
\eeq
which together with the third inequality of 
(\ref{eqn:prth118z}) 
yields that
%and (\ref{eqn:prthw}), we have
%From (\ref{eqn:prth101z}) and we have 
\beqa
\sum_{i\in S}R_i 
&\geq &\underline{J}_{S}(\theta(D,r_{\Lambda}),r_S|r_{\coS})
\mbox{ for } S\subseteq \Lambda\,.
\label{eqn:prth131z}
\eeqa
(\ref{eqn:prth101z}) and (\ref{eqn:prth131z}) imply that 
${\cal R}_{L}(D)\subseteq {\cal R}_{L}^{({\rm out})}(D)$.
\hfill \IEEEQED
%$$ 
%{\rm tr}\left[\left(\Sigma_{X_{\Lambda}}^{-1}
%+\Sigma_{N_{\Lambda}(r_{\Lambda})}^{-1}\right)^{-1}\right]
%\leq D\,,
%$$

{\it Proof of 
$R_{{\rm sum},L}(D)\geq {R}_{{\rm sum},L}^{ ({\rm l})}(D)$:}
Assume that
$(R_1, R_2,$ $\!\cdots, R_L) \in {\cal R}_{L}(D)$. 
Then, for any $\delta>0$ and any $n$ with $n\geq n_0(\delta)$, 
there exists $(\varphi_1,\varphi_2, \cdots,$ $\varphi_L,$ $\psi)\in$ 
$\!{\cal F}_{\delta}^{(n)} (R_1,R_2$ $\cdots, R_L)$
such that
$$
\sum_{i=1}^L{\rm E}||{\vc X}_i-\hat{\vc X}_i||^2\leq D+\delta\,. 
$$
For each $l=0,1,\cdots,L-1$, we use 
$(\varphi_{\tau^l{1}}, \varphi_{\tau^l(2)}, \cdots,$ 
 $\varphi_{\tau^l(L)})$ 
for the encoding of $({\vc Y}_1, {\vc Y}_2, \cdots, {\vc Y}_L)$. 
For $i\in {\Lambda}$ and for $l=0,1,\cdots,L-1$, set 
\beqno
W_{l,i}&\defeq &\varphi_{\tau^l(i)}({\vc Y}_i),
\quad
\hat{\vc X}_{l,i}
\defeq \psi_{\tau^l(i)}(\varphi_{\tau^l(i)}({\vc Y}_1)),
\\
r_{l,i}^{(n)} &\defeq& 
\frac{1}{n}I({\vc Y}_i;W_{l,i}|{\vc X}_i).
\eeqno
In particular,
$$
r_{0,i}^{(n)}=r_i^{(n)}=\frac{1}{n}I({\vc Y}_i;W_{i}|{\vc X}_i), 
\quad\mbox{for }i\in\Lambda. 
$$
Furthermore, set 
\beqno
r_{\tau^l(\Lambda)}^{(n)}
& \defeq &(r_{l,1}^{(n)},r_{l,2}^{(n)},\cdots,r_{l,L}^{(n)})\,,
\mbox{ for }l=0,1,\cdots,L-1\,, 
\\
r^{(n)}&\defeq& \frac{1}{L}\sum_{i=1}^{L}r_i^{(n)}\,.
\eeqno 
By the cyclic shift invariant property of the source $X^L$ 
and its noisy observation $Y^L=X^L+N^L$, we have 
\beqa
& &\sum_{i=1}^L{\rm E}||{\vc X}_i-\hat{\vc X}_{l,i}||^2\leq D+\delta 
\quad\mbox{for } 0\leq l\leq L-1\,, 
\label{eqn:convqq1}
\\
& &\frac{1}{L}\sum_{l=0}^{L-1}r_{l,i}^{(n)}=
   \frac{1}{L}\sum_{l=0}^{L-1}r_{\tau^l(i)}^{(n)}
   =\frac{1}{L}\sum_{j=1}^{L}r_j^{(n)}=r^{(n)}
\nonumber\\
& &\quad\mbox{ for } 1\leq i\leq L\,. 
\label{eqn:convqq2}
\eeqa
We choose $L\times L$ unitary matrix $Q=[q_{ij}]$ so that 
\beq
Q^{-1}\Sigma_{X_{\Lambda}}^{-1}Q
=
\left[
\begin{array}{cccc}
\frac{1}{\lambda_1} &            &        & \mbox{\huge 0} \\
          & \frac{1}{\lambda_2}  &        &             \\
          &          & \ddots &                 \\
\mbox{\huge 0} &     &        & \frac{1}{\lambda_L}  \\
\end{array}
\right]\,.
\label{eqn:diaggz}
\eeq
Then, we have 
\beqa
& &Q^{-1}\left(\Sigma_{X_{\Lambda}}^{-1}
+\Sigma_{N_{\Lambda}(r^{(n)})}^{-1}\right)Q
%=Q^{-1}\Sigma_{X_{\Lambda}}^{-1}Q+ \frac{1-{\rm e}^{-2r^{(n)}}}{\sigma^2}I_L
\nonumber\\
&=&
\left[
\begin{array}{cccc}
\frac{1}{\lambda_1} &            &        & \mbox{\huge 0} \\
          & \frac{1}{\lambda_2}  &        &             \\
          &          & \ddots &                 \\
\mbox{\huge 0} &     &        & \frac{1}{\lambda_L}  \\
\end{array}
\right]
+\frac{1-{\rm e}^{-2r^{(n)}}}{\sigma^2}
\left[
\begin{array}{cccc}
{1} &         &      & \mbox{\huge 0} \\
              & {1}  &        &       \\
              &      & \ddots &       \\
\mbox{\huge 0} &     &        & {1}   \\
\end{array}
\right]
%\label{eqn:diaggz}
\nonumber\\
&=&
\left[
\begin{array}{cccc}
\beta_1 &            &        & \mbox{\huge 0}\\
          & \beta_2  &        &               \\
          &          & \ddots &               \\
\mbox{\huge 0} &     &        & \beta_{L}     \\
\end{array}
\right]\,.
\nonumber
\eeqa
We set 
${\vc Z}_{\Lambda}\defeq{\vc X}_{\Lambda}Q$, 
$\hat{\vc Z}_{\tau^l(\Lambda)}\defeq \hat{\vc Z}_{\Lambda}Q$.
Furthermore, set 
\beqno
\xi_{l,i}^{(n)}&\defeq &
\frac{1}{n}{\rm E}||{\vc Z}_i-\hat{\vc Z}_{l,i}||^2\,,
\quad
\bar{\xi}_i^{(n)}\defeq 
\frac{1}{L}\sum_{l=0}^{L-1}\xi_{l,i}^{(n)}\,.
\eeqno
By the rotation invariance of the squared norm 
and (\ref{eqn:convqq1}), we have
\beqa
 \sum_{i=1}^L\bar{\xi}_i^{(n)}
&=&\sum_{i=1}^L\frac{1}{L}\sum_{l=0}^{L-1}
 \frac{1}{n}{\rm E}||{\vc Z}_i-\hat{\vc Z}_{l,i}||^2
\nonumber\\
&=&\frac{1}{L}\sum_{l=0}^{L-1}\sum_{i=1}^L
\frac{1}{n}{\rm E}||{\vc X}_i- \hat{\vc X}_{l,i}||^2
\leq D+\delta\,.
\label{eqn:prthyz}
\eeqa
On the other hand, for $i\in\Lambda$, 
we have the following chain of inequalities:
\beqa
& &\frac{n}{2}\log \left[(2\pi{\rm e})\bar{\xi}_i^{(n)}\right] 
=\frac{n}{2}\log \left[(2\pi{\rm e})
\frac{1}{L}\sum_{l=0}^{L-1}{\xi}_{l,i}^{(n)}\right] 
\nonumber\\
&\MGeq{a} &\frac{1}{L}\sum_{l=0}^{L-1}
 \frac{n}{2}\log \left[(2\pi{\rm e}) \xi_{l,i}^{(n)}\right] 
 \MGeq{b}\frac{1}{L}\sum_{l=0}^{L-1}h({\vc Z}_i|\hat{\vc Z}_{l,i}) 
\label{eqn:ssas}\\
%\nonumber\\
&\geq &\frac{1}{L}\sum_{l=0}^{L-1}
h({\vc Z}_i|{\vc Z}_{\Lambda -\{i\}}W_{\tau^l(\Lambda)}) 
\nonumber\\
&\MGeq{c}&\frac{1}{L}\sum_{l=0}^{L-1} 
\frac{n}{2}\log 
\left\{
(2\pi{\rm e})
\left[Q^{-1}
     \left(
     \Sigma_{X_{\Lambda}}^{-1}+\Sigma_{N_{\Lambda}(r_\Lambda^{(n)})}^{-1}
     \right)Q
\right]_{ii}^{-1}
\right\}
\nonumber\\
&\MEq{d}& \frac{1}{L}\sum_{l=0}^{L-1}
\frac{n}{2}\log 
\left\{
(2\pi{\rm e})
\left[
\frac{1}{\lambda_i}
+\sum_{j=1}^Lq_{ji}^2\cdot
      { \frac{1-{\rm e}^{-2r_{l,j}^{(n)}}}{\sigma^2}}
\right]^{-1}
\right\}
\nonumber\\
&\MGeq{e}& 
\frac{n}{2}\log 
\left\{
(2\pi{\rm e})
\left[\frac{1}{\lambda_i}
+\frac{1}{L}\sum_{l=0}^{L-1}
      \sum_{j=1}^Lq_{ji}^2\cdot
      { \frac{1-{\rm e}^{-2r_{l,j}^{(n)}}}{\sigma^2}}
\right]^{-1}
\right\}.
\nonumber\\
& &\label{eqn:asd}
\eeqa
Step (a) follows from the concavity of $\log t$.
Step (b) follows from Lemma \ref{lm:lm1}. 
Step (c) follows from Lemma \ref{lm:lm2}.
Step (d) follows from (\ref{eqn:diaggz}).
Step (e) follows from the convexity of $-\log t$.
From (\ref{eqn:asd}), we have  
\beqa
\bar{\xi}_i^{(n)}
&\geq&
\left[\frac{1}{\lambda_i}+\frac{1}{L}\sum_{l=0}^{L-1}
      \sum_{j=1}^Lq_{ji}^2\cdot
      { \frac{1-{\rm e}^{-2r_{l,j}^{(n)}}}{\sigma^2}}
\right]^{-1}
\nonumber\\
&\MGeq{a}& 
\left[\frac{1}{\lambda_i}+
      \sum_{j=1}^Lq_{ji}^2\cdot
      { \frac{1-{\rm e}^{-2\frac{1}{L}\sum_{l=0}^{L-1}r_{l,j}^{(n)}}}{\sigma^2}}
\right]^{-1}
\nonumber\\
&=&\left[\frac{1}{\lambda_i}+\frac{1-{\rm e}^{-2r^{(n)}}}{\sigma^2}
\right]^{-1}
\nonumber\\
&=&\beta_i^{-1}(r^{(n)}),\quad\mbox{for } i\in\Lambda\,. 
\label{eqn:prthwz}
\eeqa  
Step (a) follows from the concavity of $1-{\rm e}^{-2t}$. 
On the other hand, by (\ref{eqn:prthyz}) 
and (\ref{eqn:prthwz}), we have
\beq 
\phi(r^{(n)})=\sum_{i=1}^L
\beta_i^{-1}(r^{(n)})\leq \sum_{i=1}^L\bar{\xi}_i^{(n)} 
\leq D+\delta\,.
\label{eqn:prth101zz}
\eeq
Now we proceed to an evaluation of lower bound of the sum rate. 
In a manner quite similar to the derivation of (\ref{eqn:conv1}) 
in the proof of
${\cal R}_L(D)$ $\subseteq $ ${\cal R}_L^{ ({\rm out})}(D)$,
we have
\beqa
& &\sum_{i\in \Lambda}n(R_{\tau^l(i)}+\delta)
\nonumber\\
&\geq &I({\vc X}_{\Lambda};W_{\tau^l(\Lambda)})
+n\sum_{i\in \Lambda}r_{l,i}^{(n)} 
\quad\mbox{for }0\leq l\leq L-1.
\label{eqn:conv1z}
\eeqa
From (\ref{eqn:conv1z}), we have
\beqa
\sum_{i\in \Lambda}n(R_{i}+\delta)
&=&\frac{1}{L}\sum_{l=0}^{L-1}\sum_{i\in \Lambda}n(R_{\tau^l(i)}+\delta)
\nonumber\\
&\geq&\frac{1}{L}\sum_{l=0}^{L-1}
I({\vc X}_{\Lambda};W_{\tau^l(\Lambda)})+nLr^{(n)}\,. 
\label{eqn:conv1z0}
\eeqa
We estimate a lower bound of the first quantity in the right 
members of (\ref{eqn:conv1z0}). 
On this quantity we have the following chain of inequalities:
\beqa
& &\frac{1}{L}\sum_{l=0}^{L-1}
I({\vc X}_{\Lambda};W_{\tau^l(\Lambda)})
\nonumber\\
&=&h({\vc X}_{\Lambda})
  -\frac{1}{L}\sum_{l=0}^{L-1}h({\vc X}_{\Lambda}|W_{\tau^l(\Lambda)})
\nonumber\\
&=&h({\vc X}_{\Lambda})-\frac{1}{L}\sum_{l=0}^{L-1}
   h({\vc Z}_{\Lambda}|W_{\tau^l(\Lambda)})
\nonumber\\
&=&h({\vc X}_{\Lambda})
-\frac{1}{L}\sum_{l=0}^{L-1}\sum_{i=1}^L
h({\vc Z}_i|{\vc Z}^{i-1}W_{\tau^l(\Lambda)})
\nonumber\\
&\geq &h({\vc X}_{\Lambda})
-\sum_{i=1}^L\frac{1}{L}\sum_{l=0}^{L-1}
h({\vc Z}_i|\hat{\vc Z}_{l,i})
\nonumber\\
&\MGeq{a} &
\frac{n}{2}\log\left[(2\pi{\rm e})^{L}|\Sigma_{X_{\Lambda}}|\right]
-\sum_{i=1}^L \frac{n}{2}\log\left[(2\pi{\rm e})\bar{\xi}_i^{(n)}\right]
\nonumber\\
&=&  \frac{n}{2}\log |\Sigma_{X_{\Lambda}}|
    -\frac{n}{2}\log\left[\prod_{i=1}^L\bar{\xi}_i^{(n)}\right]\,.
\label{eqn:prthz0z}
\eeqa
Step (a) follows from (\ref{eqn:ssas}). 
Combining (\ref{eqn:conv1z0}) and (\ref{eqn:prthz0z}), 
we obtain
\beqa
\sum_{i\in \Lambda}(R_i+\delta) 
&\geq & \underline{J}\left(\prod_{i=1}^L\bar{\xi}_i^{(n)},
r^{(n)}\right)\,.
\label{eqn:prth100zz}
\eeqa
Set
\beqno
r
&\defeq&\limsup_{n\to\infty}\irb{r^{(n)}}
=\limsup_{n\to\infty}\frac{1}{L}\sum_{i=1}^L
\frac{1}{n}I(\irBw{\vc Y}_i;\irb{W_i}|{\vc X}_i), 
\\
\bar{\xi}_i
&\defeq&\limsup_{n\to\infty}\bar{\xi}_i^{(n)}
=\limsup_{n\to\infty}
 \frac{1}{L}\sum_{l=0}^{L-1}
 \frac{1}{n}{\rm E}||{\vc Z}_i-\hat{\vc Z}_{l,i}||^2.
\eeqno
By letting $n\to\infty$ in 
    (\ref{eqn:prthwz}),
    (\ref{eqn:prth101zz}), 
and (\ref{eqn:prth100zz}), we obtain 
\beq
\left.
\ba{rcl}
\bar{\xi}_i &\geq &\beta_i^{-1}(r)\mbox{ for }i\in \Lambda\,, 
\vspace{1mm}\\
\phi(r)&=&\ds\sum_{i=1}^L\beta_i^{-1}(r)\leq\sum_{i=1}^L\bar{\xi}_i 
\leq D+\delta\,,
\vspace{1mm}\\
\ds\sum_{i\in \Lambda}(R_i+\delta) 
&\geq & \ds \underline{J}\left(\prod_{i=1}^L\bar{\xi}_i,r\right)\,.
\ea
\right\}
\label{eqn:prth121zz}
\eeq
Since $\delta$ can be made arbitrary small, we have  
\beq
\left.
\ba{rcl}
\bar{\xi}_i &\geq &\beta_i^{-1}(r)\mbox{ for }i\in \Lambda\,, 
\vspace{1mm}\\
\phi(r)&=&\ds \sum_{i=1}^L\beta_i^{-1}(r)\leq\sum_{i=1}^L\bar{\xi}_i 
\leq D\,,
\vspace{1mm}\\
\ds \sum_{i\in \Lambda}R_i
&\geq & \ds \underline{J}
\left(\prod_{i=1}^L\bar{\xi}_i,r\right)\,.
\ea
\right\}
\label{eqn:prth122zz}
\eeq
From the first and second inequality of 
(\ref{eqn:prth122zz}) and Lemma \ref{lm:lm0z}, we have
$$
\prod_{i=1}^L\bar{\xi}_i\leq \theta(D,r)\,.
$$  
Hence, we have  
\beqno
\sum_{i\in \Lambda}R_i 
&\geq &\underline{J}(\theta(D,r),r)\mbox{ and }\phi(r)\leq D\,,
\eeqno
which imply that 
$R_{{\rm sum},L}(D)\geq R_{{\rm sum},L}^{\rm (l)}(D)$.
\hfill \IEEEQED

%\irb{r_i^{(n)}}\defeq\frac{1}{n}I(\irBw{\vc Y}_i;\irb{W_i}|{\vc X}_i)\,. 

\subsection{Derivation of the Inner Bound}

In this subsection we prove ${\cal R}_L^{({\rm in})}(D)$ 
$\subseteq$ ${\cal R}_L(D)$ 
stated in Theorem \ref{th:conv2}. 

{\it Proof of ${\cal R}_L^{({\rm in})}(D)$
$\subseteq$ ${\cal R}_L(D)$:} Since 
$\hat{\cal R}_L^{({\rm in})}(D)$ $\subseteq$ ${\cal R}_L(D)$
is proved by Theorem \ref{th:direct}, 
it suffices to show 
${\cal R}_L^{({\rm in})}(D)$
$\subseteq$ 
$\hat{\cal R}_L^{({\rm in})}(D)$
to prove ${\cal R}_L^{({\rm in})}(D)$
$\subseteq$ ${\cal R}_L(D)$. 
We assume that $R^L\in {\cal R}_L^{({\rm in})}(D)$. 
Then, there exists nonnegative vector $r^L$ such that
\beq
{\rm tr}\left[\left(\Sigma_{X^L}^{-1}
  +\Sigma_{N^L(r^L)}^{-1}\right)^{-1}\right]\leq D
\label{eqn:zsa0}
\eeq
and 
\beq
\sum_{i\in S} R_i\geq K(r_S|r_{\coS})
\mbox{ for any }S \subseteq \Lambda\,. 
\label{eqn:zsa1}
\eeq
Let $V_i, i\in \Lambda$ be $L$ independent Gaussian 
random variables with mean 0 and variance $\sigma_{V_i}^2$.
Define Gaussian random variables $U_i, i\in \Lambda$ by 
$
U_i=X_i+N_i+V_i. %,\:\: i\in \Lambda.
$
By definition it is obvious that
\beq
\left.
\ba{l}
U^L\to Y^L \to X^L \\
U_S\to Y_S \to X^L \to Y_{\coS}\to U_{\coS}\\
\mbox{ for any } S\subseteq \Lambda\,.  
\ea
\right\}
\label{eqn:gau00sz} 
\eeq
For given $r_i \geq 0, i\in \Lambda$ and $D>0$, choose 
$\sigma_{V_i}^2$ so that 
$\sigma_{V_i}^2=\sigma_{N_i}^2/({\baseN}^{2r_i}-1)$
when $r_i>0$. When $r_i=0,$ we choose $U_i$ so that $U_i$ take 
the constant value zero. Then, the covariance matrix 
of $N^L+V^L$ becomes $\Sigma_{N^L(r^L)}$. Choose covariance 
matrix $\Sigma_{D}$ so that
$$
{\rm tr}[\Sigma_{D}]=D\,,\quad 
\Sigma_{D}
\succeq (\Sigma_{X^L}^{-1} +\Sigma_{N^L(r^L)}^{-1})^{-1}.
$$ 
%where $A \preceq B$ means that $B-A$ is positive semi-definite. 
Since (\ref{eqn:zsa0}), the above choice of $\Sigma_{D}$ is possible.
Define the linear function $\tilde{\psi}$ of $U^L$ by 
$$
{\tilde{\psi}}\left(U^L\right) 
=U^L\Sigma_{N^L(r^L)}^{-1}
(\Sigma_{X^L}^{-1} +\Sigma_{N^L(r^L)}^{-1})^{-1}\,.
$$
Set $\hat{X}^L={\tilde{\psi}}\left(U^L\right)$ and 
\beqa 
d_{ii}
& \defeq & {\rm E}||{X}_i-\hat{X}_i||^2\,,
\nonumber\\
d_{ij}
& \defeq &
{\rm E}\left({X}_i-\hat{X}_i\right)
       \left({X}_j-\hat{X}_j\right)
       \,,
%\nonumber\\
%& &
1 \leq i\ne j \leq L.
\nonumber
\eeqa
Let $\Sigma_{{X}^L-\hat{X}^L}$ be a covariance matrix 
with $d_{ij}$ in its $(i,j)$ element. 
Then, by simple computations we can show that
\beq
\Sigma_{X^L-\hat{X}^L}=(\Sigma_{X^L}^{-1} +\Sigma_{N^L(r^L)}^{-1})^{-1}
\preceq \Sigma_D
\label{eqn:gau2zz} 
\eeq
and that for any $S\subseteq \Lambda$, 
\beqa
& &J_S(r_S|r_{\coS})=I(Y_S;U_S|U_{\coS})\,.
\label{eqn:gau1z} 
\eeqa
From (\ref{eqn:zsa0}) and (\ref{eqn:gau2zz}), we have 
\beqa
& &||X^L-{\tilde{\psi}}\left(U^L \right)||^2=||X^L-\hat{X}^L||^2
\nonumber\\
&=&{\rm tr}\left[\left(\Sigma_{X^L}^{-1}
  +\Sigma_{N^L(r^L)}^{-1}\right)^{-1}\right]
\leq {\rm tr}\left[\Sigma_D\right]=D\,.
\label{eqn:gau1sz}
\eeqa
From (\ref{eqn:gau00sz}) and (\ref{eqn:gau1sz}), we have 
$U^L\in {\cal G}(D)$. Then, from (\ref{eqn:gau1z}) 
$$
{\cal R}_L^{({\rm in})}(D)\subseteq \hat{\cal R}_L^{({\rm in})}(D)\,,
$$
completing the proof. \hfill \IEEEQED 

\section{Proofs of the Results on Matching Conditions}

In this section we prove 
Lemmas \ref{lm:lem1}-\ref{lm:pro3b} and Theorems 
\ref{th:matchTh} and \ref{th:matchTh2} 
stated in Section III. 

\subsection{
Proof of Lemma \ref{lm:lem1}
} 

In this subsection we prove Lemma \ref{lm:lem1}.
We first present a preliminary observation on 
${\cal R}_L^{(\rm out)}(D)$. For $r^L\in {\cal B}_L(D)$, 
we examine a form of the region 
\beqno
{\cal R}_L^{({\rm out})}(D,r^L)
&=&
\ba[t]{l}
  \left\{R^L \right. : 
  \ba[t]{l}
  \ds \sum_{i \in S} R_i 
  \geq {\underline{J}}_{S}\left(\theta(D,r^L),r_S|r_{\coS}\right)
  \vspace{1mm}\\
  \mbox{ for any }S \subseteq \Lambda\,. 
  \left. \right\}\,.
  \ea
\ea
\eeqno
Let $(\Lambda,f)$, $f=\{{f}_S(r_S|r_{\coS})\}_{S\subseteq \Lambda}$ 
be a co-polymatroid defined in Property \ref{pr:matroid}. 
Using $(\Lambda,f)$, ${\cal R}_L^{(\rm out)}(D,r^L)$ 
is expressed as
\beqno
{\cal R}_L^{({\rm out})}(D,r^L)
&=&
\ba[t]{l}
  \left\{R^L \right. : 
  \ba[t]{l}
  \ds \sum_{i \in S} R_i 
  \geq {f}_{S}\left(r_S|r_{\coS} \right)
  \vspace{1mm}\\
  \mbox{ for any }S \subseteq \Lambda\,. 
  \left. \right\}\,.
  \ea
\ea
\eeqno
The set ${\cal R}_L^{(\rm out)}(D,r^L$$)$ 
forms a kind of polytope which is called 
a {\it co-polymatroidal polytope} in 
the terminology of matroid theory. 
It is well known as a property of this kind of 
polytope that the polytope ${\cal R}_L^{(\rm out)}(D,r^L)$ 
consists of $L!$  end-points whose components are
given by 
\beq
\left.
\ba{rcl}
\hspace*{-4mm}& &R_{\pi(i)}\\
\hspace*{-4mm}&=&{f}_{\{\pi(i), \cdots, \pi(L)\}}
   (r_{\{\pi(i),\cdots, \pi(L)\}}|r_{\{\pi(1),\cdots, \pi(i-1)\}})
\vspace*{1mm}\\
\hspace*{-4mm}& &-{f}_{\{\pi(i+1), \cdots, \pi(L)\}}
   (r_{\{ \pi(i+1),\cdots,\pi(L)\}}|r_{\{ \pi(1), \cdots, \pi(i)\}})
\vspace*{1mm}\\
\hspace*{-4mm}& &\mbox{$\quad$ for }i=1,2,\cdots, L-1\,,
\vspace*{1mm}\\
\hspace*{-4mm}& &R_{\pi(L)}
={f}_{\{\pi(L)\}}(r_{\pi(L)}|r_{\{ \pi(1), \cdots, \pi(L-1)\}})\,,
\ea
\right\}%\,,
\label{eqn:zzza}
\eeq
where $\pi$ is an arbitrary permutation on $\Lambda$, that is
$$
\pi=\left(
\ba{cccccc}
    1 &2&\cdots&    i &\cdots&    L\\ 
\pi(1)&\pi(2)&\cdots&\pi(i)&\cdots&\pi(L)
\ea
\right)\,.
$$
For $l=1,2,\cdots,L$, set
\beqno
{\cal B}_{\pi,l}(D)
&\defeq& 
\{r^L: 
\ba[t]{l} 
r^L\in {\cal B}_{L}(D)\mbox{ and }\\
r_{\pi(i)}=0\mbox{ for } i=l+1,\cdots,L \}\,,\\
\ea
\\
\partial{\cal B}_{\pi,l}(D)
&\defeq& 
\{r^L: 
\ba[t]{l} 
r^L\in \partial{\cal B}_{L}(D)\mbox{ and }\\
r_{\pi(i)}=0\mbox{ for } i=l+1,\cdots,L \}\,. \\
\ea
\eeqno
In particular, when $\pi$ is the identity map, we omit $\pi$ to write 
${\cal B}_l(D)$ and $\partial{\cal B}_l(D)$. 
By Property \ref{pr:prz01z}, when 
$r^L \in {\cal B}_{\pi,l}(D)$, the end-point given 
by (\ref{eqn:zzza}) becomes  
\beq
\left.
\ba{rcl}
& &R_{\pi(i)}\\
&=&{f}_{\{\pi(i), \cdots, \pi(l)\}}
   (r_{\{\pi(i),\cdots, \pi(l)\}}|r_{\{\pi(1),\cdots, \pi(i-1)\}})
\vspace*{1mm}\\
& &-{f}_{\{\pi(i+1), \cdots, \pi(l)\}}
   (r_{\{ \pi(i+1),\cdots,\pi(l)\}}|r_{\{ \pi(1), \cdots, \pi(i)\}})
\vspace*{1mm}\\
& &\mbox{$\quad$ for }i=1,2,\cdots, l-1\,,
\vspace*{1mm}\\
& &R_{\pi(l)}={f}_{\{\pi(l)\}}(r_{\pi(l)}|r_{\{ \pi(1), \cdots, \pi(l-1)\}})\,,
\vspace*{1mm}\\
& &R_{\pi(i)}=0,\mbox{ for }i=l+1,\cdots, L\,.
\ea
\right\}%\,,
\label{eqn:zza0}
\eeq

{\it Proof of Lemma \ref{lm:lem1}:} 
Fix $r^L \in {\cal B}_L(D)$ arbitrary. 
Let $R^L$ be a nonnegative rate vector such that 
$L$ components of $R^L$ satisfy (\ref{eqn:zzza}). 
To prove Lemma \ref{lm:lem1}, 
it suffices to show that this nonnegative vector 
belongs to ${\cal R}_L^{(\rm in)}(D)$. 
For $l=1,2,\cdots,L$, we prove the claim that under 
the MD condition, if $r^L \in {\cal B}_{\pi,l}(D)$, 
then, the rate vector $R^L$ satisfying 
(\ref{eqn:zza0}) belongs to ${\cal R}_L^{(\rm in)}(D)$. 
We prove this claim by induction with respect to $l$. 
When $l=1$, from (\ref{eqn:zza0}), we have 
\beq
\left.
\ba{rcl}
R_{\pi(1)}&=&{f}_{\{\pi(1)\}}(r_{\pi(1)})\,,
\vspace*{1mm}\\
R_{\pi(i)}&=&0,\mbox{ for }i=2,\cdots,L\,.
\ea
\right\}
\label{eqn:aaz}
\eeq
The function ${f}_{\{\pi(1)\}}(r_{\pi(1)})$ is computed as
\beqa
\hspace*{-4mm}& &{f}_{\{\pi(1)\}}(r_{\pi(1)})
\nonumber\\
\hspace*{-4mm}&=&\underline{J}_{\{\pi(1)\}}
\left.(\theta(D,r^L),r_{\pi(1)}|r_{\{\pi(1)\}^{\rm c}})
\right|_{r_{\{\pi(1)\}^{\rm c}}={\lvc 0}}
\nonumber\\
\hspace*{-4mm}&=&
\frac{1}{2}\log^{+}
\left[\ts 
   \frac{\ds {\baseN}^{2r_{\pi(1)}}}
        {\ds \left|\Sigma_{X^L}^{-1}\right|
         \theta(D,r^L)|_{r_{\{\pi(1)\}^{\rm c}}={\lvc 0}}}
  \right]\,.
\label{eqn:za00a}
\eeqa
Since $r^L \in {\cal B}_{\pi,l}(D)$, 
we can decrease $r_{\pi(1)}$ keeping 
$r^L\in {\cal B}_{\pi,1}(D)$ 
so that it arrives at $r_{\pi(1)}^{*}=0$ or a 
positive $r_{\pi(1)}^{*}$ satisfying
\beq
(r_{\pi(1)}^{*},r_{\{\pi(1)\}^{\rm c}})
=(r_{\pi(1)}^{*},\underbrace{0,\cdots,0}_{L-1})
\in \partial{\cal B}_{\pi,1}(D)\,.
\label{eqn:z00}
\eeq
Let $(R_{\pi(1)}^{*},$ $\cdots, R_{\pi(L)}^{*})$
be a rate vector corresponding to 
$(r_{\pi(1)}^{*},$ $r_{\{\pi(1)\}^{\rm c}})$.
If $r_{\pi(1)}^{*}=0$, we have 
$r^L={\vc 0}\in {\cal B}_L(D)$.
Then, we have
$$
{\rm tr}\left[\left(\Sigma_{X^{L}}^{-1}
+\Sigma_{N^{L}(r^L)}^{-1}\right)^{-1}\right]
={\rm tr}\left[\Sigma_{X^{L}}\right]
\leq D\,.
$$
This contradicts the first assumption of 
$D<{\rm tr}\left[\Sigma_{X^{L}}\right].$
Therefore, $r_{\pi(1)}^{*}$ must be positive. 
Then, from (\ref{eqn:z00}), we have 
\beqno
%& &
(R_{\pi(1)}^{*},\cdots, R_{\pi(L)}^{*})
%\\
&=&(R_{\pi(1)}^{*},\underbrace{0,\cdots,0}_{L-1})
\in {\cal R}_L^{(\rm in)}(D)\,. 
\eeqno
By (\ref{eqn:za00a}) and the MD condition, 
${f}_{\{\pi(1)\}}(r_{\pi(1)})$ is a monotone increasing 
function of $r_{\pi(1)}$. 
Then, we have $R_{\pi(1)} \geq $ $R_{\pi(1)}^{*}$.
Hence, we have 
\beqno
%& &
(R_{\pi(1)},\cdots, R_{\pi(L)})
%\\
&=&(R_{\pi(1)},\underbrace{0,\cdots, 0}_{L-1})
   \in {\cal R}^{(\rm in)}_L(D)\,. 
\eeqno
Thus, the claim holds for $l=1$. 
We assume that the claim holds for $l-1$.
%$r_0^L \in {\cal B}_{\pi,l-1}(D)$, $(R_0,R^L)\in $ 
%${\cal R}^{(\rm in )}(r_0^L,D)$. 
Since 
$
{\rm tr}\left[
(\Sigma_{X^L}^{-1}+\Sigma_{N^L(r^L)}^{-1})^{-1}
\right]
$ 
is a monotone increasing function of 
$r_{\pi(l)}$ on ${\cal B}_{\pi,l}(D)$, we can 
decrease $r_{\pi(l)}$ keeping $r^L\in {\cal B}_{\pi,l}(D)$ 
so that it arrives at $r_{\pi(l)}^{*}=0$ or 
a positive $r_{\pi(l)}^{*}$ satisfying
\beq
(r_{\pi(l)}^{*},r_{\{\pi(l)\}^{\rm c}})
\in \partial{\cal B}_{\pi,l}(D)\,.
\label{eqn:zaaa}
\eeq
Let $(R_{\pi(1)}^{*},$ $\cdots, R_{\pi(L)}^{*})$
be a rate vector corresponding to 
$(r_{\pi(l)}^{*},$ $r_{\{\pi(l)\}^{\rm c}})$.
By Property \ref{pr:matroid} part b) and the MD 
condition, the $l$ functions   
\beqno
& &{f}_{\{\pi(i), \cdots, \pi(l)\}}
   (r_{\{\pi(i),\cdots, \pi(l)\}}|r_{\{\pi(1),\cdots, \pi(i-1)\}})
\\
& &-{f}_{\{\pi(i+1), \cdots, \pi(l)\}}
   (r_{\{ \pi(i+1),\cdots,\pi(l)\}}|r_{\{ \pi(1), \cdots, \pi(i)\}})
\\
& &\mbox{$\quad$ for }i=1,2,\cdots, l-1\,,
\vspace*{1mm}\\
& &{f}_{\{\pi(l)\}}(r_{\pi(l)}|r_{\{ \pi(1), \cdots, \pi(l-1)\}})
\eeqno
appearing in the right members of (\ref{eqn:zza0}) are monotone 
increasing functions of $r_{\pi(l)}$. Then, from (\ref{eqn:zza0}), 
we have
\beq
\left.
\ba{rcl}
R_{\pi(i)} &\geq &R_{\pi(i)}^{*} 
\mbox{ for }i=1,2,\cdots,l\,,
\\
R_{\pi(i)} &= &R_{\pi(i)}^{*}=0
\mbox{ for }i=l+1,\cdots,L\,.
\ea
\right\}
\label{eqn:zazaa}
\eeq
When $r_{\pi(l)}^{*}=0$, we have 
$
(r_{\pi(l)}^{*},r_{\{\pi(l)\}^{\rm c}})
\in {\cal B}_{\pi,l-1}(D)\,.
$
Then, by induction hypothesis we have
$$
(R_{\pi(1)}^{*},\cdots, R_{\pi(L)}^{*})
\in {\cal R}_L^{(\rm in )}(D)\,. 
$$
When $r_{\pi(l)}^{*}>0$, from (\ref{eqn:zaaa}), we have 
$$
(R_{\pi(1)}^{*},\cdots, R_{\pi(L)}^{*})
\in {\cal R}_L^{(\rm in)}(D)\,. 
$$
Hence, by $(\ref{eqn:zazaa})$, we have
\beqno
& &
(R_{\pi(1)},\cdots, R_{\pi(L)})
\\
&=&(R_{\pi(1)}, \cdots, R_{\pi(l)},
\underbrace{0,\cdots, 0}_{L-l})
\in {\cal R}^{(\rm in)}_L(D)\,. 
\eeqno
Thus, the claim is proved.   
\hfill\IEEEQED

\subsection{
Proofs of 
Lemmas \ref{lm:pro3} and \ref{lm:pro3b} 
and Theorems \ref{th:matchTh} and \ref{th:matchTh2}
}

In this subsection we prove Lemmas \ref{lm:pro3} 
and \ref{lm:pro3b} and Theorems \ref{th:matchTh} and \ref{th:matchTh2}. 

We first observe that using the eigen values 
$\alpha_k=\alpha_k(u^L),$ $k\in \Lambda$
of $\Sigma_{X^L}^{-1}$ $+\Sigma_{N^L(u^L)}^{-1}$, 
the condition  
$$
{\rm tr}
\left[ 
\left(
\Sigma_{X^L}^{-1}+\Sigma_{N^L(u^L)}^{-1} 
\right)^{-1}
\right]\leq D
$$
is rewritten as 
\beq
\sum_{i=1}^{L}\frac{1}{\alpha_{i}(u^L)}\leq D\,.
\label{eqn:aa00z}
\eeq
Next, we present a lemma necessary to prove Lemma \ref{lm:pro3}.
\begin{lm}\label{lm:eigenlm} For the eigen values 
$\alpha_k=\alpha_k(u^L),k\in \Lambda$ 
of $\Sigma_{X^L}^{-1}+\Sigma_{N^L(u^L)}^{-1}$ and 
for $u_i, i\in \Lambda$, we have the followings:
$$
\alpha_{\min}\leq u_i \leq \alpha_{\max}\,,
\:\: 
\frac{\partial \alpha_k}{\partial u_i}
\geq 0, \mbox{ for }k\in \Lambda,
\:\:
\sum_{k=1}^L\frac{\partial \alpha_k}{\partial u_i}=1\,.
%\label{zsz000}
$$
\end{lm}

Proof of this lemma needs some analytical arguments 
on the eigen values of positive semidefinite Hermitian matrix. 
Detail of the proof will be given in Appendix C. 

{\it Proof of Lemma \ref{lm:pro3}:}
Let $S$ be a set of integers that 
satisfies $\alpha_i^{-1}\geq \xi$ in the definition 
of $\theta(D,u^L)$.
Then,  $\theta(D,u^L)$ is computed as 
\beqno
\theta(D,u^L)&=&{\ts \frac{1}{(L-|S|)^{L-|S|}}}
   \left(\prod_{k\in S}\frac{1}{\alpha_k}\right)
   \left(D-\sum_{k\in S} \frac{1}{\alpha_k}\right)^{L-|S|}\,.
\eeqno
Fix $i \in \Lambda$ arbitrary. 
For simplicity of notation 
we set $A_i \defeq (a_{ii}+c_i)$ and set 
$$
\Psi\defeq \log \frac{Dc_i}{A_i-u_i}-\log \theta(D,u^L)\,.
$$
Computing the partial derivative of $\Psi$ by $u_i$, we obtain
\beq
%& &
\frac{\partial \Psi}{\partial u_i}
%\nonumber\\
=\sum_{k\in S}
\left(
\frac{\partial \alpha_k}{\partial u_i}
\right)
\left[\frac{1}{\alpha_k}
-\frac{L-|S|}
{D-{\ds \sum_{k\in S}}\frac{1}{\alpha_k}}\frac{1}{\alpha_k^2}
\right]
+\frac{1}{A_i-u_i}\,.
\label{zsz0}
\eeq
From Lemma \ref{lm:eigenlm} and (\ref{zsz0}), we obtain
$$
%& &
\hspace*{-2mm}\frac{\partial \Psi}{\partial u_i}
%\nonumber\\
\geq \sum_{k\in S}
\left(
\frac{\partial \alpha_k}{\partial u_i}
\right)
\left[\frac{1}{\alpha_k}
-\frac{L-|S|}{D-{\ds \sum_{k\in S}} \frac{1}{\alpha_k}} 
\frac{1}{\alpha_k^2}
+\frac{1}{A_i-\alpha_{\min}}
\right]\,.
$$
To examine signs of contents of the above summation we 
set  
\beqno
\Phi_k 
&\defeq& 
\left\{
D-\sum_{k\in S}{\frac{1}{\alpha_k}-\frac{L-|S|}{\alpha_k}}
\right\}(A_i-\alpha_{\min})
\\
& &+\alpha_{k}\left(D-\sum_{k\in S}{\frac{1}{\alpha_k}}\right).
\eeqno
If $|S|=L$, $\Phi_k \geq 0, k \in \Lambda$ is obvious. 
We hereafter assume $|S|\leq L-1$. Computing $\Phi_k$, we obtain
\beqa
\Phi_k&=&
  A_i\left(D-\sum_{k\in S} {\frac{1}{\alpha_k}}\right)
  -{\frac{L-|S|}{\alpha_k}}\cdot(A_i-\alpha_{\min})
\nonumber\\
& &+(\alpha_{k}-\alpha_{\min})
\left(D-\sum_{k\in S}{\frac{1}{\alpha_k}}\right)
\nonumber\\
&\geq & A_i\left(D-\sum_{k\in S}{\frac{1}{\alpha_k}}\right)
         -{\frac{L-|S|}{\alpha_k}}\cdot(A_i-\alpha_{\min})
\nonumber\\
&\MGeq{a}& A_i\sum_{k\in \Lambda-S} \frac{1}{\alpha_{k}}
-{\frac{L-|S|}{\alpha_{k}}}\cdot(A_i-\alpha_{\min})
\nonumber\\
&\geq & A_i\cdot\frac{L-|S|}{\alpha_{\max}}
-{\frac{L-|S|}{\alpha_{\min}}}\cdot(A_i-\alpha_{\min})
\nonumber\\
& = & 
{A_i}(L-|S|)
\left(\frac{1}{\alpha_{\max}}-\frac{1}{\alpha_{\min}}+\frac{1}{A_i}\right)\,.
\label{eqn:zsd0}
\eeqa
Step (a) follows from the inequality (\ref{eqn:aa00z}), that is, 
$$
D-\sum_{k=1}^L\frac{1}{\alpha_{k}(r^L)}\geq 0\,.
$$
From (\ref{eqn:zsd0}), we can see that if 
$$
 \frac{1}{\alpha_{\min}(r^L)}
-\frac{1}{\alpha_{\max}(r^L)} \leq \frac{1}{A_i} 
\mbox{ for }i\in \Lambda,
$$
then, $\Phi_k\geq 0$ for $k\in S\,.$ \hfill\IEEEQED

{\it Proof of Theorem \ref{th:matchTh}: } 
%Let $\alpha_{\max}(r^L)$ be the maximum eigen value 
%of $D(\Sigma_{X^L}^{-1}+$ $\Sigma_{N^L}^{-1})$. 
By (\ref{eqn:aa00z}), we have 
\beqno
\frac{1}{\alpha_{\min}(r^L)}
&\leq &D-\frac{L-1}{\alpha_{\max}(r^L)}
\\
&=&\frac{1}{\alpha_{\max}(r^L)}+D-\frac{L}{\alpha_{\max}(r^L)}
\,.
\eeqno
Hence, if 
$$
%\left[1- \frac{L-1}{\alpha_{\max}(r^L)}\right] 
%-\frac{1}{\alpha_{\max}(r^L)}
%=
D-\frac{L}{\alpha_{\max}(r^L)} \leq \frac{1}{a_{ii}+c_i}\,, 
$$
or equivalent to 
\beq
\left(D-\frac{1}{a_{ii}+c_i}\right) 
\alpha_{\max}(r^L) \leq L
\label{eqn:zxz}
\eeq
holds for $r^L\in {\cal B}_L(D)$ and 
$i\in \Lambda$, the condition on $\alpha_{\min}$ and $\alpha_{\max}$ 
in Lemma \ref{lm:pro3} holds. 
By Lemma \ref{lm:eigenlm}, we have 
\beq
\alpha_{\max}(r^L)\leq \alpha_{\max}^{\ast}\mbox{ for }r^L\in {\cal B}_L(D). 
\label{eqn:zsd00}
\eeq
It can be seen  from (\ref{eqn:zxz}) and (\ref{eqn:zsd00}) that 
\beq
\left(D-\frac{1}{a_{ii}+c_i}\right) 
\alpha_{\max}^{\ast} \leq L
\mbox{ for }i \in \Lambda\,.
\label{eqn:zsd}
\eeq
is a sufficient condition for (\ref{eqn:zxz}) to hold.
%Since 
By Lemma \ref{lm:eigenlm}, we have
\beq
a_{ii}+c_i\leq \alpha_{\max}^{\ast}
\mbox{ for }i\in \Lambda.
\label{eqn:zsdzz}
\eeq
From (\ref{eqn:zsd}) and (\ref{eqn:zsdzz}), we have 
$$ 
\left(D-\frac{1}{a_{ii}+c_i}\right) 
        \alpha_{\max}^{\ast}
\leq D\alpha_{\max}^{\ast}-1\,.
$$
Thus, if we have 
$
D\alpha_{\max}^{\ast}-1 \leq L
$
or equivalent to $D\leq (L+1)/\alpha_{\max}^{\ast}$, 
we have (\ref{eqn:zsd}). 
\hfill\IEEEQED

{\it Proof of Lemma \ref{lm:pro3b}:}
%Let $a$ be an identical $(i,i)$ elements of 
%$\Sigma_{X^L}^{-1}$ and 
%Set 
%\beq
%u \defeq a+c(1-{\baseN}^{-2r})
%\label{eqn:trz}
%\eeq
%From (\ref{eqn:trz}), we have 
%$$
%2r=\log \frac{c}{a+c-u}\,. 
%$$ 
%By the above transformation we regard $\theta(D,r)$
%as a function of $u$, that is, $\theta(D,r)=\theta(D,u).$ 
%We first derive expression of $\theta(D,u)$ using 
%$\beta_i, i\in \Lambda$. 
Let $S$ be a set of integers that 
satisfies $\beta_i^{-1}\geq \xi$ in the definition 
of $\theta(D,r)$.
Then %we have 
%$$
$\theta(D,r)$ is computed as 
\beqno
\theta(D,r)&=&{\ts \frac{1}{(L-|S|)^{L-|S|}}}
   \left(\prod_{k\in S}\frac{1}{\beta_k}\right)
   \left(D-\sum_{k\in S} \frac{1}{\beta_k}\right)^{L-|S|}\,.
\eeqno
Fix $i \in \Lambda$ arbitrary and set
$$
\Psi\defeq 2Lr-\log \theta(D,r)\,.
$$
Computing the derivative of $\Psi$ by $r$, we obtain
\beqno
\frac{{\rm d}\Psi}{{\rm d}r}
&=&\frac{2}{\sigma^2{\rm e}^{2r}}\sum_{k\in S}
 \left[\frac{1}{\beta_k}
-\frac{L-|S|}{D-{\ds \sum_{k\in S}}\frac{1}{\beta_k}}\frac{1}{\beta_k^2}
\right]
+{2L}
\nonumber\\
&=&
\frac{2}{\sigma^2{\rm e}^{2r}}
\sum_{k\in S}
\left[\frac{1}{\beta_k}
-\frac{L-|S|}{D-{\ds \sum_{k\in S}} \frac{1}{\beta_k}} 
\frac{1}{\beta_k^2}
+\sigma^2{\rm e}^{2r}\cdot\frac{L}{|S|}
\right]\,.
%\nonumber\\
%&\MGeq{a}&
%\sum_{k\in S}
%\left[\frac{1}{\beta_k}
%-\frac{L-|S|}{D-{\ds \sum_{k\in S}} \frac{1}{\beta_k}} 
%\frac{1}{\beta_k^2}
%+\frac{L}{|S|(a+c-\beta_{\min})}
%\right]\,.
\eeqno
%Step (a) follows from the first inequality of Lemma \ref{lm:eigenlm}. 
To examine signs of contents of the above summation we 
set  
\beqno
\Phi_k 
&\defeq& 
D-\sum_{k\in S}{\frac{1}{\beta_k}-\frac{L-|S|}{\beta_k}}
\\
& &
+\sigma^2{\rm e}^{2r}
\frac{L}{|S|}\beta_{k}\left(D-\sum_{k\in S}{\frac{1}{\beta_k}}\right).
\eeqno
If $|S|=L$, $\Phi_k \geq 0, k \in \Lambda$ is obvious. 
We hereafter assume $|S|\leq L-1$. Computing $\Phi_k$, we obtain
\beqa
& &\Phi_k
\nonumber\\
&\MGeq{a}&\sum_{k\in \Lambda-S}\frac{1}{\beta_k}
       -\frac{L-|S|}{\beta_k}
+\sigma^2{\rm e}^{2r}
    \frac{L}{|S|}\beta_{k}\sum_{k\in \Lambda-S}{\frac{1}{\beta_k}}
\nonumber\\
&\geq &  \frac{L-|S|}{\beta_{\max}}
        -\frac{L-|S|}{\beta_{\min}}
 +\sigma^2{\rm e}^{2r}\frac{L}{|S|}(L-|S|)\frac{\beta_{\min}}{\beta_{\max}}
\nonumber\\
&=& (L-|S|)
    \left[
    \frac{1}{\beta_{\max}}-\frac{1}{\beta_{\min}}
   +\sigma^2{\rm e}^{2r}\frac{L}{|S|}
     \cdot\frac{\beta_{\min}}{\beta_{\max}} 
    \right]\,.
\label{eqn:zsd0z}
\eeqa
Step (a) follows from %that is equivalent
$$
D-\sum_{k=1}^L\frac{1}{\beta_k}\geq 0
\Leftrightarrow D-\sum_{k\in S}\frac{1}{\beta_k}
\geq \sum_{k\in \Lambda-S}\frac{1}{\beta_k}\,.
$$
%(\ref{eqn:aa00z}). 
From (\ref{eqn:zsd0z}), we can see that if 
\beqa
\frac{1}{\beta_{\min}}-\frac{1}{\beta_{\max}} 
&\leq &
\sigma^2{\rm e}^{2r}\frac{L}{|S|}\cdot 
\frac{\beta_{\min}}{\beta_{\max}}\,, 
\label{eqn:zaxx}
\eeqa
then, $\Phi_k\geq 0$ for $k\in S\,.$ Since $|S|\leq L-1$, 
\beqno
\frac{1}{\beta_{\min}(r)}-\frac{1}{\beta_{\max}(r)} 
&\leq &
\sigma^2{\rm e}^{2r}\frac{L}{L-1}\cdot 
\frac{\beta_{\min}(r)}{\beta_{\max}(r)}
\eeqno
is a sufficient condition for (\ref{eqn:zaxx}) to hold. 
\hfill\IEEEQED

{\it Proof of Theorem \ref{th:matchTh2}:}
Computing $\beta_{\min}^{-1}-\beta_{\max}^{-1}$, we have 
\beqno
& &\frac{1}{\beta_{\min}(r)}-\frac{1}{\beta_{\max}(r)}
\\
&=&\frac{\lambda_{\max}-\lambda_{\min}}
{\left\{
 1+\frac{\lambda_{\max}} {\sigma^2} (1-{\rm e}^{-2r})
 \right\}
 \left\{
 1+\frac{\lambda_{\min}} {\sigma^2} (1-{\rm e}^{-2r})
 \right\}
}
\\
&\leq& \lambda_{\max}-\lambda_{\min}\,.
\eeqno
On the other hand
\beqno
{\rm e}^{2r}\frac{\beta_{\min}(r)}{\beta_{\max}(r)}
&=&
{\rm e}^{2r}\frac{1+\frac{\lambda_{\max}}{\sigma^2}(1-{\rm e}^{-2r})}
{1+\frac{\lambda_{\min}} {\sigma^2} (1-{\rm e}^{-2r})}
\cdot\frac{\lambda_{\min}}{\lambda_{\max}}
\\
&\geq&
\frac{\lambda_{\min}}{\lambda_{\max}}\,.
\eeqno
Hence, if  
\beqno
\lambda_{\max}-\lambda_{\min}
&\leq&\sigma^2\frac{L}{L-1}\cdot
\frac{\lambda_{\min}}{\lambda_{\max}}
\,,
\eeqno
or equivalent to 
\beqno
\sigma^2\geq \frac{L-1}{L}\cdot
\frac{\lambda_{\max}}{\lambda_{\min}}
(\lambda_{\max}-\lambda_{\min})\,,
\eeqno
we have
$$
 \frac{1}{\beta_{\min}(r)}
-\frac{1}{\beta_{\max}(r)}\leq 
\sigma^2{\rm e}^{2r}\frac{L}{L-1}
\cdot\frac{\beta_{\min}(r)}{\beta_{\max}(r)}
$$
for $r \geq 0$, completing the proof. \hfill\IEEEQED

\section{Conclusion}

We have considered the distributed source coding of correlated
Gaussian observation and given a partial solution to this problem by
deriving explicit outer bound of the rate distortion
region. Furthermore, we established a sufficient condition under which
this outer bound is tight.

In this paper our arguments have been concentrated on 
Problem 2, the determination problem of 
${\cal R}_L(D)$. On Problem 1, the determination problem 
of ${\cal R}_L(D^L)$, the techniques we have used to derive 
the outer bound of ${\cal R}_L(D)$ are not sufficient 
to derive an outer bound of ${\cal R}_L(D^L)$.

In \cite{oh8}, we introduced a unified approach to deal with Problems
1 and 2 and derived outer bounds of the rate distortion regions on
those two problems. For Problem 1, the outer bound of \cite{oh8} has a
form of positive semi definite programming. For Problem 2, the outer
bound of \cite{oh8} is the same as that of this paper. Recently, we 
have obtained some extentions of the results of Oohama \cite{oh8}. 
Details of those results are to be presented in a future paper. 
%
%
%However, this outer bound 
%
%whose form is not described here does not seem to be 
%tight in general.
%for any correlated $X^L$ and $N^L$.
%
%
%We do not have considered  
%
%
%
%
%On Problem 1, i.e., the determination problem of the rate distoriton region 
%in the case of vector distortion criterion 
%the author derived 
%
%
%\bibitem{oh8}Y. Oohama, 
%``Distributed source coding of correlated Gaussian observations,''
%{\it
%Proceedings of the 2008 International 
%Symposium on Information Theory and its Applications,},
%Auckland, New Zealand, December 7-10, pp. 1441-1446, 2008.
%%
%
%A complete 
%solution is still lacking.
%\end{document}
%\input{apdxFinal.tex}

\section*{\empty}
\appendix
%\newpage

\subsection{
Proof of Theorem \ref{th:exth4zz}.
}

In this appendix we prove Theorem \ref{th:exth4zz}.

{\it Proof of Theorem \ref{th:exth4zz}:}
We first observe that
\beqa
& &\sum_{i=1}^Lr_i
   +\frac{1}{2}\log
      \frac{\left|\Sigma_{X^L}^{-1}+\Sigma_{N^L(r^L)}^{-1}\right|}            
           {\left|\Sigma_{X^L}^{-1}\right|}
\nonumber\\
&=&\sum_{i=1}^Lr_i
+\frac{1}{2}\log{\left|\Sigma_{X^L}^{-1}+\Sigma_{N^L(r^L)}^{-1}\right|} 
+\frac{1}{2}\log {\left|\Sigma_{X^L}\right|}\,, 
\label{eqn:det0000}
\\
& &\left|\Sigma_{X^L}\right|=(1-\rho)^{L}
   \left\{1+\frac{\rho L}{1-\rho}\right\}\,,
\label{eqn:det1}\\
& &{\left|\Sigma_{X^L}^{-1}+\Sigma_{N^L(r^L)}^{-1}\right|}  
\nonumber\\
&=&\left(1-\sum_{i=1}^L\frac{b}{u_i+b}\right)
\prod_{i=1}^L(u_i+b)\,.
\label{eqn:det2}
\eeqa
Set
\beqno
v_i\defeq \frac{1}{u_i+b}=\{a+b+c(1-{\rm e}^{-2r_i})\}^{-1}\,. 
\eeqno
Then, we have   
\beq
\left. 
\ba{rcl}
u_i&=& v_i^{-1}-b\,,
\vspace*{2mm}\\
r_i&=&\ds \frac{1}{2}\log\frac{c}{a+b+c-v_i^{-1}}\,.
\ea
\right\}
\label{eqn:det001}
\eeq
From (\ref{eqn:det00}) in Section IV and (\ref{eqn:det001}), 
we can see that the condition $r^L \in{\cal B}_L(D)$ 
is equivalent to
\beqa
& &b\sum_{i \ne j}v_iv_j-(1+Db)\sum_{i=1}^Lv_i+D\geq 0
\nonumber\\
&\Leftrightarrow &
b\left(\sum_{i=1}^Lv_i\right)^2-b\sum_{i=1}^Lv_i^2
\nonumber\\
& & \qquad\qquad 
-(1+Db)\sum_{i=1}^Lv_i+D \geq 0 \,.
\label{eqn:det3}
\eeqa
From (\ref{eqn:det2}) and (\ref{eqn:det001}),
we have 
\beqa
& &\sum_{i=1}^L{r_i}+\frac{1}{2}
  \log {\left|\Sigma_{X^L}^{-1}+\Sigma_{N^L(r^L)}^{-1}\right|}   
\nonumber\\
&=&\sum_{i=1}^L\frac{1}{2}\log\frac{c}{(a+b+c)v_i-1}
\nonumber\\
& &\qquad\quad+\frac{1}{2}\log\left( 1-b\sum_{i=1}^Lv_i\right)
\nonumber\\
&\MGeq{{\rm a}} &
\frac{L}{2}\log\frac{c}{(a+b+c)\frac{1}{L}\sum_{i=1}^Lv_i-1}
\nonumber\\    
& &\qquad\quad+\frac{1}{2}\log \left( 1-b\sum_{i=1}^Lv_i\right)\,.
\label{eqn:det4}
\eeqa
Step (a) follows from the convexity of $-\log{t}$. 
Here, we set
$$
\gamma\defeq \left\{ \frac{1}{L}\sum_{i=1}^Lv_i \right\}^{-1}\,.
$$
Then, from (\ref{eqn:det4}), we have 
\beqa 
& & \sum_{i=1}^L{r_i}+\frac{1}{2}\log
   {\left|\Sigma_{X^L}^{-1}+\Sigma_{N^L(r^L)}^{-1}\right|}
\nonumber\\
&\geq & \frac{L}{2}\log\frac{(Dc)\gamma}{D(a+b+c)-\gamma}
       +\frac{1}{2}\log\left(1-(Db)\frac{L}{\gamma}\right)\,.
\label{eqn:det4.5}
\eeqa
Since 
$$
\sum_{i}^Lv_i^2\geq L\cdot
\left(\frac{1}{L}\sum_{i=1}^Lv_i\right)^2=L\gamma^{-2}
$$
and (\ref{eqn:det3}), we obtain 
\beqa
&&bL(L-1)\gamma^{-2}-(1+Db)L\gamma^{-1}+D\geq 0
\nonumber\\
&\Leftrightarrow&
\left(\frac{D\gamma}{L}\right)^2
-(1+Db)\left(\frac{D\gamma}{L}\right)
+Db\left(1-\frac{1}{L}\right)\geq 0.
\label{eqn:det6}
\eeqa
Since $v_i\leq b^{-1}$ for $i\in \Lambda$, 
$\gamma$ must be $\gamma\geq Lb$. 
Solving (\ref{eqn:det6}) 
under this constraint, we obtain 
\beq
D\gamma\geq 
\frac{L}{2}
\left[
1+ Db +\sqrt{(1-Db)^2+\ts\frac{4Db}{L}}
\right]=L_1\,.
\label{eqn:det8}
\eeq
Combining 
(\ref{eqn:det0000}),
(\ref{eqn:det1}), 
(\ref{eqn:det4.5}), and 
(\ref{eqn:det8}),
we have
\beqa
& & 
\sum_{i=1}^Lr_i
+\frac{1}{2}\log
      \frac{\left|\Sigma_{X^L}^{-1}+\Sigma_{N^L(r^L)}^{-1}\right|}            
           {\left|\Sigma_{X^L}^{-1}\right|}
\nonumber\\
&\geq& 
\frac{L}{2}\log
\left(\frac{(1-\rho)L_1c}{D(a+b+c)-L_1}\right)
\nonumber\\
& &\quad +\frac{1}{2}\log
\left\{\frac{1+(L-1)\rho}{1-\rho}
\left(1- \frac{LDb}{L_1}\right)
\right\}\,.
\nonumber
\eeqa
The equality holds 
$$
r_i=\frac{1}{2}\log\frac{Dc}{D(a+b+c)-L_1}\,,
\mbox{ for }i\in \Lambda\,, 
$$
completing the proof.
\hfill \IEEEQED  

\subsection{
Proof of Lemma \ref{lm:lm2}
} 

In this appendix we prove Lemma \ref{lm:lm2}.
Without loss of generality we may assume 
that $S=\{1,2,\cdots, s\}$. 
We write unitary matrix $Q_S$ as $Q_S=[q_{ij}]$, where $q_{ij}$ 
stands for the $(i,j)$ element of $Q_S$. 
The unitary matrix ${Q}_\irBr{S}$ transforms 
$X_{\irBr{S}}$ into 
$\irg{Z}_{\irBr{S}}$$=X_{\irBr{S}}{Q}_{\irBr{S}}$.
The following lemma states an important 
property on the distribution of Gaussian random vector 
$Z_S$. This lemma is a basis of the proof 
of Lemma \ref{lm:lm2}. 

\begin{lm}\label{lm:LmO}
For any $i\in S$, we have the following.
\beq 
{Z}_i=-\frac{1}{g_{ii}}\sum_{j\ne i}\nu_{ij}{Z}_j
+ \frac{1}{g_{ii}}\sum_{j=1}^{s}
\frac {q_{ji}}{\sigma_{N_j}^2}{Y}_j + \hat{N}_i\,,
\label{eqn:prlmaa00}
\eeq
where 
\beq
g_{ii}= \left[Q_S^{-1}\Sigma_{X_S}^{-1}Q_S\right]_{ii} 
+\sum_{j=1}^{s}\frac{q_{ji}^2}{\sigma_{N_j}^2}\,,
\label{eqn:defajj}
\eeq
$\nu_{ij},$ $j\in S-\{i\}$ are suitable constants 
and $\hat{N}_i$ is a zero mean Gaussian random 
variables with variance $\frac{1}{g_{ii}}$. 
For each $i\in S$, $\hat{N}_i$ is independent 
of ${Z}_j,j\in S-\{i\}$ and ${Y}_j,j\in S$. 
\end{lm}

{\it Proof:} Without loss of generality we may assume $i=1$.
Let $\Sigma_{X_SY_S}$ be a covariance matrix
on the pair of the Gaussian random vectors $X_S$ 
and $Y_S$. Since $Y_S=X_S+N_S$, we have 
\beqno
\Sigma_{X_SY_S}=
\left[
\ba{cc} 
\Sigma_{X_S} &\Sigma_{X_S}\\
\Sigma_{X_S} &\Sigma_{X_S}+\Sigma_{N_S}
\ea
\right]\,.
\eeqno
Since $Z_S=X_SQ_S$, we have
\beqno
\Sigma_{Z_SY_S}=
\left[
\ba{cc}
Q_S^{-1}\Sigma_{X_S}Q_S &      Q_S^{-1}\Sigma_{X_S}\\
        \Sigma_{X_S}Q_S  & \Sigma_{X_S}+\Sigma_{N_S}
\ea
\right]\,.
\eeqno
The density function $p_{Z_SY_S}(z_S,y_S)$ of  
$(Z_S,Y_S)$ is given by 
$$
p_{Z_SY_S}(z_S,y_S)=
\frac{1}{(2\pi{\rm e})^{s}\left|\Sigma_{Z_SY_S}\right|^{\frac{1}{2}}}
{\rm e}^{\scs 
-\frac{1}{2}
[z_S y_S]\Sigma_{Z_SY_S}^{-1}
\mbox{\scriptsize 
$\left[\ba{c}{}^{t}z_S\\{}^{t}y_S\ea\right]$}},
$$
where $\Sigma_{Z_SY_S}^{-1}$ has the following form:
\beqno
\Sigma_{Z_SY_S}^{-1}=
\left[
\ba{cc}
   Q_S^{-1}(\Sigma_{X_S}^{-1}+\Sigma_{N_S}^{-1})Q_S 
& -Q_S^{-1} \Sigma_{N_S}^{-1}\\
 -\Sigma_{N_S}^{-1}Q_S & \Sigma_{N_S}^{-1}
\ea
\right]\,.
\eeqno
Set
\beq
\left.
\ba{rcl}
\nu_{ij}  &\defeq &\ds 
 \left[Q_S^{-1}(\Sigma_{X_S}^{-1}+\Sigma_{N_S}^{-1})Q_S\right]_{ij}
\vspace*{1mm}\\
&=&\ds\left[Q_S^{-1}\Sigma_{X_S}^{-1}Q_S\right]_{ij}
+\sum_{k=1}^s\frac{q_{ki}q_{kj}}{\sigma_{N_k}^2}\,,
\vspace*{1mm}\\
\beta_{ij}&\defeq &\ds 
-\left[Q_S^{-1}\Sigma_{N_S}^{-1}\right]_{ij}=
-\frac{q_{ji}}{\sigma_{N_j}^2}\,.
\ea
\right\}
\label{eqn:prlmz2z} 
\eeq
Now, we consider the following partition of 
$\Sigma_{Z_SY_S}^{-1}$: 
\beqno
\Sigma_{Z_SY_S}^{-1}
&=&
\left[
\ba{cc}
   Q_S^{-1}(\Sigma_{X_S}^{-1}+\Sigma_{N_S}^{-1})Q_S 
& -Q_S^{-1} \Sigma_{N_S}^{-1}\\
 -\Sigma_{N_S}^{-1}Q_S & \Sigma_{N_S}^{-1}
\ea
\right]
\nonumber\\
&=&
\left[
\ba{c|c}    
      g_{11} & g_{12} \\\hline
      {}^{\rm t}g_{12} & G_{22}    
\ea
\right]\,,
\label{eqn:prlmz3} 
\eeqno 
where $g_{11}$, $g_{12}$, and $G_{22}$ are scalar, 
$2s-1$ dimensional vector, and $(2s-1)\times (2s-1)$ 
matrix, respectively. It is obvious from 
the above partition of $\Sigma_{Z_SY_S}^{-1}$ 
that we have 
\beq
\left.
\ba{rcl}
g_{11}&=&\ds\nu_{11}=
   \left[Q_S^{-1}\Sigma_{X_S}^{-1}Q_S\right]_{11}
   +\sum_{k=1}^s\frac{q_{k1}^2}{\sigma_{N_k}^2}\,,
\vspace{1mm}\\
g_{12}&=&\left[\nu_{12}\cdots\nu_{1s}
               \beta_{11}\beta_{12}\cdots\beta_{1s}\right]\,.
\ea
\right\}
\label{eqn:prlmz5} 
\eeq
It is well known that 
$\Sigma_{Z_SY_S}^{-1}$ has the following expression:
\beqno
\Sigma_{Z_SY_S}^{-1}
&=&\left[\ba{c|c}    
      g_{11} & g_{12} \\\hline
      {}^{\rm t}g_{12} & G_{22}    
      \ea\right]
\nonumber\\
&=&\left[
\ba{c|c}    
      1 & 0_{12} \\\hline
      \frac{1}{g_{11}}{}^{\rm t}g_{12} & I_{L-1}    
\ea
\right]
\left[
\ba{c|c}    
      g_{11} & 0_{12} \\\hline
      {}^{\rm t}0_{12}&G_{22}-\frac{1}{g_{11}}{}^{\rm t}g_{12}g_{12}    
\ea
\right]
\nonumber\\
& &\qquad\qquad \qquad \times
\left[
\ba{c|c}    
      1 & \frac{1}{g_{11}}g_{12} \\\hline
      {}^{\rm t}0_{12} & I_{L-1}    
\ea
\right]\,.
%\label{eqn:prlmz6} 
\eeqno
Set
\beq
\left.\ba{rcl}
\hat{n}_1
&\defeq& 
\left[z_1|z_{S-\{1\}}y_S\right]
\left[\ba{c}1\\\hline \frac{1}{g_{11}}{}^{\rm t}g_{12}\ea\right]
\vspace{1mm}\\
&=&z_1+\frac{1}{g_{11}}\left[z_{S-\{1\}}y_S\right]{}^{\rm t}g_{12}\,.
\ea\right\}
\label{eqn:prlmz64} 
\eeq
Then, we have
\beqa
\hspace*{-6mm}& &[z_Sy_S]\Sigma_{Z_SY_S}
    \left[\ba{c}{}^{\rm t}z_S\\{}^{\rm t}y_S\ea\right]
\nonumber\\
\hspace*{-6mm}&=&[ z_1 | z_{S-\{1\}}y_S]
   \left[\ba{c|c}    
      g_{11} & g_{12} \\\hline
      {}^{\rm t}g_{12} & G_{22}    
      \ea\right]
  \left[\ba{c}z_1\\\hline{}^{\rm t}z_{S-\{1\}}\\{}^{\rm t}y_S\ea\right]  
\nonumber\\
\hspace*{-6mm}&=&[\hat{n}_1|z_{S-\{1\}}y_S]
\left[
\ba{c|c}    
      g_{11} & 0_{12}\\\hline
      {}^{\rm t}0_{12}&G_{22}-\frac{1}{g_{11}}{}^{\rm t}g_{12}g_{12}    
\ea
\right]
\left[\ba{c}\hat{n}_1\\\hline{}^{\rm t}z_{S-\{1\}}\\{}^{\rm t}y_S\ea\right]\,.  
\label{eqn:zasds}
\eeqa
From (\ref{eqn:prlmz2z})-%, 
     %(\ref{eqn:prlmz5}), and  
     (\ref{eqn:prlmz64}), we have 
\beqa 
\hat{n}_1&=&z_1
+\frac{1}{g_{11}}\sum_{j=2}^s\nu_{1j}z_j
+\frac{1}{g_{11}}\sum_{j=1}^s\beta_{1j}y_j
\nonumber\\
&=&z_1
+\frac{1}{g_{11}}\sum_{j=2}^s\nu_{1j}z_j
-\frac{1}{g_{11}}\sum_{j=1}^s
\frac{q_{j1}}{\sigma_{N_j}^2}y_j\,.
\label{eqn:prlmz66} 
\eeqa
It can be seen from (\ref{eqn:zasds}) and (\ref{eqn:prlmz66}) 
that the random variable $\hat{N}_1$ defined by
$$
\hat{N}_1\defeq 
Z_1+\frac{1}{g_{11}}\sum_{j=2}^s\nu_{1j}Z_j
   -\frac{1}{g_{11}}\sum_{j=1}^s\frac{q_{j1}}{\sigma_{N_j}^2}Y_j
$$
is a zero mean Gaussian random variable with variance 
$\frac{1}{{g}_{11}}$ and is independent 
of $Z_{S-\{1\}}$ and $Y_{S}$. This completes the proof 
of Lemma \ref{lm:LmO}.
\hfill\IEEEQED

The followings are two variants of the entropy power 
inequality.

\begin{lm}\label{lm:lm5zz} Let ${\vc U}_i,i=1,2,3$ 
be $n$ dimensional 
random vectors with densities and let 
$T$ be a random variable taking values in a finite set. 
We assume that ${\vc U}_3$ is independent of ${\vc U}_1$, 
${\vc U}_2$, and $T$. Then,  we have 
\beqno
\EP{{\lvc U}_2+{\lvc U}_3|{\lvc U}_1T}
\geq 
\EP{{\lvc U}_2|{\lvc U}_1T}+\EP{{\lvc U}_3}\,.
\eeqno
\end{lm}
\begin{lm}\label{lm:lm5zzb} Let ${\vc U}_i$, $i=1,2,3$ be 
$n$ random vectors with densities. Let $T_1, T_2$ 
be random variables taking values in finite sets. 
We assume that those five random variables
form a Markov chain 
$
(T_1,{\vc U}_1) \to {\vc U}_3 \to (T_2,{\vc U}_2) 
$
in this order. Then, we have    
\beqno
& &\EP{{\lvc U}_1+{\lvc U}_2|{\lvc U}_3T_1T_2}
\\
&\geq& \EP{{\lvc U}_1|{\lvc U}_3T_1}
      +\EP{{\lvc U}_2|{\lvc U}_3T_2}\,.
\eeqno
\end{lm}

{\it Proof of Lemma \ref{lm:lm2}:}
%Without loss of generality we may assume 
%that $S=\{1,2,\cdots, s\}$. 
%We write unitary matrix $Q_S$ as $Q_S=[q_{ij}]$, where $q_{ij}$ 
%stands for the $(i,j)$ element of $Q_S$. By an elementary 
%computation, for any $i\in S$, we have the following.
By Lemma \ref{lm:LmO}, we have
\beq 
{\vc Z}_i=-\frac{1}{g_{ii}}\sum_{j\ne i}\nu_{ij}{\vc Z}_j
+ \frac{1}{g_{ii}}\sum_{j=1}^{s}
\frac {q_{ji}}{\sigma_{N_j}^2}{\vc Y}_j + \hat{\vc N}_i\,,
\label{eqn:prlmaa}
\eeq
where 
%\beq
%g_{ii}= \left[Q_S^{-1}\Sigma_{X_S}^{-1}Q_S\right]_{ii} 
%+ \sum_{j=1}^{s}\frac{q_{ji}^2}{\sigma_{N_j}^2}\,,
%\label{eqn:defajj}
%\eeq
%$\nu_{ij},$ $j\in S-\{i\}, $ are suitable constants 
%and 
$\hat{\vc N}_i$ is a vector of $n$ independent 
copies of zero mean Gaussian random 
variables with variance $\frac{1}{g_{ii}}$. 
For each $i\in S$, $\hat{\vc N}_i$ is independent of 
${\vc Z}_j, j \in S-\{i\}$
and ${\vc Y}_j, j \in S$. 
Set 
\beqno
h^{(n)} &\defeq& \frac{1}{n}h({\vc Z}_i|{\vc Z}_{S-\{i\}},W_S)\,.
\eeqno
Furthermore, for $k\in \Lambda$, define
\beqno
&&S_k \defeq \{k,k+1,\cdots,s\}\,,
%\\ 
%& &
\Psi_k=\Psi_k({\vc Y}_{S_k}) 
\defeq \sum_{j=k}^{s} \frac {q_{ji}}{\sigma_{N_j}^2}{\vc Y}_j\,.
\eeqno
Applying Lemma \ref{lm:lm5zz} to (\ref{eqn:prlmaa}), we have
\beq  
\frac{{\baseN}^{2h^{(n)}}}{2\pi{\rm e}}
\geq \frac{1}{(g_{ii})^2}
\frac{1}{2\pi{\rm e}}{\baseN}^{\frac{2}{n}h(\Psi_1|{\svc Z}_{S-\{i\}},W_S)} 
+\frac{1}{g_{ii}}\,.
\label{eqn:PrConvLm1}
\eeq
On the quantity $h(\Psi_1|{\svc Z}_{S-\{i\}},W_S)$ 
in the right member of (\ref{eqn:PrConvLm1}), we have 
the following chain of equalities:
\beqa
& &h(\Psi_1|{\vc Z}_{S-\{i\}},W_S)
\nonumber\\
&=&I(\Psi_1;{\vc X}_S|{\vc Z}_{S-\{i\}},W_S)
   +h(\Psi_1|{\vc X}_S,{\vc Z}_{S-\{i\}},W_S)
 \nonumber\\
&\MEq{a}&
   I(\Psi_1;{\vc Z}_S|{\vc Z}_{S-\{i\}},W_S)  
   +h(\Psi_1|{\vc X}_S,W_S)
\nonumber\\
&=& I(\Psi_1;{\vc Z}_i|{\vc Z}_{S-\{i\}},W_S) 
   +h(\Psi_1|{\vc X}_S,W_S)
\nonumber\\
&=&h({\vc Z}_i|{\vc Z}_{S-\{i\}},W_S)
    -h({\vc Z}_i|\Psi_1,{\vc Z}_{S-\{i\}},W_S) 
\nonumber\\
& &+h(\Psi_1|{\vc X}_S,W_S)
\nonumber\\
&\MEq{b}& nh^{(n)}-h({\vc Z}_i|\Psi_1,{\vc Z}_{S-\{i\}}) 
           +h(\Psi_1|{\vc X}_S,W_S)
\nonumber\\
&=&nh^{(n)}-\frac{n}{2}\log\left[{2\pi{\rm e}}(g_{ii})^{-1}\right]
   +h(\Psi_1|{\vc X}_S,W_S)\,. 
\label{eqn:PrConvLm2}
\eeqa
Step (a) follows from that ${\vc Z}_S$ can be obtained from 
${\vc X}_S$ by the invertible matrix $Q$. 
Step (b) follows from the Markov chain 
$${\vc Z}_i\to (\Psi_1,{\vc Z}_{S-\{i\}})\to {\vc Y}_S\to W_S.$$ 
From (\ref{eqn:PrConvLm2}), we have
\beq
\frac{1}{2\pi{\rm e}}{\baseN}^{\frac{2}{n}h(\Psi_1|{\svc Z}_{S-\{i\}},W_S)}
=\frac{{\baseN}^{2h^{(n)}}}{2\pi{\rm e}}g_{ii}\cdot
 \frac{1}{2\pi{\rm e}}{\baseN}^{\frac{2}{n}h(\Psi_1|{\svc X}_S,W_S)}.
\label{eqn:PrConvLm3}
\eeq
Substituting (\ref{eqn:PrConvLm3}) into (\ref{eqn:PrConvLm1}), 
we obtain  
\beq
\frac{{\baseN}^{2h^{(n)}}}{2\pi{\rm e}}
\geq \frac{{\baseN}^{2h^{(n)}}}{2\pi{\rm e}}\frac{1}{g_{ii}}
     \cdot\frac{1}{2\pi{\rm e}}{\baseN}^{\frac{2}{n}h(\Psi_1|{\svc X}_S,W_S)} 
     +\frac{1}{g_{ii}}\,.
\label{eqn:PrConvLm4}
\eeq
Solving (\ref{eqn:PrConvLm4}) with respect to 
$\frac{{\baseN}^{2h^{(n)}}}{2\pi{\rm e}}$, we obtain
%and the technique that Oohama \cite{oh4}, \cite{oh2} developed 
%to prove the converse coding theorem, we obtain 
\beq
\frac{{\baseN}^{2h^{(n)}}}{2\pi{\rm e}}
\geq
  \left[g_{ii} 
 -\frac{1}{2\pi{\rm e}}{\baseN}^{\frac{2}{n}h(\Psi_1|{\svc X}_S,W_S)} 
  \right]^{-1}\,.
\label{eqn:prlmz}
\eeq
Next, we evaluate a lower bound of  
$
{\baseN}^{\frac{2}{n}h(\Psi_1|{\svc X}_S,W_S)}\,.
$ 
Note that for $j=1,2,\cdots,s-1$ we have 
the following Markov chain:
\beq
\left(W_{S_{j+1}},\Psi_{j+1}({\vc Y}_{S_{j+1}})\right)\to {\vc X}_S 
\to \left(W_j,\ts \frac{q_{ji}}{\sigma_{N_j}^2}{\vc Y}_j\right)\,. 
\label{eqn:Markov}
\eeq
Based on (\ref{eqn:Markov}), we apply 
Lemma \ref{lm:lm5zzb} to 
$
\frac{1}{2\pi{\rm e}}  
{\baseN}^{\frac{2}{n}h(\Psi_j|{\svc X}_S,W_S)}
$
for $j=1,2,\cdots,s-1$. Then, for $j=1,2,$ $\cdots,s-1$, 
we have the following chains of inequalities : 
\beqa
& &\frac{1}{2\pi{\rm e}} %\cdot 
  {\baseN}^{\frac{2}{n}h(\Psi_j|{\svc X}_S,W_S)}
\nonumber\\
&=&\frac{1}{2\pi{\rm e}} %\cdot 
  {\baseN}^{\frac{2}{n}
  h\left(\left.\Psi_{j+1}+\frac{q_{ji}}{\sigma_{N_1}^2}{\svc Y}_j
         \right|{\svc X}_S,W_{S_{j+1}},W_j\right)}
\nonumber\\
&\geq & \frac{1}{2\pi{\rm e}} 
        {\baseN}^{\frac{2}{n}
        h\left(\left.\Psi_{j+1}\right|{\svc X}_S,W_{S_{j+1}}\right)}
        +\frac{1}{2\pi{\rm e}} 
         {\baseN}^{
         \frac{2}{n}
         h\left(\left.\frac{q_{ji}}{\sigma_{N_j}^2}{\svc Y}_j\right|
         {\svc X}_S,W_j\right)
         }
\nonumber\\
&=& \frac{1}{2\pi{\rm e}} 
        {\baseN}^{\frac{2}{n}
         h\left(\left.
         \Psi_{j+1}\right|{\svc X}_S,W_{S_{j+1}}\right)}
        +q_{ji}^2\frac{{\baseN}^{-2r_j^{(n)}}}{\sigma_{N_j}^2}\,.
\label{eqn:prlmz1}
\eeqa
Using (\ref{eqn:prlmz1}) iteratively for
$j=1,2,\cdots, s-1$, we have
\beq
\frac{1}{2\pi{\rm e}} %\cdot 
  {\baseN}^{\frac{2}{n}h(\Psi_1|{\svc X}_S,W_S)}
\nonumber\\
\geq 
\sum_{j=1}^{s}q_{ji}^2\frac{{\baseN}^{-2r_j^{(n)}}}{\sigma_{N_j}^2}\,.
\label{eqn:prlmz2}
\eeq
Combining (\ref{eqn:defajj}), (\ref{eqn:prlmz}), 
and (\ref{eqn:prlmz2}), we have 
\beqa
\frac{{\baseN}^{2h^{(n)}}}{2\pi{\rm e}}
&\geq&
  \left\{
  \left[Q_S^{-1}\Sigma_{X_S}^{-1}Q_S\right]_{ii} 
  +\sum_{j=1}^{s}q_{ji}^2
   \frac{1-{\baseN}^{-2r_j^{(n)}}}{\sigma_{N_j}^2}
   \right\}^{-1}
\nonumber\\
&=&\left[Q_S^{-1}
         (\Sigma_{X_S}^{-1}+\Sigma_{N_S(r_S^{(n)})}^{-1})
         Q_S\right]_{ii}^{-1}\,,
\label{eqn:prlmz4}
\eeqa
completing the proof. 
\hfill \IEEEQED 

\subsection{ 
%Proof of Lemma \ref{lm:eigenlm}
%Some 
%Properties on the 
Eigen Values 
of $\Sigma_{X^L}^{-1}+\Sigma_{N^L(u^L)}^{-1}$ 
} 
%Let $A$ and $B$ 
%be two Hermitian matrices with

In this appendix we prove some properties on 
eigen values of 
$\Sigma_{X^L}^{-1}$ $+\Sigma_{N^L(u^L)}^{-1}$.
Using those properties, we prove Lemma \ref{lm:eigenlm}. 

We first consider the case treated in section IV, where
$\Sigma_{X^L}^{-1}$ $+\Sigma_{N^L(u^L)}^{-1}$ has the 
identical value $b$ of non diagonal elements.  
Using (\ref{eqn:formula}), we can show that 
$\alpha_i, i=1,2,$ $\cdots, L$ are $L$ solutions to the 
following eigen value equation:
\beq
\left( 
1-\sum_{i=1}^{L}\frac{b}{u_i+b - \alpha}
\right)
\prod_{i=1}^L(u_i+b - \alpha)
=0\,.
\label{eqn:eigeq}
\eeq
Let $m$ be the number of distinct values of 
$u_1$,$u_2$,$\cdots$,$u_L$ 
and let 
$u_{i_1}<$
$u_{i_2}<$ 
$\cdots<$ 
$u_{i_m}$
be the ordered list of those values.  
For each $j=1,2,$ $\cdots,m$, 
set ${\cal L}_j$ $\defeq $ $\{l: u_l=u_{i_j}\}$ 
and $l_j\defeq |{\cal L}_j|$. 
%For each $j=1,2,$ $\cdots,m$, 
%the quantity $l_j$ stands for the multiplicity of 
%the eigen value  
Then, the eigen value equation (\ref{eqn:eigeq}) becomes   
\beq
\left( 
1-\sum_{j=1}^{m}\frac{bl_j}{u_{i_j}+b - \alpha}
\right)
\prod_{j=1}^m(u_{i_j}+b -\alpha)^{l_j}=0\,.
\label{eqn:eigeq2}
\eeq
From (\ref{eqn:eigeq2}), we obtain the following proposition.
\begin{pro}\label{pro:pro0} 
Eigen values of $\Sigma_{X^L}^{-1}+\Sigma_{N^L(u^L)}^{-1}$
satisfies the following two properties.
\begin{itemize}
\item[a)] 
The matrix 
$\Sigma_{X^L}^{-1}+\Sigma_{N^L(u^L)}^{-1}$ 
has $m$ positive eigen values, which are the $m$ 
distinct solutions of the nonlinear scalar equation
\beq
\hspace*{-2mm}
1=g(\alpha)\defeq
\sum_{j=1}^{m}\frac{bl_j}{u_{i_j}+b - \alpha}\,.
\label{eqn:za00z}
\eeq 
%Let ${\cal E}_0$ be the set of solutions 
%of (\ref{eqn:za00}) and 
Let 
${\alpha}_{1}<$
${\alpha}_{2}<$ 
$\cdots<{\alpha}_{m}$ be the 
ordered list of solutions of (\ref{eqn:za00z}). 
Then, we have 
\beqa
& &0<\alpha_{1}<u_{{i}_1}+b<
\alpha_{2}<u_{{i}_2}+b<\cdots
\nonumber\\
&&\quad <\alpha_{m}<u_{{i}_m}+b\,.
\label{eqn:zaaoaa}
\eeqa
The multiplicity of those eigen values is 1.   
\item[b)] When $l_j\geq 2$, the matrix 
$\Sigma_{X^L}^{-1}$ 
$+\Sigma_{N^L(u^L)}^{-1}$ has the eigen value 
$u_{i_j}$$+b$ with the multiplicity $l_j-1$. 
\end{itemize}
\end{pro}
\begin{figure}[t]
\bc
\includegraphics[width=8.5cm]{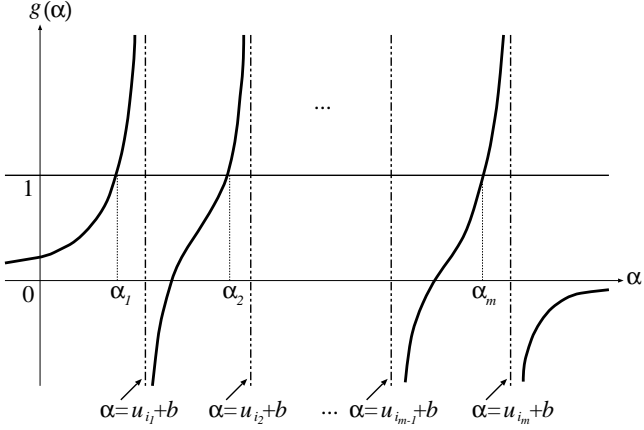}
\caption{Shape of $g(\alpha)$.} 
\label{fig:graphG}
\ec
\end{figure}

{\it Proof: } 
We first prove the part a). 
From (\ref{eqn:eigeq2}), 
we can see that every solution of the equation 
$1=g(\alpha)$ is an eigen value of 
$\Sigma_{X^L}^{-1}$$+$$\Sigma_{N^L(u^L)}^{-1}$. 
Since 
$$
g^{\prime}(\alpha)=
\sum_{j=1}^{m}\frac{bl_j}{(u_{{i}_j}+b-\alpha)^2}>0\,, 
$$
$g(\alpha)$ is differentiable and monotone increasing 
in each of the $m$ open intervals
$(-\infty, $ 
$u_{{i}_1}+b)$,
$(u_{i_1}+b,$ $u_{i_2}+b)$,
$\cdots$,  
$(u_{{i}_{m-1}}+b,$ $u_{{i}_{m}}+b)$.
Since $g(\alpha)$ is unbounded in each of these intervals, 
it has positive and negative values there, 
and thus $1=g(\alpha)$ has a unique solution in each 
of these $m$ disjoint intervals. 
In particular, since  
\beqno
\left|
\Sigma_{X^L}^{-1}+\Sigma_{N^L(u^L)}^{-1}
\right|
&=&
\left(1-g(0)\right)\prod_{i=1}^L(u_i+b)>0, 
\eeqno
we have $0<g(0)<1$. This implies that 
$1=g(\alpha)$ has a unique solution in 
the interval $(0,$ $u_{{i}_1}+b)$.
Furthermore, since 
\beqno
& &\lim_{\alpha \downarrow u_{{i}_{m}}+b}g(\alpha)=-\infty\,,
\quad \lim_{\alpha \to +\infty}g(\alpha)=0\,,
\eeqno
there is no eigen value in the open interval 
$(u_{{i}_{m}}+b,+\infty)$.
Summarizing the above arguments, 
we obtain $(\ref{eqn:zaaoaa})$.
For convenience we show the shape of $g(\alpha)$ 
in Fig. \ref{fig:graphG}. 
The part b) is obvious 
from (\ref{eqn:eigeq2}). 
\hfill\IEEEQED
 
Next, we consider the case where $X^L$ is a general 
covariance matrix. Set
$$
{\arraycolsep 1mm
 \Sigma_{X^L}^{-1}+\Sigma_{N^L(u^L)}^{-1}
=\left[
\ba{c|c}    
      u_1 & b_{12} \\\hline
      {}^{\rm t}b_{12} & B_{22}    
\ea
\right]\,.
}
$$
Let 
$\eta_1$,
$\eta_2$, $\cdots$, 
$\eta_{L-1}$ be $L-1$ eigen values of $B_{22}$. 
Since $B_{22}$ is positive definite, those $L-1$ 
eigen values are positive. Let $p$ be the number of 
distinct eigen values of $B_{22}$  
and let 
$\eta_{k_1}<$
$\eta_{k_2}<$ 
$\cdots<$ 
$\eta_{k_p}$
be the ordered list of eigen values of $B_{22}$.
For each $j=1,2,$ $\cdots,p$, set
${\cal T}_j$ $\defeq $ 
$\{l:\eta_l=\eta_{k_j}\}$ and 
$t_j\defeq |{\cal T}_j|$. 
For each $j=1,2,$ $\cdots,p$, the quantity $t_j$ is  
the multiplicity of the eigen value $\eta_{k_j}$.
Choose the $(L-1)\times(L-1)$ unitary matrix $Q_{22}$ so that 
$$
{}^{\rm t}Q_{22}B_{22}Q_{22}=Q_{22}^{-1}B_{22}Q_{22}=\left[
\begin{array}{cccc}
  \eta_1 &         &        & \mbox{\LARGE 0}\\
          & \eta_2 &        &          \\
          &         & \ddots &          \\
\mbox{\LARGE 0}
          &         &        & \eta_{L-1}\\
\end{array}
\right]
\label{eqn:diag2}
$$
and set 
$$
\tilde{b}_{12}=[\tilde{b}_1\tilde{b}_2\cdots\tilde{b}_{L-1}]
\defeq b_{12}Q_{22}\,.
$$
Then, we have the following lemma.
\begin{lm} \label{lm:eigenlm2}
\beqno
& &\left|\Sigma_{X^L}^{-1}+\Sigma_{N^L(u^L)}^{-1}-\alpha I_L\right|
\nonumber\\
&=&(u_1-\alpha)\prod_{l=1}^{L-1}(\eta_l-\alpha)
   -\sum_{l=1}^{L-1}\tilde{b}_j^2 \prod_{j\ne l}(\eta_j-\alpha)
\nonumber\\
&=&\left(
   u_1-\alpha-\sum_{l=1}^{L-1} \frac{\tilde{b}_j^2}{\eta_l-\alpha}
   \right)
   \prod_{l=1}^{L-1}(\eta_l-\alpha)\,.
\eeqno
\end{lm}

{\it Proof:}
Set
$$
{\arraycolsep 1mm
Q
\defeq 
\left[
\ba{c|c}    
      1 &  0_{12} \\\hline
      {}^{\rm t}0_{12} & Q_{22}    
\ea
\right]\,.
}
$$
Then, we have 
\beqa
&&Q^{-1}(\Sigma_{X^L}^{-1}+\Sigma_{N^L(u^L)}^{-1}-\alpha I_L)Q
\nonumber\\
&=&{}^{\rm t}Q(\Sigma_{X^L}^{-1}+\Sigma_{N^L(u^L)}^{-1}-\alpha I_L)Q
\nonumber\\
&=& 
\left[
\ba{c|c}    
      1 & {0}_{12} \\\hline
      {}^{\rm t}0_{12} & {}^{\rm t}Q_{22}    
\ea
\right]
\left[
\ba{c|c}    
      u_1-\alpha & b_{12}\\\hline
      {}^{\rm t}b_{12} & 
      B_{22}-\alpha I_{L-1}    
\ea
\right]
\left[
\ba{c|c}    
      1 & 0_{12} \\\hline
      {}^{\rm t}{0_{12}} & Q_{22}    
\ea
\right]
\nonumber\\
&=&
\left[
\ba{c|c}    
      u_1-\alpha & b_{12}Q_{22} \\\hline
      {}^{\rm t}Q_{22}{}^{\rm t}b_{12} & 
      {}^{\rm t}Q_{22}(B_{22}-\alpha I_{L-1})Q_{22}    
\ea
\right]
\nonumber\\
&=&
\left[
\ba{c|c}     
      u_1-\alpha & \tilde{b}_{12}\\\hline
      {}^{\rm t}\tilde{{b}}_{12} & 
      \begin{array}{cccc}
      \eta_1-\alpha &        &      &\mbox{\LARGE 0}\\
              & \eta_2-\alpha&      &\\
              &              &\ddots&\\
      \mbox{\LARGE 0}
              &              &      &\eta_{L-1}-\alpha\\
\end{array}
\ea
\right]\,.
\label{eqn:asazz}
\eeqa
By (\ref{eqn:asazz}), we have
\beqno
&&\left|\Sigma_{X^L}^{-1}+\Sigma_{N^L(u^L)}^{-1}-\alpha I_L\right|
\nonumber\\
&=&
\left|
Q^{-1}(\Sigma_{X^L}^{-1}+\Sigma_{N^L(u^L)}^{-1}-\alpha I_L)Q
\right|
\nonumber\\
&=&
\left|
\ba{c|c}     
      u_1-\alpha & \tilde{b}_{12}\\\hline
      {}^{\rm t}\tilde{{b}}_{12} & 
      \begin{array}{cccc}
      \eta_1-\alpha &         &        & \mbox{\LARGE 0}\\
              & \eta_2-\alpha &        & \\
              &              & \ddots & \\
      \mbox{\LARGE 0}
              &              &        & \eta_{L-1}-\alpha \\
\end{array}
\ea
\right|
\nonumber\\
&=&(u_1-\alpha)\prod_{l=1}^{L-1}(\eta_l-\alpha)
   -\sum_{l=1}^{L-1}\tilde{b}_j^2 \prod_{j\ne l}(\eta_j-\alpha)
\nonumber\\
&=&
   \left(
   u_1-\alpha-\sum_{l=1}^{L-1} \frac{\tilde{b}_j^2}{\eta_l-\alpha}
   \right)
\prod_{l=1}^{L-1}(\eta_l-\alpha)\,,
\eeqno
completing the proof. 
\hfill \IEEEQED 

From Lemma \ref{lm:eigenlm2}, we obtain the following proposition.
The first two parts in this proposition are known results (cf. \cite{db}).  
\begin{pro}\label{pro:pro1} 
Set $\epsilon_j\defeq\sum_{l\in{\cal T}_j}\tilde{b}_l^2$
and   
\beqno
{\cal C}_1&\defeq&\{j: 1\leq j\leq p,\epsilon_j>0\}\,,\\
{\cal C}_2&\defeq&\{j: 1\leq j\leq p,\epsilon_j=0\}\,.
\eeqno
Then, eigen values of $\Sigma_{X^L}^{-1}+\Sigma_{N^L(u^L)}^{-1}$
satisfies the following three properties.
\begin{itemize}
\item[a)] Set $w=|{\cal C}_1|$. Let 
$j_1<$
$j_2<$
$\cdots<$
$j_w$
be the ordered list of ${\cal C}_1$. 
For $i=1,2,\cdots,w$, 
set $k_{j_i}\defeq \tilde{k}_i$. 
Then, the matrix 
$\Sigma_{X^L}^{-1}+\Sigma_{N^L(u^L)}^{-1}$ 
has $(w+1)$ eigen values, which are the $(w+1)$ 
distinct solutions of the nonlinear scalar equation
\beq
\hspace*{-2mm}
u_1=\tilde{g}(\alpha)\defeq
\alpha-\sum_{j\in {C_1}}
\frac{\epsilon_j}{\alpha-\eta_{{k}_j}}
=\alpha-\sum_{i=1}^{w}
\frac{\epsilon_{j_i}}{\alpha-\eta_{\tilde{k}_i}}.
\label{eqn:eigeqzz}
\eeq 
Let ${\cal E}_0$ be the set of solutions 
of (\ref{eqn:eigeqzz}) and let
$\alpha_{1}<$
$\alpha_{2}<$ 
$\cdots<\alpha_{w+1}$ be its ordered list. 
Then, we have 
\beqa
0&<&\alpha_{1}<\eta_{\tilde{k}_1}<
\alpha_{2}<\eta_{\tilde{k}_2}<\cdots
\nonumber\\
& &\qquad\qquad <\alpha_{w}<\eta_{\tilde{k}_w}<\alpha_{w+1}\,,
\label{eqn:zaa}
\\
& &\alpha_1<u_1<\alpha_{w+1}\,.
\label{eqn:zaa2}
\eeqa
\item[b)] 
Set
\beqno  
{\cal E}_1&\defeq & \{ \eta_{k_j}: t_j\geq 2, j\in {\cal C}_1\}\,,
{\cal E}_2 \defeq \{ \eta_{k_j}: j\in {\cal C}_2\}\,.
\eeqno
By the above definition and $(\ref{eqn:zaa})$, we have
$
{\cal E}_0 \cap {\cal E}_1={{\cal E}}_1
\cap{\cal E}_2=\emptyset\,.
$
The set of all distinct eigen values of 
$\Sigma_{X^L}^{-1}+\Sigma_{N^L(u^L)}^{-1}$ is given by
${\cal E}_0\cup$ ${\cal E}_1\cup$ ${\cal E}_2$. 
For each $\eta_{k_j} \in {{\cal E}}_1$, the multiplicity 
of $\eta_{k_j}$ becomes $t_j-1$.   
For each $\eta_{k_j} $$\in {\cal E}_2$ $\cap ({\cal E}_0)^{\rm c}$, 
the multiplicity of $\eta_{k_j}$ remains $t_j$.   
For each $\eta_{k_j} $ $\in {\cal E}_2$$\cap {\cal E}_0$, 
the multiplicity of $\eta_{k_j}$ becomes $t_j+1$.     
The multiplicity of $\alpha$ $\in {\cal E}_0$ $\cap ({\cal E}_2)^{\rm c}$ 
is 1.
\item[c)] Every eigen value of 
$\Sigma_{X^L}^{-1}+\Sigma_{N^L(u^L)}^{-1}$ 
is a monotone increasing function of $u_1$.
\end{itemize} 
\end{pro}
\begin{figure}[t]
\bc
\includegraphics[width=8.5cm]{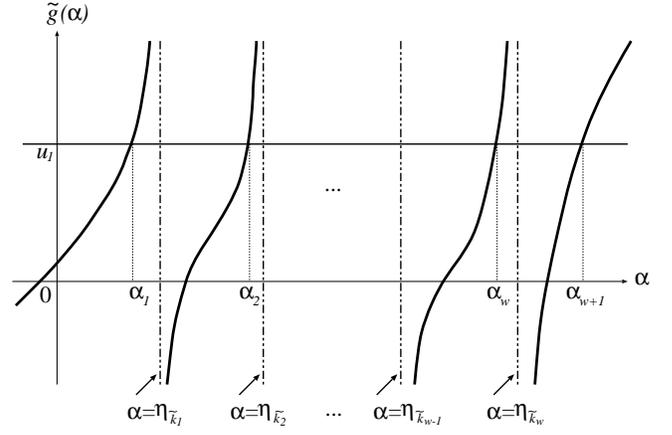}
\caption{Shape of $\tilde{g}(\alpha)$.} 
\label{fig:graphTiG}
\ec
\end{figure}

{\it Proof: } By Lemma \ref{lm:eigenlm2}, the eigen value 
equation of $\Sigma_{X^L}^{-1}$$+$$\Sigma_{N^L(u^L)}^{-1}$ 
is    
\beqa
& &\left(
u_1-\alpha-\sum_{l=1}^{L-1}\frac{\tilde{b}_l^2}{\eta_l-\alpha}
\right)
\prod_{j=1}^{L-1}(\eta_j-\alpha)
\nonumber\\
&=& 
\left(u_1-\alpha-\sum_{j=1}^{p}
\frac{\epsilon_j}{\eta_{k_j}-\alpha}\right)
\prod_{j=1}^{p}(\eta_{k_j}-\alpha)^{t_j}
\nonumber\\
&=&\left(u_1-\alpha-\sum_{j\in {\cal C}_1}
\frac{\epsilon_j}{\eta_{k_j}-\alpha}\right)
\nonumber\\
& &\times 
\left\{\prod_{j\in {\cal C}_1}(\eta_{k_j}-\alpha)^{t_j}\right\}
\left\{\prod_{j\in {\cal C}_2}(\eta_{k_j}-\alpha)^{t_j}\right\}
=0\,.
\label{eqn:za0a}
\eeqa
We first prove the part a). 
From (\ref{eqn:za0a}), we can see that every solution 
of the equation $u_1=g(\alpha)$
is an eigen value of 
$\Sigma_{X^L}^{-1}$$+$$\Sigma_{N^L(u^L)}^{-1}$. 
Since 
$$
\tilde{g}^{\prime}(\alpha)=
1+\sum_{i=1}^{w}\frac{\epsilon_{j_k}}{(\alpha-\eta_{\tilde{k}_i})^2}>0\,, 
$$
$\tilde{g}(\alpha)$ is differentiable and monotone increasing 
in each of the $(w+1)$ open intervals
$(-\infty, \eta_{\tilde{k}_1})$,
$(\eta_{\tilde{k}_1},\eta_{\tilde{k}_2})$,
$\cdots$,  
$(\eta_{\tilde{k}_{w}},\infty)$.
Since $\tilde{g}(\alpha)$ is unbounded in each of these intervals, 
it has positive and negative values there, and thus 
$u_1=\tilde{g}(\alpha)$ has a unique solution in each of these
$(w+1)$ disjoint intervals. In particular, since  
\beqno
\left|
\Sigma_{X^L}^{-1}+\Sigma_{N^L(u^L)}^{-1}
\right|
&=&
\left(u_1-\tilde{g}(0)\right)
\prod_{j=1}^{L-1}\eta_j>0, 
\eeqno
we have $0<\tilde{g}(0)<u_1$. This implies 
that $u_1=\tilde{g}(\alpha)$ has a unique solution 
in the interval $(0,\eta_{\tilde{k}_1})$.
Hence, $(\ref{eqn:zaa})$ is proved. 
It remains to prove $(\ref{eqn:zaa2})$. 
Since $u_1=\tilde{g}(\alpha_1)=\tilde{g}(\alpha_{w+1})$, 
we have 
$$
\left.
\ba{rcl}
%u_1- \alpha_{\min}&=& 
u_1- \alpha_1&=&\ds \sum_{i=1}^{w}
\frac{\epsilon_{j_i}}{\eta_{\tilde{k}_i}-\alpha_1}\MgL{a}0\,,
\vspace{1mm}\\
%u_1- \alpha_{\max}&=&\ds 
u_1- \alpha_{w+1}&=&\ds \sum_{i=1}^{w}
\frac{\epsilon_{j_i}}{\eta_{\tilde{k}_i}-\alpha_{w+1}}\MsL{b}0\,.
\ea
\right\}
$$
Steps (a) and (b) follow from $(\ref{eqn:zaa})$. 
For convenience, the shape of $\tilde{g}(\alpha)$ is shown in 
Fig. \ref{fig:graphTiG}. 
Thus, the proof of  the part a) is completed. 
%Summarizing the above arguments, 
%we obtain $(\ref{eqn:zaaoaa})$.
The part b) is obvious 
from (\ref{eqn:za0a}). 
Finally, we show the part c). Taking the derivative of 
(\ref{eqn:eigeqzz}) 
with respect to $u_1$, we obtain 
$$
1=\tilde{g}^{\prime}(\alpha)\frac{{\rm d}\alpha}{{\rm d}u_1}=
\left(1+
\sum_{i=1}^{w}\frac{\epsilon_{j_i}}{(\alpha-\eta_{\tilde{k}_i})^2}
\right) 
\frac{{\rm d}\alpha}{{\rm d}u_1}
\,,
$$   
from which we obtain
$$
\frac{{\rm d}\alpha}{{\rm d}u_1}=
\left(1+\sum_{i=1}^{w}\frac{\epsilon_{j_i}}{(\alpha-\eta_{\tilde{k}_i})^2}
\right)^{-1}>0\,.
$$
Hence, every eigen value belonging to ${\cal E}_0$ is 
monotone increasing function of $u_1$. If the eigen value 
does not belong to ${\cal E}_0$, it does not depend on $u_1$. 
Thus, the part c) is proved.  
\hfill\IEEEQED

{\it Proof of Lemma \ref{lm:eigenlm}:} It suffices 
to prove the claim of Lemma \ref{lm:eigenlm} for 
$i=1$, that is,
\beqa
& & \alpha_{\max}\geq u_1 \geq \alpha_{\min}\,,
\label{eqn:zaNa}\\ 
& &\frac{\partial \alpha_k}{\partial u_1}\geq 0, \mbox{ for }k\in \Lambda,
\label{eqn:zaNb}
\\
& &\sum_{k=1}^L\frac{\partial \alpha_k}{\partial u_1}=1\,. 
\label{eqn:zaNc}
\eeqa 
Inequalities (\ref{eqn:zaNa}) and (\ref{eqn:zaNb}) follow from Proposition 
\ref{pro:pro1} parts a) and c), respectively. 
It remains to prove (\ref{eqn:zaNc}). 
Since for any matrix its trace is equal to 
the sum of its eigen values, we have   
\beq
\sum_{k=1}^L\alpha_k = \sum_{k=1}^Lu_k\,.
\label{eqn:sumeq} 
\eeq
Taking partial derivative of both sides of 
(\ref{eqn:sumeq}) with respect to $u_1$, 
we obtain (\ref{eqn:zaNc}).\hfill\IEEEQED

\newcommand{\Skip}{}
{%\footnotesize

}

\end{document}